\documentclass[11pt,a4paper]{article}
\pdfoutput=1

\usepackage{jheppub}

\usepackage{amsmath}
\usepackage{bbm}
\usepackage{float}
\usepackage{graphicx}	
\usepackage{hyperref}
\usepackage{mathtools}
\usepackage{cancel}

\usepackage[normalem]{ulem}

\usepackage{multirow}
\usepackage{rotating}
\usepackage{subcaption}

\def\lsim{\raise0.3ex\hbox{$\;<$\kern-0.75em\raise-1.1ex
\hbox{$\sim\;$}}}
\def\gsim{\raise0.3ex\hbox{$\;>$\kern-0.75em\raise-1.1ex
\hbox{$\sim\;$}}}

\def\thetitle{ 
Symmetry in neutrino oscillation in matter: \\ 
New picture and the $\nu$SM - non-unitarity interplay \\
 \vspace{- 6mm}
}
\title{\thetitle}
\hypersetup{pdftitle={\thetitle}}
\author{Hisakazu Minakata}
\affiliation{
Center for Neutrino Physics, Department of Physics, Virginia Tech, Blacksburg, Virginia 24061, USA \\
}

\emailAdd{hisakazu.minakata@gmail.com}
\date{\today}

\abstract{We update and summarize the present status of our understanding of the reparametrization symmetry with $i \leftrightarrow j$ state exchange in neutrino oscillation in matter. We introduce a systematic method called {\em ``Symmetry Finder''} (SF) to uncover such symmetries, demonstrate its efficient hunting capability, and examine their characteristic features. Apparently they have a local nature: The 1-2 and 1-3 state exchange symmetries exist at around the solar- and atmospheric-resonances, respectively, with the level-crossing states exchanged. However, this view is not supported, to date, in the globally valid Denton {\it et al.}~(DMP) perturbation theory, which possesses the 1-2 exchange symmetry but not the 1-3. It is probably due to lack of our understanding, and we find a clue for a larger symmetry structure than that we know. In the latter part of this article, we introduce non-unitarity, or unitarity violation (UV) into the $\nu$SM neutrino paradigm, a low-energy description of beyond $\nu$SM new physics at high (or low) scale. Based on the analyses of UV extended versions of the atmospheric-resonance and the DMP perturbation theories, we argue that the reparametrization symmetry has a diagnostics capability for the theory with the $\nu$SM and UV sectors. A speculation is given on the topological nature of the identity which determines the transformation property of the UV $\alpha$ parameters. 
} 

\newcommand{\Dmsqren}{\Delta m^2_{ \text{ren} }}

\begin{document} 

\maketitle

\section{Introduction}
\label{sec:introduction} 

Symmetry is one of the deepest subjects in physics. When one picks up a field theory textbook from bookshelf, say ref.~\cite{Itzykson:1980rh}, one finds the description of various symmetries, space-time symmetries, internal symmetries, $CP$, $T$ and $CPT$, discrete symmetries, symmetry in the hadron spectrum, $O(4)$ in the Coulomb problem, not talking about gauge symmetry for constructing the Standard Model (SM). Most likely, even the several big monographs would not be sufficient for full-coverage of the subjects because of its profound consequences and evolving nature. Fortunately, a set of beautiful lectures on symmetry in particle physics delivered in the last decades in the 20th century is left for us~\cite{Coleman:1985rnk}. 

In this paper, we discuss the reparametrization symmetry in neutrino oscillation in matter. It indeed has quite different character from those described in refs.~\cite{Itzykson:1980rh,Coleman:1985rnk}. Invariance under the reparametrization merely implies that there is another way of parametrizing the equivalent solution of the theory. Consequently, a general view on such symmetry would be that it might be useful, but no conceptually deep notion is likely to be involved. Recently, however, we have been accumulating new experiences about the reparametrization symmetry~\cite{Minakata:2021dqh,Minakata:2021goi,Minakata:2022zua}, which may introduce a new perspective on this view.
Therefore, in this paper we present our self-contained global picture of the symmetry in neutrino oscillation in matter, with a hope of bringing the subject to the readers' attention and for a new judgement. If successful, we could possibly overturn the above prejudice about the reparametrization symmetry. 

What does the symmetry look like in neutrino oscillation in matter? Let us give a simple and concrete example. In a perturbative framework valid at around the atmospheric resonance~\cite{Minakata:2015gra}, which will be dubbed as the ``helio-perturbation theory'' in this paper, it is noticed~\cite{Martinez-Soler:2019nhb} that the expression of the oscillation probability is invariant under the transformations 
\begin{eqnarray} 
&&
\lambda_{1} \leftrightarrow \lambda_{3},
\hspace{8mm} 
\cos \phi \rightarrow \mp \sin \phi, 
\hspace{8mm} 
\sin \phi \rightarrow \pm \cos \phi. 
\label{symmetry-IA-helioP}
\end{eqnarray}
The notations are such that $\phi$ is the $\theta_{13}$ in matter, and $\lambda_{1} > \lambda_{3}$ denote the two eigenvalues which participate the level crossing at around the atmospheric resonance in the inverted mass ordering~\cite{Minakata:2015gra}, see section~\ref{sec:formulation}.\footnote{
In the normal mass ordering $\lambda_{3} > \lambda_{2}$ are the two eigenvalues which have the level crossing~\cite{Minakata:2015gra}. }
The similar symmetry as in eq.~\eqref{symmetry-IA-helioP} but with replacing the 1-3 exchange by the 1-2 exchange using $\psi$ as the matter-dressed $\theta_{12}$ ($\theta_{12}$ in matter) was observed earlier in the Denton {\it et al.} (DMP) perturbation theory~\cite{Denton:2016wmg}. Precise meaning of the term ``matter-dressed $\theta_{12}$'' is explained after eq.~\eqref{cos-sin-2varphi}, and similarly $\theta_{13}$ in matter by eq.~\eqref{phi-def}. 

Recently we have developed a systematic way of finding the reparametrization symmetry in neutrino oscillation in matter, termed as {\em ``Symmetry Finder''} (SF)~\cite{Minakata:2021dqh,Minakata:2021goi,Minakata:2022zua}. We will review this machinery and its powerfulness, and try to show the readers where we are in our journey of uncovering and understanding the symmetry. It is interesting in its own right, serving for example for keeping consistency of the calculations of the observables, a ``bread and butter'' item but an important task for the theorists. 
Eventually we are going to suggest, in the active three neutrino framework extended to include unitarity violation (UV),\footnote{
We are aware that in physics literatures UV usually means ``ultraviolet''. But, in this paper UV is used as an abbreviation for ``unitarity violation'' or ``unitarity violating''. }
that the reparametrization symmetry distinguishes between the $\nu$SM (neutrino-mass-embedded SM) and the UV sectors of the theory, offering a useful tool for diagnosing such theories~\cite{Minakata:2022zua}. 
We hope that SF, a systematic approach, provides an efficient digging-out machinery for the symmetries in neutrino oscillation in matter and their deeper understanding. We believe that it follows the spirit of the early analyses on symmetries and strengthen their impacts~\cite{Fogli:1996nn,Fogli:2001wi,deGouvea:2000pqg,Altarelli:2010gt,Fogli:1996pv,Minakata:2001qm,Minakata:2010zn,Coloma:2016gei,deGouvea:2008nm,Zhou:2016luk}.

\subsection{Local character of the reparametrization symmetry}
\label{sec:local-character}

\begin{figure}[h!]
\begin{center}
\vspace{2mm}
\includegraphics[width=0.46\textwidth]{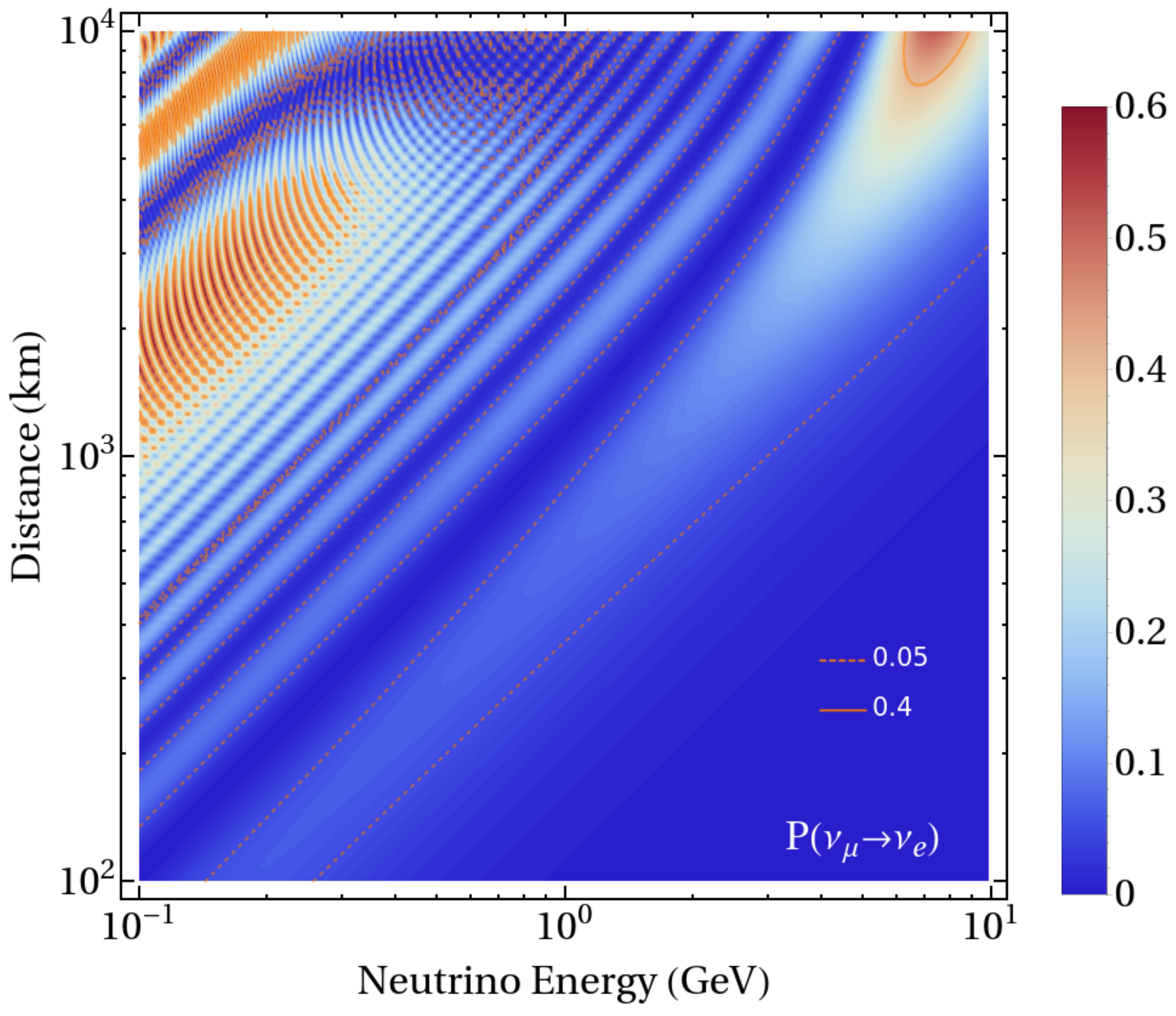}
\end{center}
\vspace{-3mm}
\caption{The equi-probability contour of $P(\nu_{\mu} \rightarrow \nu_{e})$ is presented~\cite{Minakata:2019gyw} in region of energy-baseline that covers the atmospheric neutrino observation by Super-Kamiokande. The two peaks are visible, the solar-scale ($E \sim 200$ MeV, $L \sim 2000$ km), and the atmospheric-scale ($E \sim 8$ GeV, $L \sim 10^{4}$ km) enhanced oscillations. The matter density is taken to be a constant, $\rho = 4.0~\text{g/cm}^3$, which gives only a bold approximation to the Earth matter density. 
} 
\label{fig:global-oscillation}
\end{figure}

To our current understanding, the reparametrization symmetry of neutrino oscillation takes different forms depending upon where we are, i.e., which regions of neutrino energy $E$, baseline $L$, and the matter density $\rho$ along the neutrino trajectory in the kinematical phase space. Therefore, let us first introduce the matter effect \cite{Wolfenstein:1977ue} and draw a global picture of neutrino oscillation in the earth matter environment. The matter potential will be defined in eq.~\eqref{matt-potential} in section~\ref{sec:3nu-unitarity}. 
In Fig.~\ref{fig:global-oscillation}, the equi-probability contour of $P(\nu_{\mu} \rightarrow \nu_{e})$ is presented~\cite{Minakata:2019gyw} in region of energy-baseline that roughly covers the Super-Kamiokande's atmospheric neutrino observation 0.1 GeV$\lsim E \lsim 10$ GeV, see Fig.~3 in ref.~\cite{Super-Kamiokande:2017yvm}. It also overlaps with the regions for all the ongoing and planned  long-baseline accelerator neutrino experiments. The two peaks are visible, the solar-scale ($E \sim 200$ MeV, $L \sim 2000$ km), and the atmospheric-scale ($E \sim 8$ GeV, $L \sim 10^{4}$ km) enhanced oscillations. For brevity we refer these respective regions as the solar-resonance and the atmospheric-resonance regions hereafter. 
In this article, the term ``resonance'' should be understood in this less strict sense than usual, see refs.~\cite{Wolfenstein:1977ue,Mikheyev:1985zog,Barger:1980tf,Smirnov:2016xzf}. 

Now, what we are telling the readers is that the reparametrization symmetry takes the different form around each peak. That is, 
\begin{itemize}
\item 
In the solar-resonance region the reparametrization symmetry of the 1-2 state exchange type exists, to be discussed in section~\ref{sec:Symmetry-SRP}.

\item 
In the atmospheric-resonance region the reparametrization symmetry of the 1-3 state exchange type exists~\cite{Minakata:2021goi}. 

\end{itemize}
\noindent
Since the symmetry in the framework with local validity in the solar-resonance region~\cite{Martinez-Soler:2019nhb} has never been investigated in the literature, we will fill the gap in this article. 

In fact, the features described in the above itemized statements appeal to our intuition. The 1-2 and 1-3 state level crossings, respectively, are the key to the solar-scale and the atmospheric-scale resonances, and they are the dominant players in these respective regions. The symmetry type specified by the exchanged states mentioned above just reflects the main players in each region. See e.g., refs.~\cite{Arafune:1996bt,Cervera:2000kp,Freund:2001pn,Akhmedov:2004ny} for the earlier versions of the atmospheric-resonance perturbation theory. Given the existence of the various versions, not to trigger any confusion, we discuss in this paper the particular version in ref.~\cite{Minakata:2015gra} under the name of ``helio-perturbation theory''\footnote{
The term ``helio-perturbation'', shorthand of the ``helio-terrestrial-ratio perturbation'', is invented because it perturbs the dominant effect of the atmospheric resonance by the small solar-scale effect of order $\sim \Delta m^2_{21} / \Delta m^2_{31} \approx 0.03$. }
to discuss the reparametrization symmetry in the atmospheric-resonance region. 

\subsection{Globally valid vs. locally valid frameworks} 
\label{sec:globally-valid} 

But, it turned out that the things are not so simple. Progress in perturbative treatment of neutrino oscillation in matter now allows us to have a limited number of the ``globally valid'' frameworks, the DMP~\cite{Denton:2016wmg} and Agarwalla {\it et al.} (AKT)~\cite{Agarwalla:2013tza} theories. By ``globally valid'' we mean that the framework is valid throughout the terrestrial region depicted in Fig.~\ref{fig:global-oscillation}.\footnote{
In fact, the region of validity of the globally valid frameworks is likely to extend to much higher energies which is explored e.g., by IceCube-DeepCore~\cite{IceCube:2020tka}. See the related discussions in ref.~\cite{Minakata:2021nii}. }
As opposed to the above mentioned ``locally valid'' theories, globally valid one is able to describe the both solar and atmospheric resonances. The secret for such greater capability is in usage of the Jacobi method. See ref.~\cite{Agarwalla:2013tza} for a concise exposition of the Jacobi method. 

Does the globally valid framework allow to formulate the reparametrization symmetry in the both solar- and atmospheric-resonance regions? So far the answer is No.\footnote{
While the present author is suspicious about this conclusion, it is the current status of our understanding of the reparametrization symmetry in neutrino oscillation in matter. A conjecture is given toward generalization of the SF formalism to accommodate much more generic reparametrization symmetry~\cite{Minakata:2022zua}. }
According to the result of ref.~\cite{Minakata:2021dqh}, only the reparametrization symmetry of the 1-2 state exchange type is obtained. Or, in other word, the better way of formulating SF such that the potential of the globally valid frameworks is fully utilized remains to be discovered. Remember, however, that the problem has been examined only in the DMP theory so far, and it is interesting to know how the problem looks like in the AKT perturbation theory~\cite{Agarwalla:2013tza}. 

Here is additional (not so pedagogical) comments on the globally valid vs. locally valid frameworks of neutrino oscillation. One may argue that the wider coverage indicates superiority of the globally valid framework over the local ones. Moreover, one can show that the DMP preserves the Naumov identity~\cite{Naumov:1991ju}, at least approximately~\cite{Minakata:2021goi}. This is a necessary condition that has to be satisfied for the globally valid framework, while the helio-perturbation theory does not support this property for a good reason~\cite{Minakata:2021goi}. Nonetheless, we would like to emphasize that it is only one side of the coin. From the viewpoint of our symmetry discussion, one to one correspondence between the resonant level crossing and the existing symmetry type is revealing and looks physically appealing. 

\subsection{Paper plan: Part I and II} 
\label{sec:plan} 

This paper has the two parts. Part I spans from sections~\ref{sec:introduction} to \ref{sec:Symmetry-SRP}, and Part II from sections~\ref{sec:Symmetry-UV} to \ref{sec:hamiltonian-symmetry}. In Part I we define the target of our discussion, the reparametrization symmetry of the state exchange type, and introduce the concept of  ``Symmetry Finder'' (SF) a systematic way of hunting the reparametrization symmetry, see section~\ref{sec:introducing-SF}. We briefly describe the 1-2 state exchange symmetry in the DMP perturbation theory~\cite{Minakata:2021dqh}, as a prototype of such symmetry we discuss in this article. 
In section~\ref{sec:SRP}, we review the solar-resonance perturbation (SRP) theory~\cite{Martinez-Soler:2019nhb} and introduce so called the $V$ matrix method~\cite{Minakata:1998bf}. In section~\ref{sec:Symmetry-SRP}, we formulate SF for the SRP theory and analyze the SF equation to obtain the 1-2 exchange symmetry. The SF treatment of the SRP theory has not been done before so that we are going to add something new in this subject. Our discussion will be pedagogical in most part of Part I, aiming at facilitating the readers' understanding of the subject. For this purpose we restrict ourselves into the $\nu$SM symmetry in Part I. 

In Part II we focus on the reparametrization symmetry in theory with non-unitary flavor mixing matrix, or UV. See refs.~\cite{Antusch:2006vwa,Escrihuela:2015wra,Blennow:2016jkn,Fong:2016yyh,Fong:2017gke}, for example, with more references coming later. As overviewed in section~\ref{sec:Symmetry-UV}, looking for new physics beyond the $\nu$SM is the vigorously pursued subject in particle physics, and non-unitarity is one of the promising ways for its low-energy description. We are interested in such a possibility that the symmetry can be used as a diagnostics tool for the theory with the $\nu$SM and UV sectors. For this purpose, we feel, non-unitarity would provide a useful testing ground for such possibility because its principle and the relation to the high- or low-energy new physics is relatively well defined~\cite{Antusch:2006vwa,Escrihuela:2015wra,Blennow:2016jkn,Fong:2016yyh,Fong:2017gke}. 
From sections~\ref{sec:helio-UV-P} to \ref{sec:hamiltonian-symmetry} we give a self-contained treatment of the 1-3 state exchange symmetry in the helio-UV perturbation theory~\cite{Martinez-Soler:2018lcy}, a UV-extension of the helio-perturbation theory~\cite{Minakata:2015gra}. Since the SF treatment of the helio-UV perturbation theory has never been done in the literature, the derivation and discussion of such symmetry in this theory is all new. 

In summary, our goals and the motivating force in this paper are: 
\begin{itemize}

\item 
Part I: We summarize the current status of our understanding of the reparametrization symmetry of the state exchange type in neutrino oscillation in matter. To the best of our knowledge, no one expected that so many symmetries are hidden in the DMP and the helio-perturbation theories~\cite{Minakata:2021dqh,Minakata:2021goi}. A tantalizing question is: What is the nature and implications of the symmetry? 

\item 
Part II: We introduce non-unitarity and analyze the symmetry in the UV-extended frameworks of neutrino evolution. We realize possible utility of the symmetry as a diagnostics tool for theories with the $\nu$SM and UV sectors. A part of the reparametrization symmetry acts only on the $\nu$SM variables, not UV ones, distinguishing between the two sectors of the theory~\cite{Minakata:2022zua}. Can we observe the whole picture of this? 

\end{itemize}

\section{Introducing Symmetry Finder}
\label{sec:introducing-SF} 

Is there a systematic way of uncovering reparametrization symmetry in neutrino oscillation? Our answer is {\em Yes}: Symmetry Finder (SF) does the job in vacuum~\cite{Parke:2018shx} and in matter~\cite{Minakata:2021dqh,Minakata:2021goi,Minakata:2022zua}''. 

Let us consider that the expression of the flavor basis state (i.e., wave function) $\nu$ in terms of the mass eigenstate $\bar{\nu}$ in vacuum or in matter in the following two different ways, 
\begin{eqnarray} 
&&
\nu = U (\theta_{23}, \theta_{13}, \theta_{12}, \delta) \bar{\nu} 
= U (\theta_{23}^{\prime}, \theta_{13}^{\prime}, \theta_{12}^{\prime}, \delta^{\prime}) \bar{\nu}^{\prime}, 
\label{SF-eq-general}
\end{eqnarray}
where the quantities with ``prime'' imply the transformed ones, and $\bar{\nu}^{\prime}$ may involve eigenstate exchanges and/or rephasing of the wave functions. 
If it is in matter the mixing angles and the CP phase can be elevated to the matter-dressed variables. Since the SF equation represents the same flavor state by the two different sets of the physical parameters, it implies a symmetry. 

\subsection{The PDG and SOL conventions of the flavor mixing matrix} 
\label{sec:PDG-SOL} 

To discuss the 1-2 state exchange symmetries in vacuum and in matter, which we will do in Part I, it is convenient to introduce the flavor mixing matrix $U \equiv U_{\text{\tiny MNS}}$~\cite{Maki:1962mu} in a convention called ``SOL''~\cite{Parke:2018shx,Martinez-Soler:2018lcy}, which is slightly different from the usual Particle Data Group (PDG) convention~\cite{Zyla:2020zbs}, 
\begin{eqnarray} 
&&
\hspace{-5mm}
U_{\text{\tiny SOL}} 
\equiv 
\left[
\begin{array}{ccc}
1 & 0 &  0  \\
0 & e^{ - i \delta} & 0 \\
0 & 0 & e^{ - i \delta} \\
\end{array}
\right] 
U_{\text{\tiny PDG}} 
\left[
\begin{array}{ccc}
1 & 0 &  0  \\
0 & e^{ i \delta} & 0 \\
0 & 0 & e^{ i \delta} \\
\end{array}
\right] 
= 
\left[
\begin{array}{ccc}
1 & 0 &  0  \\
0 & c_{23} & s_{23} \\
0 & - s_{23} & c_{23} \\
\end{array}
\right] 
\left[
\begin{array}{ccc}
c_{13}  & 0 & s_{13} \\
0 & 1 & 0 \\
- s_{13} & 0 & c_{13} \\
\end{array}
\right] 
\left[
\begin{array}{ccc}
c_{12} & s_{12} e^{ i \delta}  &  0  \\
- s_{12} e^{- i \delta} & c_{12} & 0 \\
0 & 0 & 1 \\
\end{array}
\right] 
\nonumber \\
&\equiv&
U_{23} (\theta_{23}) U_{13} (\theta_{13}) U_{12} (\theta_{12}, \delta). 
\label{SOL-def} 
\end{eqnarray}
In this paper, hereafter, we use the abbreviated notations $c_{ij} \equiv \cos \theta_{ij}$, $s_{ij} \equiv \sin \theta_{ij}$ etc., where $ij = 12, 13, 23$. 
In eq.~\eqref{SOL-def} $\delta$ denotes the lepton analogue of the quark Kobayashi-Maskawa (KM) CP violating phase~\cite{Kobayashi:1973fv}, and the second line defines the notations for the three rotation matrices in the SOL convention. $U_{\text{\tiny PDG}}$ denotes the $U$ matrix in the PDG convention~\cite{Zyla:2020zbs} 
\begin{eqnarray} 
U_{\text{\tiny PDG}} 
&=& 
\left[
\begin{array}{ccc}
1 & 0 &  0  \\
0 & c_{23} & s_{23} \\
0 & - s_{23} & c_{23} \\
\end{array}
\right] 
\left[
\begin{array}{ccc}
c_{13}  & 0 & s_{13} e^{- i \delta} \\
0 & 1 & 0 \\
- s_{13} e^{ i \delta} & 0 & c_{13}  \\
\end{array}
\right] 
\left[
\begin{array}{ccc}
c_{12} & s_{12}  &  0  \\
- s_{12} & c_{12} & 0 \\
0 & 0 & 1 \\
\end{array}
\right] 
\nonumber \\
&\equiv&
U_{23} (\theta_{23}) U_{13} (\theta_{13}, \delta) U_{12} (\theta_{12}), 
\label{MNS-PDG}
\end{eqnarray}
with the second line the rotation matrices in the PDG convention. 
The reason for our terminology of $U_{\text{\tiny SOL}}$ in eq.~\eqref{SOL-def} is because the CP phase factor $e^{ \pm i \delta}$ is attached to (sine of) the ``solar angle'' $\theta_{12}$ in $U_{\text{\tiny SOL}}$, respectively. Whereas in $U_{\text{\tiny PDG}}$, $e^{ \pm i \delta}$ is attached to $s_{13}$. Notice that the convention change from the PDG to SOL conventions does not alter the oscillation probability because the rephasing factors in eq.~\eqref{SOL-def} can be absorbed into the neutrino wave functions, leaving no effect in the observables. 
In Part II we will use the PDG convention for convenience to treat the 1-3 exchange symmetry in the helio-perturbation theory~\cite{Minakata:2021goi} and its UV extension. 

\subsection{Symmetry Finder (SF) in vacuum}
\label{sec:SF-vacuum} 

The idea of SF has a simple realization in vacuum where the flavor eigenstate $\nu$ is related to the mass eigenstate $\bar{\nu}$ using the SOL convention $U$ matrix~\eqref{SOL-def} 
\begin{eqnarray} 
&& 
\nu 
= U \bar{\nu} 
= 
U_{23} (\theta_{23}) U_{13} (\theta_{13}) 
U_{12} (\theta_{12}, \delta) \bar{\nu}. 
\label{flavor-mass-vacuum}
\end{eqnarray}
Then, one can easily prove the relation 
\begin{eqnarray} 
&& 
U_{12} ( \theta_{12}, \delta ) 
\left[
\begin{array}{c}
\nu_{1} \\
\nu_{2} \\
\nu_{3} \\
\end{array}
\right] 
= U_{12} \left( \theta_{12} + \frac{\pi}{2}, \delta \right) 
\left[
\begin{array}{c}
- e^{ i \delta } \nu_{2}  \\
e^{ - i \delta } \nu_{1} \\
\nu_{3} \\
\end{array}
\right] 
= 
U_{12} \left( \frac{\pi}{2} - \theta_{12}, \delta \pm \pi \right) 
\left[
\begin{array}{c}
e^{ i \delta } \nu_{2}  \\
- e^{ - i \delta } \nu_{1} \\
\nu_{3} \\
\end{array}
\right].
\nonumber \\
\label{symmetry-finder-vacuum}
\end{eqnarray}
Hereafter, the state $\nu_{1}$ denotes the one with the largest $\nu_{e}$ component. The state $\nu_{2}$ is the one that is separated from the state $\nu_{1}$ by the mass squared difference \\ 
$m^2_{2} - m^2_{1} \equiv \Delta m^2_{ \text{solar} } \simeq 7.5 \times 10^{-5}$ eV$^2 > 0$. 

As we stated above, the relation~\eqref{symmetry-finder-vacuum} implies symmetry~\cite{Parke:2018shx}. The first equality means that use of $\theta_{12}^{\prime} = \theta_{12} + \frac{\pi}{2}$ and the exchanged (and rephased) mass eigenstates $1 \leftrightarrow 2$ produces the same oscillation probability. Since rephasing does not affect the observables, the first equality in eq.~\eqref{symmetry-finder-vacuum} implies the 1-2 exchange symmetry under the transformation 
\begin{eqnarray} 
&& 
\text{Symmetry IA-vacuum:}
\hspace{6mm}
m^2_{1} \leftrightarrow m^2_{2}, 
\hspace{8mm}
c_{12} \rightarrow - s_{12},
\hspace{8mm}
s_{12} \rightarrow c_{12}. 
\label{Symmetry-IA-vacuum}
\end{eqnarray} 
Existence of an alternative choice, $c_{12} \rightarrow s_{12}$ and $s_{12} \rightarrow - c_{12}$ ($\theta_{12} \rightarrow \theta_{12} - \frac{\pi}{2}$), should be understood. Similarly, the second equality in ~\eqref{symmetry-finder-vacuum} implies the symmetry of the probability under the transformation 
\begin{eqnarray} 
&&
\text{Symmetry IB-vacuum:}
\hspace{6mm}
m^2_{1} \leftrightarrow m^2_{2}, 
\hspace{8mm}
c_{12} \leftrightarrow s_{12}, 
\hspace{8mm}
\delta \rightarrow \delta \pm \pi.
\label{Symmetry-IB-vacuum}
\end{eqnarray} 

\subsection{Symmetry Finder in matter}
\label{sec:SF-matter} 

In the exact diagonalization scheme of Zaglauer and Schwarzer (ZS)~\cite{Zaglauer:1988gz}, the Hamiltonian is formally identical to that in vacuum apart from replacements of the mixing angles $\theta_{ij}$ to the matter-dressed ones $\tilde{\theta}_{ij}$, $\delta \rightarrow \tilde{\delta}$, and the eigenvalues $m^2_{i} \rightarrow \lambda_{i}$ ($i,j=1,2,3$). Therefore, the symmetries~\eqref{Symmetry-IA-vacuum} and \eqref{Symmetry-IB-vacuum} are easily elevated to Symmetry IA-ZS and IB-ZS, with the fully matter-dressed variables~\cite{Parke:2018shx,Minakata:2021dqh}. 

Let us move into the more manageable approximate frameworks. Within the $\nu$SM, so far, the following two types of the reparametrization symmetry are identified and analyzed. 
\begin{itemize}

\item 
Eight reparametrization symmetries of the 1-2 state exchange type in DMP~\cite{Minakata:2021dqh}.

\item 
Sixteen reparametrization symmetries of the 1-3 state exchange type in the helio-perturbation theory~\cite{Minakata:2021goi}. 

\end{itemize}
The list will be enriched after section~\ref{sec:Symmetry-SRP} below by 
\begin{itemize}

\item 
Eight reparametrization symmetries of the 1-2 state exchange type in the solar-resonance perturbation (SRP) theory. 

\end{itemize}
\noindent 
Notice that the obtained symmetries are so numerous, including the ZS symmetries above, that it cannot occur by accident. A certain nontrivial structure must be behind the appearance of the so many symmetries. It then suggests that existence of the reparametrization symmetry is universal in neutrino oscillation in matter. We shall introduce the various symmetries that appear in several different theories in a step by step manner and try to understand their structure in this article. 

\subsection{Reparametrization symmetry in DMP}
\label{sec:DMP-symmetry} 

To formulate SF in matter we need the the expression of the flavor basis state $\nu$ in terms of the mass eigenstate $\bar{\nu}$ analogous to eq.~\eqref{flavor-mass-vacuum} in vacuum. It can be computed with so called the $V$ matrix method~\cite{Minakata:1998bf}, which will be explained in section~\ref{sec:V-matrix} for the SRP theory. 
Here we just want to show the ``atmosphere'' of how SF works to find symmetry in matter. Hence, we just quote the expression of the flavor basis state in terms of the mass eigenstate to first order in the DMP perturbation theory computed in ref.~\cite{Minakata:2021dqh}: 
\begin{eqnarray} 
\left[
\begin{array}{c}
\nu_{e} \\
\nu_{\mu} \\
\nu_{\tau} \\
\end{array}
\right] 
&=& 
U_{23} (\theta_{23}) U_{13} (\phi) U_{12} (\psi, \delta)
\nonumber \\
&\times&
\left\{ 
1 + 
\epsilon c_{12} s_{12} \sin ( \phi - \theta_{13} ) 
\left[
\begin{array}{ccc}
0 & 0 & - s_\psi 
\frac{ \Delta m^2_{ \text{ren} } }{ \lambda_{3} - \lambda_{1} } \\
0 & 0 & c_\psi e^{ - i \delta} 
\frac{ \Delta m^2_{ \text{ren} } }{ \lambda_{3} - \lambda_{2} } \\
s_\psi 
\frac{ \Delta m^2_{ \text{ren} } }{ \lambda_{3} - \lambda_{1} } & 
- c_\psi e^{ i \delta} 
\frac{ \Delta m^2_{ \text{ren} } }{ \lambda_{3} - \lambda_{2} } & 0 \\
\end{array}
\right]
\right\} 
\left[
\begin{array}{c}
\nu_{1} \\
\nu_{2} \\
\nu_{3} \\
\end{array}
\right], ~~~~
\label{SFeq-DMP-1st}
\end{eqnarray}
where $\psi$ and $\phi$ denote, respectively, $\theta_{12}$ and $\theta_{13}$ in matter. $c_{\psi}$ and $s_{\psi}$ are shorthand notations for $\cos \psi$ and $\sin \psi$, respectively. $\epsilon$ is the unique expansion parameter in the DMP perturbation theory and is defined as 
\begin{eqnarray} 
&&
\epsilon \equiv \frac{ \Delta m^2_{21} }{ \Delta m^2_{ \text{ren} } }, 
\hspace{10mm}
\Delta m^2_{ \text{ren} } \equiv \Delta m^2_{31} - s^2_{12} \Delta m^2_{21},
\label{epsilon-Dm2-ren-def}
\end{eqnarray}
where $\Delta m^2_{ \text{ren} }$ is the ``renormalized'' atmospheric $\Delta m^2$ used in ref.~\cite{Minakata:2015gra}. 
The same quantity is known as the effective $\Delta m^2_{ \text{ee} }$ in the $\nu_{e} \rightarrow \nu_{e}$ channel in vacuum~\cite{Nunokawa:2005nx}. While we prefer usage of $\Delta m^2_{ \text{ren} }$ in the context of the present paper, the question of which symbol should be appropriate to use here is under debate~\cite{Minakata:2015gra}. The authors of ref.~\cite{Denton:2021vtf} make the choice alternative to ours. 

With eq.~\eqref{SFeq-DMP-1st} we write down the equation similar to eq.~\eqref{symmetry-finder-vacuum} in vacuum. The added first-order structure in eq.~\eqref{SFeq-DMP-1st} leads to a proliferation of the reparametrization symmetries, the eight DMP symmetries~\cite{Minakata:2021dqh}, as tabulated in Table~\ref{tab:DMP-UV-symmetry}. To see how the SF equation is actually formulated and solved, please wait until sections~\ref{sec:SF-solRP} and \ref{sec:SRP-1st-2nd} in which the SRP theory is treated. 

Table~\ref{tab:DMP-UV-symmetry} must be used with care. Here we must focus on the first three columns of Table~\ref{tab:DMP-UV-symmetry} in which the informations of the $\nu$SM DMP symmetry are tabulated~\cite{Minakata:2021dqh}. The fourth column is added to display the $\widetilde{\alpha}$ parameter transformation property for the DMP-UV perturbation theory~\cite{Minakata:2022zua}, a UV-extended version of the DMP theory~\cite{Minakata:2021nii} to be discussed later. The $\alpha$ parametrization~\cite{Escrihuela:2015wra} will be used to describe non-unitary mixing matrix, as defined in eq.~\eqref{alpha-matrix-def} in section~\ref{sec:3nu-non-unitarity}, and $\widetilde{\alpha}$ denotes the $\alpha$ parameters in the SOL convention. For the definition of $\widetilde{\alpha}$, see eq.~\eqref{alpha-SOL-def} in Appendix~\ref{sec:3-conventions}. 
The fourth column will be useful for comparison with the case of UV extended helio-perturbation theory, which will be discussed in sections~\ref{sec:SF-helio-UV} and \ref{sec:SF-solution-helioP-UV}. 
%
\begin{table}[h!]
\vglue 0.2cm
\begin{center}
\caption{Summary of the reparametrization symmetries of the 1-2 state exchange type in the DMP and DMP-UV perturbation theories. The column ``Type'' shows the symmetry type. For the $\nu$SM DMP, look at the first three columns only:~The symmetry denoted as ``X'' in the Type column is referred to as ``Symmetry X-DMP'' in the text.  For the DMP-UV perturbation theory, the fourth column must be included in addition to the first three columns to show the $\widetilde{\alpha}$ parameters' transformation, and the symmetry is denoted as ``Symmetry X-DMP-UV''. 
}
\label{tab:DMP-UV-symmetry}
\vglue 0.2cm
\begin{tabular}{c|c|c|c}
\hline 
Type & 
Vacuum parameter transf. & 
Matter parameter transf. & 
UV parameter transf. 
\\
\hline 
\hline 
IA & 
none & 
$\lambda_{1} \leftrightarrow \lambda_{2}$, 
$c_{\psi} \rightarrow \mp s_{\psi}$, 
$s_{\psi} \rightarrow \pm c_{\psi}$ &
none 
\\
\hline 
IB & 
$\theta_{12} \rightarrow - \theta_{12}$, 
$\delta \rightarrow \delta + \pi$. & 
$\lambda_{1} \leftrightarrow \lambda_{2}$, 
$c_{\psi} \rightarrow \pm s_{\psi}$, 
$s_{\psi} \rightarrow \pm c_{\psi}$ & 
none
\\
\hline
IIA & 
$\theta_{23} \rightarrow - \theta_{23}$, 
$\theta_{12} \rightarrow - \theta_{12}$. & 
$\lambda_{1} \leftrightarrow \lambda_{2}$, 
$c_{\psi} \rightarrow \pm s_{\psi}$, 
$s_{\psi} \rightarrow \pm c_{\psi}$ & 
$\widetilde{\alpha}_{\mu e} \rightarrow - \widetilde{\alpha}_{\mu e}$, 
$\widetilde{\alpha}_{\tau \mu} \rightarrow - \widetilde{\alpha}_{\tau \mu}$ 
\\
\hline 
IIB & 
$\theta_{23} \rightarrow - \theta_{23}$, 
$\delta \rightarrow \delta + \pi$. & 
$\lambda_{1} \leftrightarrow \lambda_{2}$, 
$c_{\psi} \rightarrow \mp s_{\psi}$, 
$s_{\psi} \rightarrow \pm c_{\psi}$ &
same as IIA
\\
\hline 
IIIA & 
$\theta_{13} \rightarrow - \theta_{13}$, 
$\theta_{12} \rightarrow - \theta_{12}$. & 
$\lambda_{1} \leftrightarrow \lambda_{2}$, 
$\phi \rightarrow - \phi$ & 
$\widetilde{\alpha}_{\mu e} \rightarrow - \widetilde{\alpha}_{\mu e}$, 
$\widetilde{\alpha}_{\tau e} \rightarrow - \widetilde{\alpha}_{\tau e}$
\\ 
 & & 
$c_{\psi} \rightarrow \pm s_{\psi}$, 
$s_{\psi} \rightarrow \pm c_{\psi}$ & 
\\
\hline 
IIIB & 
$\theta_{13} \rightarrow - \theta_{13}$, 
$\delta \rightarrow \delta + \pi$. & 
$\lambda_{1} \leftrightarrow \lambda_{2}$, 
$\phi \rightarrow - \phi$ & 
same as IIIA
\\ 
 & & 
$c_{\psi} \rightarrow \mp s_{\psi}$, 
$s_{\psi} \rightarrow \pm c_{\psi}$ &
\\
\hline 
IVA & 
$\theta_{23} \rightarrow - \theta_{23}$, 
$\theta_{13} \rightarrow - \theta_{13}$. & 
$\lambda_{1} \leftrightarrow \lambda_{2}$, 
$\phi \rightarrow - \phi$ & 
$\widetilde{\alpha}_{\tau e} \rightarrow - \widetilde{\alpha}_{\tau e}$, 
$\widetilde{\alpha}_{\tau \mu} \rightarrow - \widetilde{\alpha}_{\tau \mu}$ 
\\ 
 & & 
$c_{\psi} \rightarrow \mp s_{\psi}$, 
$s_{\psi} \rightarrow \pm c_{\psi}$ &
\\
\hline 
IVB & 
$\theta_{23} \rightarrow - \theta_{23}$, 
$\theta_{13} \rightarrow - \theta_{13}$, & 
$\lambda_{1} \leftrightarrow \lambda_{2}$, 
$\phi \rightarrow - \phi$ &
same as IVA
\\ 
 &
$\theta_{12} \rightarrow - \theta_{12}$, $\delta \rightarrow \delta + \pi$. 
 &
$c_{\psi} \rightarrow \pm s_{\psi}$, $s_{\psi} \rightarrow \pm c_{\psi}$ &
\\
\hline 
\end{tabular}
\end{center}
\vglue -0.4cm 
\end{table}

Readers may be anxious to know how Table~\ref{tab:DMP-UV-symmetry} is obtained. If it is a burning question for a reader he/she can go to ref.~\cite{Minakata:2021dqh} to reproduce the results in the first three columns, or ref.~\cite{Minakata:2022zua} for all the four columns. But, instead, we move on to the SRP theory to see the new symmetry results, where we meet the very similar structure as DMP. As our intuition told us in section~\ref{sec:introduction}, the theory will have the symmetries of the 1-2 state exchange type. 

\section{Solar-resonance perturbation theory} 
\label{sec:SRP} 

The solar-resonance perturbation (SRP) theory~\cite{Martinez-Soler:2019nhb}, one of the locally-valid theories, aims at describing physics around the solar-scale resonance. In this section, we briefly review the SRP theory toward investigation of the reparametrization symmetry in the theory. 

\subsection{Three active-neutrino system with unitary flavor mixing matrix}
\label{sec:3nu-unitarity} 
 
We start by defining the standard three neutrino evolution system in matter. 
It is defined by the Schr\"odinger equation in the vacuum mass eigenstate basis, the ``check basis'', 
\begin{eqnarray}
i \frac{d}{dx} \check{\nu} = 
\frac{1}{2E} 
\left\{  
\left[
\begin{array}{ccc}
0 & 0 & 0 \\
0 & \Delta m^2_{21} & 0 \\
0 & 0 & \Delta m^2_{31} \\
\end{array}
\right] + 
U^{\dagger} \left[
\begin{array}{ccc}
a - b & 0 & 0 \\
0 & -b & 0 \\
0 & 0 & -b \\
\end{array}
\right] U 
\right\} 
\check{\nu} 
\equiv 
\check{H} \check{\nu}. 
\label{evolution-check-basis} 
\end{eqnarray}
In eq.~\eqref{evolution-check-basis}, which defines the check basis Hamiltonian $\check{H}$, $U$ denotes the $3 \times 3$ unitary flavor mixing matrix which relates the flavor-basis neutrino state $\nu$ to the vacuum mass eigenstates as 
\begin{eqnarray}
\nu_{\alpha} = U_{\alpha i} \check{\nu}_{i}. 
\label{U-def}
\end{eqnarray}
Hereafter, the subscript Greek indices $\alpha$, $\beta$, or $\gamma$ run over $e, \mu, \tau$, and the Latin indices $i$, $j$ run over the mass eigenstate indices $1,2,$ and $3$. $E$ is neutrino energy and $\Delta m^2_{ji} \equiv m^2_{j} - m^2_{i}$. The usual phase redefinition of neutrino wave function is done to leave only the mass squared differences. Notice, however, that doing or undoing this phase redefinition does not affect our symmetry discussion in this article. 

The functions $a(x)$ and $b(x)$ in eq.~(\ref{evolution-check-basis}) denote the
Wolfenstein matter potentials~\cite{Wolfenstein:1977ue} due to charged current (CC) and neutral current (NC) reactions, respectively, 
\begin{eqnarray} 
a(x) &=&  
2 \sqrt{2} G_F N_e E \approx 1.52 \times 10^{-4} \left( \frac{Y_e \rho}{\rm g\,cm^{-3}} \right) \left( \frac{E}{\rm GeV} \right) {\rm eV}^2, 
\nonumber \\
b(x) &=& \sqrt{2} G_F N_n E = \frac{1}{2} \left( \frac{N_n}{N_e} \right) a, 
\label{matt-potential}
\end{eqnarray}
where $G_F$ is the Fermi constant. 
$N_e$ and $N_n$ are the electron and neutron number densities in matter. $\rho$ and $Y_e$ denote, respectively, the matter density and number of electrons per nucleon in matter. These quantities except for $G_F$ are, in principle, position dependent. Until reaching section~\ref{sec:hamiltonian-symmetry}, however, we take the uniform matter density approximation. 

In the $\nu$SM unitary three-neutrino system, the NC potential $b (x)$ does not affect the neutrino flavor change, because it comes as the unit matrix in flavor space. But, it is included in eq.~\eqref{evolution-check-basis} for use in our discussion of the system with non-unitary in section~\ref{sec:helio-UV-P} in Part II, and to show the relationship between the NC and the CC matter potentials. In discussions in Part I we simply set $b (x)=0$. 

\subsection{Region of validity of the solar-resonance perturbation (SRP) theory}
\label{sec:region-SRP} 

The SRP theory~\cite{Martinez-Soler:2019nhb} aims at describing physics around the solar-scale enhancement, or the resonance. See e.g., refs.~\cite{Peres:2003wd,Peres:2009xe,Akhmedov:2008qt,Razzaque:2014vba} for the pioneering discussions on physics in this region. 
Given the formula 
\begin{eqnarray}
\frac{ \Delta m^2_{21} L}{4 E} 
&=&
0.953
\left(\frac{\Delta m^2_{21}}{7.5 \times 10^{-5}\mbox{eV}^2}\right)
\left(\frac{L}{1000 \mbox{km}}\right)
\left(\frac{E}{100 \mbox{MeV}}\right)^{-1}, 
\label{kinematic1}
\end{eqnarray}
the SRP theory will be valid in a region around neutrino energy $E=( 1 - 5 ) \times 100$ MeV and baseline $L= ( 1 - 10 ) \times 1000$ km. See Fig.~\ref{fig:global-oscillation}. In this region, the matter potential $a$ defined in eq.~\eqref{matt-potential} is comparable in size to the vacuum effect represented by $\Delta m^2_{21}$, 
\begin{eqnarray} 
r_{a} \equiv 
\frac{ a }{ \Delta m^2_{21} } 
&=& 
%
0.609 
\left(\frac{ \Delta m^2_{21} }{ 7.5 \times 10^{-5}~\mbox{eV}^2}\right)^{-1}
\left(\frac{\rho}{3.0 \,\text{g/cm}^3}\right) \left(\frac{E}{200~\mbox{MeV}}\right) 
\sim \mathcal{O} (1). 
\label{a/Dm2solar}
\end{eqnarray}
Therefore, this perturbative framework must be able to describe the solar-scale resonance~\cite{Wolfenstein:1977ue,Mikheyev:1985zog,Barger:1980tf}. 

\subsection{Solar-resonance perturbation theory in brief}
\label{sec:SRP-brief} 

For convenience in discussion of the 1-2 exchange symmetry, we use the SOL convention of the $U$ matrix~\eqref{SOL-def} to construct the SRP theory, as in ref.~\cite{Martinez-Soler:2019noy}. 
We transform to the tilde basis $\widetilde{\nu} = U_{12} (\theta_{12}, \delta) \check{\nu} = U_{13}^{\dagger} (\theta_{13}) U_{23}^{\dagger} (\theta_{23}) \nu$ with the Hamiltonian $\widetilde{H} = U_{12} (\theta_{12}, \delta) \check{H} U_{12}^{\dagger} (\theta_{12}, \delta)$. We decompose $\widetilde{H}$ as 
\begin{eqnarray} 
\widetilde{H} 
&=&
\frac{1}{2E}
\left[
\begin{array}{ccc}
s^2_{12} \Delta m^2_{21} + c^2_{13} a & 
c_{12} s_{12} e^{ i \delta} \Delta m^2_{21} & 
0 \\
c_{12} s_{12} e^{- i \delta} \Delta m^2_{21} & 
c^2_{12} \Delta m^2_{21} & 0 \\
0 & 0 & \Delta m^2_{31} + s^2_{13} a \\
\end{array}
\right]
+ 
\frac{1}{2E}
\left[
\begin{array}{ccc}
0 & 0 & c_{13} s_{13} a \\
0 & 0 & 0 \\
c_{13} s_{13} a & 0 & 0 \\
\end{array}
\right] 
\nonumber \\
&\equiv& 
\widetilde{H}_{0} + \widetilde{H}_{1}, 
\label{tilde-H-SRP}
\end{eqnarray}
where we have defined $\widetilde{H}_{0}$ ($\widetilde{H}_{1}$) as the first (second) term in eq.~\eqref{tilde-H-SRP}. We then transform to the ``hat basis'' 
\begin{eqnarray} 
\hat{\nu} 
= U_{12}^{\dagger} (\varphi, \delta) \widetilde{\nu} 
= U_{12}^{\dagger} (\varphi, \delta) U_{13}^{\dagger} (\theta_{13}) U_{23}^{\dagger} (\theta_{23}) \nu, 
\label{hat-basis}
\end{eqnarray}
%
%
where $U_{12} (\varphi, \delta)$ is parametrized as 
\begin{eqnarray} 
U_{12} (\varphi, \delta) = 
\left[
\begin{array}{ccc}
\cos \varphi & e^{ i \delta} \sin \varphi & 0 \\
- e^{ - i \delta} \sin \varphi & \cos \varphi & 0 \\
0 & 0 & 1 \\
\end{array}
\right], 
\label{U-varphi-def}
\end{eqnarray}
and is determined so that $\hat{H}_{0} = U_{12}^{\dagger} (\varphi, \delta) \widetilde{H}_{0} U_{12} (\varphi, \delta)$ is diagonal. 
The condition entails 
\begin{eqnarray} 
\cos 2 \varphi &=& 
\frac{ \cos 2\theta_{12} - c^2_{13} r_{a} }
{ \sqrt{ \left( \cos 2\theta_{12} - c^2_{13} r_{a} \right)^2 +  \sin^2 2\theta_{12} } }, 
\nonumber \\
\sin 2 \varphi &=& 
\frac{ \sin 2\theta_{12} }
{ \sqrt{ \left( \cos 2\theta_{12} - c^2_{13} r_{a} \right)^2 +  \sin^2 2\theta_{12} } }, 
\label{cos-sin-2varphi}
\end{eqnarray}
where $r_{a}$ is defined in eq.~\eqref{a/Dm2solar}. This equation defines the matter-dressed $\theta_{12}$, the effective mixing angle which governs the 1-2 space rotation in matter. 

To organize perturbative expansion in an intelligent way, we decompose the hat-basis Hamiltonian $\hat{H}$ into the zeroth-order and the first-order terms, $\hat{H}_{0}$ and $\hat{H}_{1}$ respectively, as $\hat{H} = \hat{H}_{0} + \hat{H}_{1}$, where 
\begin{eqnarray} 
&& \hat{H}_{0} 
= 
\frac{1}{2E}
\left[
\begin{array}{ccc}
\lambda_{1} & 0 & 0 \\
0 & \lambda_{2} & 0 \\
0 & 0 & \lambda_{3} \\
\end{array}
\right], 
\hspace{6mm}
\hat{H}_{1} 
= 
\frac{1}{2E}
\left[
\begin{array}{ccc}
0 & 0 & c_{\varphi} c_{13} s_{13} a \\
0 & 0 & s_{\varphi} c_{13} s_{13} e^{ - i \delta} a \\
c_{\varphi} c_{13} s_{13} a & s_{\varphi} c_{13} s_{13} e^{ i \delta} a & 0 \\
\end{array}
\right], ~~~~
\label{hat-H-0th-1st-SRP}
\end{eqnarray}
where 
$c_{\varphi} \equiv \cos \varphi$ and $s_{\varphi} \equiv \sin \varphi$. The zeroth-order eigenvalues are given by 
\begin{eqnarray} 
\lambda_{1} 
&=& 
\sin^2 \left( \varphi - \theta_{12} \right) \Delta m^2_{21} + \cos^2 \varphi c^2_{13} a, 
\nonumber \\
\lambda_{2} 
&=&
\cos^2 \left( \varphi - \theta_{12} \right) \Delta m^2_{21} + \sin^2 \varphi c^2_{13} a, 
\nonumber \\
\lambda_{3} 
&=& 
\Delta m^2_{31} + s^2_{13} a. 
\label{eigenvalues}
\end{eqnarray}
The SRP theory is defined as the perturbation theory with the unperturbed Hamiltonian $\hat{H}_{0}$ which is perturbed by the first-order Hamiltonian 
$\hat{H}_{1}$. For more about the unique feature of the SRP theory, see section~\ref{sec:exp-parameter}. 

\subsection{$V$ matrix method} 
\label{sec:V-matrix}

The formulation of Symmetry Finder (SF)~\cite{Minakata:2021dqh,Minakata:2021goi,Minakata:2022zua} heavily relies on so called the $V$ matrix method~\cite{Minakata:1998bf}. Therefore, we start from exposition of the method. The $V$ matrix method stands also as one of the ways of computing the oscillation probability. See for example refs.~\cite{Minakata:2015gra,Denton:2016wmg}. Once we have the expression of the flavor eigenstate $\nu_{\alpha}$ in terms of the mass eigenstate basis $\bar{\nu}_{i}$ in matter as 
\begin{eqnarray}
&&
\nu_{\alpha} = V_{\alpha i} \bar{\nu}_{i}
\label{Vmatrix-def}
\end{eqnarray}
the oscillation probability can readily be calculated in complete parallelism with the case in vacuum by replacing the $U$ matrix by the $V$ matrix as 
\begin{eqnarray}
&& 
P(\nu_\beta \rightarrow \nu_\alpha) = 
\left| ~\sum_i V_{\alpha i} V^*_{\beta i}  ~e^{-i\frac{\lambda_i x}{2E}} ~\right|^2
\nonumber\\
&= & 
\delta_{\alpha \beta}  - 4 \sum_{j > i} 
{\mbox Re}[V_{\alpha i} V_{\beta i}^* V_{\alpha j}^* V_{\beta j}]
\sin^2 \frac{ (\lambda_{j} - \lambda_{ i}) x}{ 4E } 
- 2 \sum_{j > i} 
{\mbox Im}[V_{\alpha i} V_{\beta i}^* V_{\alpha j}^* V_{\beta j}] 
\sin \frac{ (\lambda_{j} - \lambda_{i}) x}{ 2E } 
\nonumber\\
\label{P-ba-general}
\end{eqnarray}
assuming adiabaticity of the neutrino evolution in matter, where $x$ denotes the baseline. 

Let us compute the $V$ matrix elements to first order in the SRP theory. In perturbation theory to first order in $\hat{H}_{1}$, the mass eigenstate in matter can be written as $\bar{\nu}_{i} = \hat{\nu}_{i}^{(0)} + \hat{\nu}_{i}^{(1)}$, and hence $\bar{\nu}_{i} = \hat{\nu}_{i}^{(0)}$ in the lowest order. Inverting the state relationship in eq.~\eqref{hat-basis}, we obtain at the zeroth order 
\begin{eqnarray} 
\nu = 
U_{23} (\theta_{23}) U_{13} (\theta_{13}) U_{12} (\varphi, \delta) \hat{\nu}^{(0)} 
\equiv V^{(0)} \hat{\nu}^{(0)}, 
\label{V-matrix-0th}
\end{eqnarray}
which defines the zeroth order $V$ matrix.  

We calculate the first order correction to the hat basis wave functions. Using the familiar perturbative formula for the wave functions, we have 
\begin{eqnarray}
\hat{\nu}_{i}^{(1)} = 
\sum_{j\neq i} \frac{ 2E ( \hat{H}_{1} )^{*}_{ji} }{ \lambda_i - \lambda_j }
\hat{\nu}_{j}^{(0)}
\label{nu-hat-first-order}
\end{eqnarray}
with $\hat{H}_{1}$ in \eqref{hat-H-0th-1st-SRP}, and the $\lambda_i$ are the eigenvalues of $2E \hat{H}_{0}$, see \eqref{eigenvalues}. See ref.~\cite{Minakata:2022zua} for a clarifying remark on this computation. Using the result of $\hat{\nu}_{i}^{(1)}$ from eq.~\eqref{nu-hat-first-order}, the mass eigenstate is given to first order in SRP theory as 
\begin{eqnarray} 
&& 
\left[
\begin{array}{c}
\hat{\nu}_{1} \\
\hat{\nu}_{2} \\
\hat{\nu}_{3} \\
\end{array}
\right] 
= 
\left[
\begin{array}{ccc}
1 & 0 & - \frac{ c_{\varphi} c_{13} s_{13} a }{ \lambda_3 - \lambda_1 } \\
0 & 1 & - \frac{ s_{\varphi} c_{13} s_{13} a }{ \lambda_3 - \lambda_2 } e^{ - i \delta} \\
\frac{ c_{\varphi} c_{13} s_{13} a }{ \lambda_3 - \lambda_1 } & 
\frac{ s_{\varphi} c_{13} s_{13} a }{ \lambda_3 - \lambda_2 } e^{ i \delta} & 1 \\
\end{array}
\right] 
\left[
\begin{array}{c}
\hat{\nu}_{1}^{(0)} \\
\hat{\nu}_{2}^{(0)} \\
\hat{\nu}_{3}^{(0)} \\
\end{array}
\right] 
\nonumber \\
&=& 
\left[
\begin{array}{ccc}
1 & 0 & - \frac{ c_{\varphi} c_{13} s_{13} a }{ \lambda_3 - \lambda_1 } \\
0 & 1 & - \frac{ s_{\varphi} c_{13} s_{13} a }{ \lambda_3 - \lambda_2 } e^{ - i \delta} \\
\frac{ c_{\varphi} c_{13} s_{13} a }{ \lambda_3 - \lambda_1 } & 
\frac{ s_{\varphi} c_{13} s_{13} a }{ \lambda_3 - \lambda_2 } e^{ i \delta} & 1 \\
\end{array}
\right] 
\left[ U_{23} (\theta_{23}) U_{13} (\theta_{13}) U_{12} (\varphi, \delta)  \right]^{\dagger} 
\left[
\begin{array}{c}
\nu_{e} \\
\nu_{\mu} \\
\nu_{\tau} \\
\end{array}
\right] 
\equiv 
V^{\dagger} 
\left[
\begin{array}{c}
\nu_{e} \\
\nu_{\mu} \\
\nu_{\tau} \\
\end{array}
\right], 
\nonumber
\end{eqnarray}
using eq.~\eqref{V-matrix-0th} in the second line. Inverting this relation we obtain 
\begin{eqnarray} 
&& 
\left[
\begin{array}{c}
\nu_{e} \\
\nu_{\mu} \\
\nu_{\tau} \\
\end{array}
\right] 
=
U_{23} (\theta_{23}) U_{13} (\theta_{13}) U_{12} (\varphi, \delta) 
\biggl\{
1 + 
\mathcal{W} ( \varphi, \delta; \lambda_{1}, \lambda_{2} ) 
\biggr\} 
\left[
\begin{array}{c}
\hat{\nu}_{1} \\
\hat{\nu}_{2} \\
\hat{\nu}_{3} \\
\end{array}
\right] 
\equiv 
V 
\left[
\begin{array}{c}
\hat{\nu}_{1} \\
\hat{\nu}_{2} \\
\hat{\nu}_{3} \\
\end{array}
\right], 
%
\label{Vmatrix-SRP}
\end{eqnarray}
where $\mathcal{W} (\theta_{13}, \varphi, \delta; \lambda_{1}, \lambda_{2} )$ is defined by 
\begin{eqnarray} 
&&
\mathcal{W} (\theta_{13}, \varphi, \delta; \lambda_{1}, \lambda_{2} ) 
\equiv 
c_{13} s_{13} 
\left[
\begin{array}{ccc}
0 & 0 & \frac{ c_{\varphi} a }{ \lambda_3 - \lambda_1 } \\
0 & 0 & \frac{ s_{\varphi} a }{ \lambda_3 - \lambda_2 } e^{ - i \delta} \\
- \frac{ c_{\varphi} a }{ \lambda_3 - \lambda_1 } & 
- \frac{ s_{\varphi} a }{ \lambda_3 - \lambda_2 } e^{ i \delta} & 0 \\
\end{array}
\right]. 
\label{mathcal-W-SRP}
\end{eqnarray}
Equation~\eqref{Vmatrix-SRP} defines the $V$ matrix to first order in expansion. 

Notice the remarkable similarity between the $V$ matrix expressions of the flavor state in eq.~\eqref{SFeq-DMP-1st} (DMP) and eq.~\eqref{Vmatrix-SRP} (SRP). Not so surprisingly the symmetry structure of the SRP theory is akin to that of DMP, as shown in Table~\ref{tab:SRP-symmetry}. 
In fact, the DMP and SRP are essentially identical in their leading order apart from yes/no of the matter dressing of $\theta_{13}$, and we have no surprise on the very similar symmetry structures, apart from a difference of presence or absence of the $\phi$ transformations. But, it is automatically enforced in DMP because $\sin 2\phi \propto \sin 2\theta_{13}$~\cite{Denton:2016wmg}. 

We have computed the oscillation probability by inserting the $V$ matrix elements obtained by using eqs.~\eqref{Vmatrix-SRP} and \eqref{mathcal-W-SRP} into eq.~\eqref{P-ba-general}. The calculated results reproduce the formulas obtained by using the $S$ matrix method to first order in the SRP theory~\cite{Martinez-Soler:2019nhb}. 

\subsection{Framework-generated effective expansion parameter}
\label{sec:exp-parameter} 

We remarks that the SRP theory has an exceptional feature as a perturbation theory. Look at the hat basis Hamiltonian, eq.~\eqref{hat-H-0th-1st-SRP}. What is peculiar is that the 3-3 element of $\hat{H}$ is of order $\lambda_{3} \sim \Delta m^2_{31}$, whereas all the other elements, the eigenvalues $\lambda_{1}$ and $\lambda_{2}$ as well as the only non-vanishing off-diagonal elements in $\hat{H}_{1}$, are of order $\Delta m^2_{21} \sim a$. Therefore, the perturbative Hamiltonian is {\em not} small compared to the first $2 \times 2$ block of the unperturbed Hamiltonian, both in $\hat{H}$ and $\widetilde{H}$. 

Then, the question is: How does the SRP work as the perturbation theory? The answer is that it works because of the new effective expansion parameter which emerges from the framework itself. Recall the $V$ matrix computation in section~\ref{sec:V-matrix}. All the first order correction terms are inversely proportional to $\lambda_{3} - \lambda_{1}$ or $\lambda_{3} - \lambda_{2}$. The feature stems from the special structure of $\hat{H}_{1}$ whose non-vanishing elements exist only in the 3-i and i-3 ($i=1,2$) elements, as seen in eq.~\eqref{hat-H-0th-1st-SRP}. Then, the large denominator with $\lambda_{3} \simeq \Delta m^2_{31} \gg a$ acts as a suppression factor, the propagator suppression. The suppression factor can be read off from the first-order $V$ matrix as 
\begin{eqnarray}
A_{ \text{exp} } 
&\equiv& 
c_{13} s_{13} 
\biggl | \frac{ a }{ \Delta m^2_{31} } \biggr | 
= 2.78 \times 10^{-3} 
\left(\frac{ \Delta m^2_{31} }{ 2.4 \times 10^{-3}~\mbox{eV}^2}\right)^{-1}
\left(\frac{\rho}{3.0 \,\text{g/cm}^3}\right) \left(\frac{E}{200~\mbox{MeV}}\right).
\nonumber \\
\label{expansion-parameter}  
\end{eqnarray}
$A_{ \text{exp} }$ acts as an effective expansion parameter, which is a factor of 10 smaller than \\
$\Delta m^2_{21} / \Delta m^2_{31} \simeq 0.03$. As a consequence the agreement with only the leading-order term in the probability is shown to be quite good~\cite{Martinez-Soler:2019nhb}. 

\section{Symmetry in the solar-resonance perturbation theory}
\label{sec:Symmetry-SRP} 

We investigate the reparametrization symmetry in the SRP theory. We derive the SF equation, a powerful machinery to identify the symmetries, and obtain the solutions given in Table~\ref{tab:SRP-symmetry}. 
What we do first is to embody the general statement of symmetry in eq.~\eqref{SF-eq-general} in the SRP theory. 

\subsection{Symmetry Finder equation in the SRP theory} 
\label{sec:SF-solRP} 

To prepare the first state in the right-hand side of eq.~\eqref{SF-eq-general} we define an alternative but physically equivalent state to that in eq.~\eqref{Vmatrix-SRP}, 
\begin{eqnarray} 
&& 
\hspace{-6mm}
F \left[
\begin{array}{c}
\nu_{e} \\
\nu_{\mu} \\
\nu_{\tau} \\
\end{array}
\right] 
= 
F 
U_{23} (\theta_{23}) U_{13} (\theta_{13}) U_{12} (\varphi, \delta) 
%
R^{\dagger} R 
\biggl\{
1 + 
\mathcal{W} (\theta_{13}, \varphi, \delta; \lambda_{1}, \lambda_{2} ) 
\biggr\} 
R^{\dagger} R 
\left[
\begin{array}{c}
\nu_{1} \\
\nu_{2} \\
\nu_{3} \\
\end{array}
\right], ~~~
\label{SFeq-ansatz-solRP}
\end{eqnarray}
where $\mathcal{W} (\theta_{13}, \varphi, \delta; \lambda_{1}, \lambda_{2} )$ is defined in eq.~\eqref{mathcal-W-SRP}. In eq.~\eqref{SFeq-ansatz-solRP} we have introduced the flavor-state rephasing matrix $F$ and the generalized 1-2 state exchange matrix $R$,
which are defined by 
\begin{eqnarray} 
&&
F \equiv 
\left[
\begin{array}{ccc}
e^{ i \tau } & 0 & 0 \\
0 & e^{ i \sigma } & 0 \\
0 & 0 & 1 \\
\end{array}
\right], 
\hspace{8mm}
R \equiv 
\left[
\begin{array}{ccc}
0 & - e^{ i ( \delta + \alpha) } & 0 \\
e^{ - i ( \delta + \beta) } & 0 & 0 \\
0 & 0 & 1 \\
\end{array}
\right]. 
\label{F-R-SRP-def}
\end{eqnarray}
The flavor-state rephasing $F$ does not affect the observables because it can be absorbed by the neutrino states, and inserting unity, $R^{\dagger} R=1$, is of course harmless. But, in fact, the $F$ matrix actually plays a role: Without it we would miss several symmetries we are going to uncover with $F$~\cite{Minakata:2021dqh,Minakata:2021goi,Minakata:2022zua}. Moreover, the state exchange matrix $R$ and $F$ form a complex system composed of the phases $\tau$, $\sigma$, $\alpha$, and $\beta$, and they come in to the SF equation to produce the coupled nontrivial solutions. 
Notice that the rephasing matrices, the both $F$ and $R$ in eq.~\eqref{F-R-SRP-def} takes the nonvanishing, nontrivial (not unity) elements in the 1-2 sub-sector because we restrict ourselves into the 1-2 state exchange symmetry in this theory. 

Now, we demand that the state defined in eq.~\eqref{SFeq-ansatz-solRP} must be written by the flavor state but with the transformed parameters which are denoted with the primed symbol. That is, the transformations are such that $\varphi \rightarrow \varphi^{\prime}$, $\theta_{23} \rightarrow \theta_{23} ^{\prime}$, $\theta_{13} \rightarrow \theta_{13} ^{\prime}$, and $\delta \rightarrow \delta^{\prime}$, which becomes symmetry transformations if the SF equation has a solution. This is equivalent to prepare the second state in the right-hand side of eq.~\eqref{SF-eq-general}. Using the notation $\delta^{\prime} = \delta + \xi$, and the abbreviated notations $s_{13}^{\prime} \equiv \sin \theta_{13}^{\prime}$ and $c_{13}^{\prime} \equiv \cos \theta_{13}^{\prime}$ and the same for $\theta_{23}$, and $\varphi$ for later use, the SF equation in the SRP theory reads: 
\begin{eqnarray} 
&&
\left[
\begin{array}{ccc}
e^{ i \tau } & 0 & 0 \\
0 & e^{ i \sigma } & 0 \\
0 & 0 & 1 \\
\end{array}
\right]
\left[
\begin{array}{c}
\nu_{e} \\
\nu_{\mu} \\
\nu_{\tau} \\
\end{array}
\right] 
\nonumber \\
&& 
\hspace{-10mm} 
=
\left[
\begin{array}{ccc}
1 & 0 &  0  \\
0 & c_{23} & s_{23} e^{ i \sigma } \\
0 & - s_{23} e^{ - i \sigma } & c_{23} \\
\end{array}
\right] 
\left[
\begin{array}{ccc}
c_{13}  & 0 & s_{13} e^{ i \tau } \\
0 & 1 & 0 \\
- s_{13} e^{ - i \tau } & 0 & c_{13} \\
\end{array}
\right] 
F U_{12} (\varphi, \delta) R^{\dagger} 
R 
\biggl\{
1 + 
\mathcal{W} (\theta_{13}, \varphi, \delta; \lambda_{1}, \lambda_{2} ) 
\biggr\} 
R^{\dagger}
R 
\left[
\begin{array}{c}
\nu_{1} \\
\nu_{2} \\
\nu_{3} \\
\end{array}
\right] 
\nonumber \\
&=& 
\left[
\begin{array}{ccc}
1 & 0 &  0  \\
0 & c_{23}^{\prime} & s_{23}^{\prime} \\
0 & - s_{23}^{\prime} & c_{23}^{\prime} \\
\end{array}
\right] 
\left[
\begin{array}{ccc}
c_{13}^{\prime} & 0 & s_{13}^{\prime} \\
0 & 1 & 0 \\
- s_{13}^{\prime} & 0 & c_{13}^{\prime} \\
\end{array}
\right] 
U_{12} ( \varphi^{\prime}, \delta + \xi) 
\biggl\{
1 + 
\mathcal{W} (\theta_{13}^{\prime}, \varphi^{\prime}, \delta + \xi; \lambda_{2}, \lambda_{1} ) 
\biggr\} 
\left[
\begin{array}{c}
- e^{ i ( \delta + \alpha) } \nu_{2} \\
e^{ - i ( \delta + \beta) } \nu_{1} \\
\nu_{3} \\
\end{array}
\right].
\nonumber \\
\label{SFeq-SRP-final}
\end{eqnarray}
As became explicit in eq.~\eqref{SFeq-SRP-final}, the vacuum angles $\theta_{23}$ and $\theta_{13}$, in general, transform under the symmetry transformations after the phase redefinition $F$ in the flavor eigenstate is introduced. It is one of the most interesting features of the SF equation in matter~\cite{Minakata:2021dqh,Minakata:2021goi,Minakata:2022zua}. 

\subsection{The first and the second conditions and their solutions} 
\label{sec:SRP-1st-2nd} 

We look for the solution of the SF equation under the ansatz $s_{23} e^{ i \sigma } = s_{23}^{\prime}$ and $s_{13} e^{ i \tau } = s_{13}^{\prime}$, because apparently we have no other choice within the present SF formalism. The ansatz leads to the two consequences: 
(1) The possible values of $\tau$ and $\sigma$ are restricted to integer multiples of $\pi$. 
(2) The SF equation~\eqref{SFeq-SRP-final} can be decomposed into the following first (first line) and the second (second line) conditions. 
\begin{eqnarray} 
F U_{12} (\varphi, \delta) R^{\dagger} 
&=&  
U_{12} ( \varphi^{\prime}, \delta + \xi), 
\nonumber \\
R \mathcal{W} (\theta_{13}, \varphi, \delta; \lambda_{1}, \lambda_{2} ) R^{\dagger} 
&=& 
\mathcal{W} (\theta_{13}^{\prime}, \varphi^{\prime}, \delta + \xi; \lambda_{2}, \lambda_{1} ).
\label{1st-2nd-conditions}
\end{eqnarray}
One can show that the first condition can be reduced to 
\begin{eqnarray} 
&&
c_{\varphi^{\prime}} 
= - s_{\varphi} e^{ - i ( \alpha - \tau ) } 
= - s_{\varphi} e^{ i ( \beta + \sigma ) }, 
\hspace{8mm}
s_{\varphi^{\prime}} 
= c_{\varphi} e^{ i ( \beta + \tau - \xi ) } 
= c_{\varphi} e^{ - i ( \alpha - \sigma - \xi ) }.
\label{1st-condition-SRP}
\end{eqnarray}
We note that under the above restriction of $\tau$ and $\sigma$ being integer multiples of $\pi$, eq.~\eqref{1st-condition-SRP} implies that all the rest of the phase parameters, $\xi$, $\alpha$, and $\beta$, must also be integer multiples of $\pi$~\cite{Minakata:2021dqh}. This is the key property that comes out from the first condition which restricts the solution space in the SF framework in its current form. The solutions of the first condition~\eqref{1st-condition-SRP} are summarized in Table~\ref{tab:SF-solutions}.\footnote{
The readers might be puzzled by ``Symmetry Xf'' in the first column in which ``f'' implies flipping the sign of $s_{12}$ because it was absent in the DMP symmetries in Table~\ref{tab:DMP-UV-symmetry}. It is a characteristically new feature of the 1-3 exchange symmetry in the helio-perturbation theory~\cite{Minakata:2021goi}, as will be explained in section~\ref{sec:SF-solution-helioP-UV}. }

\begin{table}[h!]
\vglue -0.2cm
\begin{center}
\caption{The universal solution of the first conditions which are common to the SRP (section~\ref{sec:SRP-1st-2nd}), DMP~\cite{Minakata:2021dqh}, the helio- and helio-UV perturbation theories (section~\ref{sec:1st-2nd-condition-helioP}). For the former two symmetries, the symmetry symbols with ``f'' ($s_{12}$ sign flip) such as Symmetry Xf, must be ignored in the first column. 
}
\label{tab:SF-solutions}
\vglue 0.2cm
\begin{tabular}{c|c|c}
\hline 
Symmetry & 
$\tau, \sigma, \xi$ & 
$\alpha, \beta$
\\
\hline 
\hline 
IA, IAf & 
$\tau = \sigma = 0$, $\xi = 0$ & 
$\alpha = \beta = 0$ (upper) \\ 
& &
$\alpha = \pi, \beta = - \pi$ (lower)  \\
\hline
IB, IBf & 
$\tau = \sigma = 0$, $\xi = \pi$ & 
$\alpha = \pi, \beta = - \pi$ (upper) \\
& & $\alpha = \beta = 0$ (lower) \\
\hline 
IIA, IIAf & 
$\tau = 0, \sigma = - \pi$, $\xi = 0$ & 
$\alpha = \pi, \beta = 0$ (upper) \\
& & $\alpha = 0, \beta = \pi$ (lower)  \\
\hline 
IIB, IIBf & 
$\tau = 0, \sigma = - \pi$, $\xi = \pi$ & 
$\alpha = 0, \beta = \pi$ (upper) \\ 
& & $\alpha = \pi, \beta = 0$ (lower) \\
\hline 
IIIA, IIIAf & 
$\tau = \pi, \sigma = 0$, $\xi = 0$ & 
$\alpha = 0, \beta = \pi$ (upper) \\ 
& & $\alpha = \pi, \beta = 0$ (lower) \\
\hline 
IIIB, IIIBf & 
$\tau = \pi, \sigma = 0$, $\xi = \pi$ & 
$\alpha = \pi, \beta = 0$ (upper)  \\ 
 & & 
$\alpha = 0, \beta = \pi$ (lower)   \\
\hline 
IVA, IVAf & 
$\tau = \sigma = \pi$, $\xi = 0$ & 
$\alpha = \pi, \beta = - \pi$ (upper) \\ 
& &
$\alpha = \beta = 0$ (lower)  \\
\hline 
IVB, IVBf & 
$\tau = \sigma = \pi$, $\xi = \pi$ & 
$\alpha = \beta = 0$ (upper) \\ 
& &
$\alpha = \pi, \beta = - \pi$ (lower)  \\
\hline 
\end{tabular}
\end{center}
\vglue -0.4cm 
\end{table}

In fact the solutions of the first condition possess an interesting universal properties. Because only the $\nu$SM part of the theory comes in to the first condition, the solution is universal in the SRP, DMP~\cite{Minakata:2021dqh}, and the helio-perturbation theories~\cite{Minakata:2021goi}. The property holds also in the DMP-UV~\cite{Minakata:2022zua} and the helio-UV perturbation theories whose latter is to be discussed in sections~\ref{sec:helio-UV-P},~\ref{sec:SF-helio-UV}, and~\ref{sec:SF-solution-helioP-UV}. The fact that the universal solution applies to the helio- and the helio-UV perturbation theories is nontrivial because the 1-3 exchange is involved. But, via a smart choice of the $R$ matrix etc., one can make the solutions really identical among these theories~\cite{Minakata:2021goi}. 
Thus, Table~\ref{tab:SF-solutions} serves not only for the SRP but also for the DMP and helio-perturbation theories including their UV extensions. The symmetry with symbol ``f'' ($s_{12}$ sign flip) applies only to the one in the helio- and helio-UV perturbation theories. 

The second condition in eq.~\eqref{1st-2nd-conditions} reads: 
\begin{eqnarray} 
&&
c_{13} s_{13}
\left[
\begin{array}{ccc}
0 & 0 & 
- e^{ i \alpha } s_{\varphi} \frac{ a }{ \lambda_3 - \lambda_2 } \\
0 & 0 & 
e^{ - i ( \delta + \beta) } c_{\varphi} \frac{ a }{ \lambda_3 - \lambda_1 } \\
e^{ - i \alpha } s_{\varphi} \frac{ a }{ \lambda_3 - \lambda_2 } & 
- e^{ i ( \delta + \beta) } c_{\varphi} \frac{ a }{ \lambda_3 - \lambda_1 } & 0 \\
\end{array}
\right] 
\nonumber \\
&&
\hspace{16mm} 
= c_{13}^{\prime} s_{13}^{\prime}
\left[
\begin{array}{ccc}
0 & 0 & c_{\varphi}^{\prime} \frac{ a }{ \lambda_3 - \lambda_2 } \\
0 & 0 & e^{ - i ( \delta + \xi ) } s_{\varphi}^{\prime} \frac{ a }{ \lambda_3 - \lambda_1 } \\
- c_{\varphi}^{\prime} \frac{ a }{ \lambda_3 - \lambda_2 } & 
- s_{\varphi}^{\prime} e^{ i ( \delta + \xi ) } \frac{ a }{ \lambda_3 - \lambda_1 } & 0 \\
\end{array}
\right]. 
\label{2nd-condition-SRP-full}
\end{eqnarray}
One can examine the solutions of eq.~\eqref{2nd-condition-SRP-full} one by one for the given solutions of the first condition in Table~\ref{tab:SF-solutions}. This really straightforward calculation is left for the interested readers. The solutions obtained in such exercise consist of the symmetries tabulated in Table~\ref{tab:SRP-symmetry}.

\begin{table}[h!]
\begin{center}
\caption{All the reparametrization symmetries of the 1-2 state exchange type found in the solar-resonance perturbation theory are tabulated as ``Symmetry X'', a shorthand of ``Symmetry X-SRP''.  
In this table the notations are such that $\lambda_{j}$ ($j=1,2$) are the first two eigenvalues of $2E H$, and $\varphi$ denotes $\theta_{12}$ in matter.
}
\label{tab:SRP-symmetry}
\vglue 0.2cm
\begin{tabular}{c|c|c}
\hline 
SRP Symmetry & 
Vacuum parameter transformations & 
Matter parameter transformations
\\
\hline 
\hline 
Symmetry IA & 
none & 
$\lambda_{1} \leftrightarrow \lambda_{2}$, 
$c_{\varphi} \rightarrow \mp s_{\varphi}$, 
$s_{\varphi} \rightarrow \pm c_{\varphi}$. \\
\hline 
Symmetry IB & 
$\theta_{12} \rightarrow - \theta_{12}$, 
$\delta \rightarrow \delta + \pi$. & 
$\lambda_{1} \leftrightarrow \lambda_{2}$, 
$c_{\varphi} \rightarrow \pm s_{\varphi}$, 
$s_{\varphi} \rightarrow \pm c_{\varphi}$. \\
\hline
Symmetry IIA & 
$\theta_{23} \rightarrow - \theta_{23}$, 
$\theta_{12} \rightarrow - \theta_{12}$. & 
$\lambda_{1} \leftrightarrow \lambda_{2}$, 
$c_{\varphi} \rightarrow \pm s_{\varphi}$, 
$s_{\varphi} \rightarrow \pm c_{\varphi}$. \\
\hline 
Symmetry IIB & 
$\theta_{23} \rightarrow - \theta_{23}$, 
$\delta \rightarrow \delta + \pi$. & 
$\lambda_{1} \leftrightarrow \lambda_{2}$, 
$c_{\varphi} \rightarrow \mp s_{\varphi}$, 
$s_{\varphi} \rightarrow \pm c_{\varphi}$. \\
\hline 
Symmetry IIIA & 
$\theta_{13} \rightarrow - \theta_{13}$, 
$\theta_{12} \rightarrow - \theta_{12}$. & 
$\lambda_{1} \leftrightarrow \lambda_{2}$, 
$c_{\varphi} \rightarrow \pm s_{\varphi}$, 
$s_{\varphi} \rightarrow \pm c_{\varphi}$. \\
\hline 
Symmetry IIIB & 
$\theta_{13} \rightarrow - \theta_{13}$, 
$\delta \rightarrow \delta + \pi$. & 
$\lambda_{1} \leftrightarrow \lambda_{2}$, 
$c_{\varphi} \rightarrow \mp s_{\varphi}$, 
$s_{\varphi} \rightarrow \pm c_{\varphi}$. \\
\hline 
Symmetry IVA & 
$\theta_{23} \rightarrow - \theta_{23}$, 
$\theta_{13} \rightarrow - \theta_{13}$ & 
$\lambda_{1} \leftrightarrow \lambda_{2}$, 
$c_{\varphi} \rightarrow \mp s_{\varphi}$, 
$s_{\varphi} \rightarrow \pm c_{\varphi}$. \\
\hline 
Symmetry IVB & 
$\theta_{23} \rightarrow - \theta_{23}$, 
$\theta_{13} \rightarrow - \theta_{13}$, & 
$\lambda_{1} \leftrightarrow \lambda_{2}$, \\ 
 &
$\theta_{12} \rightarrow - \theta_{12}$, $\delta \rightarrow \delta + \pi$. 
 &
$c_{\varphi} \rightarrow \pm s_{\varphi}$, $s_{\varphi} \rightarrow \pm c_{\varphi}$. \\
\hline 
\end{tabular}
\end{center}
\vglue -0.6cm 
\end{table}

\subsection{Symmetries of the 1-2 and 1-3 state exchange types in $\nu$SM}
\label{sec:12-13-symmetry} 

Together with the result obtained in ref.~\cite{Minakata:2021goi}, we have confirmed our physical picture that the symmetries of the 1-2 and 1-3 state exchange types exist in the solar- and atmospheric-resonance regions, respectively. Notice that ref.~\cite{Minakata:2021goi} is, so far, the unique case in which the 1-3 state exchange symmetry is found and discussed. 

What is the relationship between the 1-2 symmetry in DMP and the 1-3 symmetry in the helio-perturbation theory? It is shown that there exists a limiting procedure, the ATM limit, by which DMP approaches to the helio-perturbation theory~\cite{Minakata:2020oxb}. Then, the natural question would be: What happens in the 1-2 exchange symmetry under such limit in DMP? Does it have something to do with the 1-3 exchange symmetry in the helio-perturbation theory? The answer is {\em No}, to our understanding. 
In taking such a limit in DMP, we enter into the regions $c_{\psi} \rightarrow 0$ ($s_{\psi} \rightarrow 0$) for the normal mass ordering (inverted mass ordering). In the both cases $\sin 2\psi \rightarrow 0$, and $\psi$ degrees of freedom is frozen. Therefore, by the ATM limit the whole DMP theory turns into the helio-perturbation theory and no remnant of the $\psi$ symmetry is left. This is related to that the ATM limit is called as the ``operational limit'' in ref.~\cite{Minakata:2021nii}. 

After the comparative treatment of the 1-2 exchange symmetries in the DMP and SRP theories whose symmetry results are summarized in Tables~\ref{tab:DMP-UV-symmetry} and~\ref{tab:SRP-symmetry}, respectively, we must go on to discuss the 1-3 symmetry in the helio-perturbation theory. But, we will do it in an extended framework which include non-unitarity, a promising way for discussing physics beyond the $\nu$SM at high or low scales. 

\section{Symmetry in three neutrino system with non-unitarity}
\label{sec:Symmetry-UV} 

Now, we enter into Part II in which we change the gear. So far we have discussed the reparametrization symmetry within the $\nu$SM frameworks. From now on, we jump-in to the theory of neutrino oscillation in matter with non-unitary flavor mixing matrix. 

It is a very popular idea that the $\nu$SM provides only an incomplete picture of our world. A well known concrete model describing the departure from the $\nu$SM is the existence of low mass neutral leptons, the sterile neutrinos, see e.g., refs.~\cite{Dasgupta:2021ies,Conrad:2013mka,Dentler:2018sju}. In more generic context possible deviation from the $\nu$SM is extensively discussed in the framework called non-standard interactions (NSI)~\cite{Wolfenstein:1977ue,Valle:1987gv,Guzzo:1991hi,Roulet:1991sm,Grossman:1995wx,Gonzalez-Garcia:2001snt,Berezhiani:2001rs}, and non-unitarity, neutrino evolution with non-unitary flavor mixing matrix~\cite{Antusch:2006vwa,Escrihuela:2015wra,Fong:2016yyh,Fong:2017gke}. See e.g. refs.~\cite{Ohlsson:2012kf,Miranda:2015dra,Farzan:2017xzy,Proceedings:2019qno} for reviews of NSI, refs.~\cite{Davidson:2003ha,Antusch:2008tz,Biggio:2009nt,Esteban:2018ppq} for constraints on NSI, and refs.~\cite{Martinez-Soler:2018lcy,Martinez-Soler:2019noy,Fernandez-Martinez:2007iaa,Goswami:2008mi,Antusch:2009pm,Antusch:2009gn,Antusch:2014woa,Ge:2016xya,Fernandez-Martinez:2016lgt,Dutta:2016vcc,Parke:2015goa,Ellis:2020hus,Coloma:2021uhq} for a limited list of the remaining articles on non-unitarity. 

In this paper we focus on non-unitarity approach to new physics beyond $\nu$SM. For our purpose of understanding the implications of the symmetry in neutrino oscillation, we feel that non-unitarity is a better framework to try first. It is because the generic NSI are much less constrained frameworks than non-unitarity. It has typically 25 parameters in addition to the $\nu$SM ones, by including the production, propagation, and detection NSI.\footnote{
The precise number of degrees of freedom is model-dependent, such as doing independent counting of neutron and proton NSI or not, and is not the real concern here. The one given above for NSI is based on: 8 from propagation and 9+9-1 (overall phase) from production and detection. } 
Whereas non-unitarity has only 9, see section~\ref{sec:3nu-non-unitarity}. By construction, the way of modification of the active three-flavor neutrino sector due to new physics at high or low scales is not arbitrary, but determined by a UV producing new physics sector. See discussions in e.g., refs.~\cite{Antusch:2006vwa,Fong:2017gke}. 

An interesting question would then be whether consideration of the reparametrization symmetry affects our understanding of the theory with non-unitarity. To address such a question in a reliable manner, we need to do: 
(1) To establish the theoretical framework of neutrino oscillation in matter to include the effect of UV, and 
(2) To perform the SF analysis in such a way that the internal consistency between the constraints coming from the ``genuine non-unitary'' and ``unitary evolution'' parts is met~\cite{Martinez-Soler:2018lcy}. See section~\ref{sec:SF-solution-helioP-UV}. 
The first task is carried out by formulating the ``DMP-UV'' perturbation theory~\cite{Minakata:2021nii} and the ``helio-UV'' perturbation theory~\cite{Martinez-Soler:2018lcy} corresponding to their $\nu$SM versions.\footnote{
In the sterile neutrino model, the DMP-UV perturbation theory does not necessarily provide adequate description of such models in the whole kinematical region. For example, if the sterile mass is $\sim$ eV scale there exist resonances at energy of $\mathcal{O}(1)$ TeV~\cite{Yasuda:2000xs,Nunokawa:2003ep}, which are outside the validity of the framework. This problem can be avoided if we remain in $\vert \rho E \vert \lsim 100 (\text{g/cm}^3)$ GeV, as discussed in ref.~\cite{Minakata:2021nii}. In a related but different approach, an extended DMP-like theory with sterile neutrino is formulated in ref.~\cite{Parke:2019jyu}, but the $\mathcal{O}(1)$ TeV resonance is not covered in its current treatment. }
Moreover, the consistent SF analysis for the symmetry in the DMP-UV perturbation theory is carried out, and the results are reported in ref.~\cite{Minakata:2022zua}. The resulting eight symmetries, Symmetry X-DMP-UV (X=IA, IB, $\cdot \cdot \cdot $, IVB), are copied from this reference to Table~\ref{tab:DMP-UV-symmetry}. 

Therefore, what is lacking in symmetry discussion in the UV-extended theories, within the our present scope of SF, is to analyze the reparametrization symmetry in the helio-UV perturbation theory. It will be the remaining goal in this article, to which we will devote the rest of this paper.

Since the symmetry structure of the SPR theory is so akin to the one in DMP, we do not try to extend our study to the SRP-UV theory. But since such UV extended SRP theory is formulated in ref.~\cite{Martinez-Soler:2019noy} one can easily proceed toward the task whenever the demand exists. 

\section{The helio-UV perturbation theory}
\label{sec:helio-UV-P} 

This section is meant to be a brief review of the helio-UV perturbation theory with non-unitary flavor mixing matrix~\cite{Martinez-Soler:2018lcy}, a UV extended version of the helio-perturbation theory~\cite{Minakata:2015gra}. Throughout Part II, we use the PDG convention~\cite{Zyla:2020zbs} for the $U$ matrix, because we are going to discuss the 1-3 exchange symmetry~\cite{Minakata:2021goi}. 

Despite the difference in the theory treated and the state exchange types in the symmetries, many of the features of the discussions from section~\ref{sec:helio-UV-P} through section~\ref{sec:hamiltonian-symmetry} are so similar to the ones in ref.~\cite{Minakata:2022zua} in which the symmetry of the DMP-UV-perturbation theory is discussed.\footnote{
The merit of such similarity is that by going through section~\ref{sec:helio-UV-P} through section~\ref{sec:hamiltonian-symmetry} in this paper the readers not only understand symmetry in the helio-UV perturbation theory, but also can have a very good idea on what is the DMP-UV symmetry, and vice versa. } 
Nonetheless, we go through the whole SF formulation in the helio-UV perturbation theory because a factor of two larger number of symmetries necessitates an independent SF analysis, and the detailed differences often matter. It entails the intriguing differences between the helioP-UV and DMP-UV symmetries in the $\alpha$ and the $\widetilde{\alpha}$ transformation properties and the structure of the rephasing matrices. See Table~\ref{tab:helioP-UV-symmetry}, Table~\ref{tab:DMP-UV-symmetry}, and Appendix~\ref{sec:DMP-UV-summary}. 

\subsection{Three active-neutrino system with non-unitary flavor mixing matrix}
\label{sec:3nu-non-unitarity} 

While discussion for the theoretical basis of the system of three active neutrinos propagating under the influence of non-unitary flavor mixing matrix is highly nontrivial, we believe that by now there is a standard way~\cite{Blennow:2016jkn,Fong:2017gke}. That is, we start from the Schr\"odinger equation in the vacuum mass eigenstate basis, the ``check basis'', $i \frac{d}{dx} \check{\nu} = \check{H} \check{\nu}$, where $\check{H}$ is given by replacing the $U$ matrix eq.~\eqref{evolution-check-basis} by the non-unitary $N$ matrix, 
\begin{eqnarray} 
\check{H}
= 
\frac{1}{2E} 
\left\{  
\left[
\begin{array}{ccc}
0 & 0 & 0 \\
0 & \Delta m^2_{21} & 0 \\
0 & 0 & \Delta m^2_{31} \\
\end{array}
\right] + 
N^{\dagger} \left[
\begin{array}{ccc}
a - b & 0 & 0 \\
0 & -b & 0 \\
0 & 0 & -b \\
\end{array}
\right] N 
\right\}. 
\label{checkH-UV-def} 
\end{eqnarray}
$N$ is the $3 \times 3$ non-unitary flavor mixing matrix which relates the flavor neutrino states to the vacuum mass eigenstates as 
\begin{eqnarray}
\nu_{\alpha} = N_{\alpha i} \check{\nu}_{i}. 
\label{N-def}
\end{eqnarray}
The properties of the Greek and Latin indices are as before. The CC and NC matter potentials~\cite{Wolfenstein:1977ue} $a(x)$ and $b(x)$, respectively, are defined in eq.~\eqref{matt-potential}. We use the uniform matter density approximation until reaching section~\ref{sec:hamiltonian-symmetry}.  

To parametrize the non-unitarity mixing matrix $N$, we use so called the $\alpha$ parametrization~\cite{Escrihuela:2015wra,Schechter:1980gr}: 
\begin{eqnarray} 
N 
&=& 
\left( \bf{1} - \alpha \right) U_{\text{\tiny PDG}} 
= 
\left\{ 
\bf{1} - 
\left[ 
\begin{array}{ccc}
\alpha_{ee} & 0 & 0 \\
\alpha_{\mu e} & \alpha_{\mu \mu}  & 0 \\
\alpha_{\tau e}  & \alpha_{\tau \mu} & \alpha_{\tau \tau} \\
\end{array}
\right] 
\right\} 
U_{\text{\tiny PDG}}, 
\label{alpha-matrix-def}
\end{eqnarray}
where $U_{\text{\tiny PDG}}$ denotes the PDG convention $U$ matrix defined in eq.~\eqref{MNS-PDG}. Notice that the $\alpha$ matrix defined as in eq.~\eqref{alpha-matrix-def} is $U$ matrix convention dependent~\cite{Martinez-Soler:2018lcy}, and hence the one in eq.~\eqref{alpha-matrix-def} is defined under the PDG convention. See Appendix~\ref{sec:3-conventions}. 
The diagonal $\alpha_{\beta \beta}$ parameters are real, and the off-diagonal ones $\alpha_{\beta \gamma}$ ($\beta \neq \gamma$) are complex so that the $\alpha$ matrix bring in the nine degrees of freedom in addition to the $\nu$SM ones. 

\subsection{Formulating the helio-UV perturbation theory}
\label{sec:formulation}

The renormalized helio-UV perturbation theory has two kind of the expansion parameters, $\epsilon$ and the UV $\alpha$ parameters. $\epsilon = \Delta m^2_{21} / \Delta m^2_{ \text{ren} }$ is the one used in the helio-perturbation theory~\cite{Minakata:2015gra}, as well as in the DMP perturbation theory as in eq.~\eqref{epsilon-Dm2-ren-def}. The other expansion parameters are the $\alpha$ matrix elements $\alpha_{\beta \gamma}$ in eq.~\eqref{alpha-matrix-def} which represent the UV effect. 

We start our formulation by transforming to the tilde basis with the Hamiltonian 
\begin{eqnarray} 
&&
\widetilde{H} 
=
( U_{13} U_{12} ) \check{H} ( U_{13} U_{12} )^{\dagger} 
= 
\widetilde{H} _{ \nu\text{SM}} 
+ \widetilde{H}_\text{ UV }^{(1)} 
+ \widetilde{H}_\text{ UV }^{(2)}, 
\label{tilde-H-helioUV}
\end{eqnarray}
where $U_{13}$ and $U_{12}$ without arguments imply the rotation matrices in vacuum~\eqref{MNS-PDG}, and 
\begin{eqnarray} 
&&
\widetilde{H} _{ \nu\text{SM}} 
= 
\frac{ \Delta m^2_{ \text{ren} } }{ 2E }
\left\{
\left[
\begin{array}{ccc}
\frac{ a(x) }{ \Delta m^2_{ \text{ren} }} + s^2_{13} + \epsilon s^2_{12} & 
0 & c_{13} s_{13} e^{- i \delta} \\
0 & \epsilon c^2_{12} & 0 \\
c_{13} s_{13} e^{ i \delta}  & 0 & c^2_{13} + \epsilon s^2_{12} 
\end{array}
\right] 
+ 
\epsilon c_{12} s_{12} 
\left[
\begin{array}{ccc}
0 & c_{13} & 0 \\
c_{13} & 0 & - s_{13} e^{- i \delta}  \\
0 & - s_{13} e^{ i \delta} & 0 
\end{array}
\right]
\right\}.
\nonumber \\
\label{tilde-H-helioP}
\end{eqnarray}
In $\widetilde{H} _{ \nu\text{SM}}$ in eq.~\eqref{tilde-H-helioP}, the rephasing to remove the NC potential is understood~\cite{Martinez-Soler:2018lcy}. We call the first and second terms in eq.~\eqref{tilde-H-helioP} as 
$\widetilde{H} _{ \nu\text{SM}}^{(0)}$ and $\widetilde{H} _{ \nu\text{SM}}^{(1)}$, respectively. 

Here is an important note for our nomenclature of the various bases. In the both SRP and helio-UV perturbation theories, we use the notation ``hat basis'' for the one with the diagonalized unperturbed Hamiltonian. See $\hat{H}$ in eq.~\eqref{hat-H-0th-1st-SRP} for the SRP and the one in eq.~\eqref{hatH-0th-1st-helioUV} for the helio-UV perturbation theory. 
The basis which is one step before the hat basis, i.e., the one to be diagonalized by a single rotation, is termed as the ``tilde basis'' in the both theories. 
Therefore, the definition of the tilde basis Hamiltonian differs between the SRP and helio-UV perturbation theories. The former is defined as $\widetilde{H} = U_{12} \check{H} U_{12}^{\dagger}$ and given in eq.~\eqref{tilde-H-SRP}, and the latter in eqs.~\eqref{tilde-H-helioUV} and \eqref{tilde-H-helioP}. Notice that the check basis is the vacuum mass eigenstate basis which is common to the both theories, apart from the difference of with and without the UV effects. 

The UV part in eq.~\eqref{tilde-H-helioUV} has the first and second order terms in the $\alpha$ parameters 
\begin{eqnarray} 
&&
\widetilde{H}_\text{ UV }^{(1)} = 
\frac{b}{2E} 
U_{23}^{\dagger} A U_{23}, 
\hspace{8mm}
\widetilde{H}_\text{ UV }^{(2)} = 
- \frac{b}{2E} 
U_{23}^{\dagger} A^{(2)} U_{23},
\label{tilde-H-UV}
\end{eqnarray}
where 
\begin{eqnarray}
A &\equiv&
\left[ 
\begin{array}{ccc}
2 \alpha_{ee} \left( 1 - \frac{ a (x) }{ b (x) } \right) & \alpha_{\mu e}^* & \alpha_{\tau e}^* \\
\alpha_{\mu e} & 2 \alpha_{\mu \mu}  & \alpha_{\tau \mu}^* \\
\alpha_{\tau e}  & \alpha_{\tau \mu} & 2 \alpha_{\tau \tau} \\
\end{array}
\right], 
\nonumber \\
A^{(2)} 
&\equiv& 
\left[
\begin{array}{ccc}
\alpha_{ee}^2 \left( 1 - \frac{ a (x) }{ b (x) } \right) + |\alpha_{\mu e}|^2 + |\alpha_{\tau e}|^2 & 
\alpha_{\mu e}^* \alpha_{\mu \mu} + \alpha_{\tau e}^* \alpha_{\tau \mu} & 
\alpha_{\tau e}^* \alpha_{\tau \tau} \\
\alpha_{\mu e} \alpha_{\mu \mu} + \alpha_{\tau e} \alpha_{\tau \mu}^* & 
\alpha_{\mu \mu}^2 + |\alpha_{\tau \mu}|^2 & 
\alpha_{\tau \mu}^* \alpha_{\tau \tau} \\
\alpha_{\tau e} \alpha_{\tau \tau} & 
\alpha_{\tau \mu} \alpha_{\tau \tau} & 
\alpha_{\tau \tau}^2 \\
\end{array}
\right].
\label{A-A2-def}
\end{eqnarray}
For a consistent nomenclature $A$ must carry the superscript as $A^{(1)}$, but for simplicity of the expressions we omit it throughout this paper. 
In what follows we keep omitting the superscript $^{(1)}$ for many of the quantities in first order in the $\alpha$ parameters, because our treatment will be free from the second order terms apart from in section~\ref{sec:hamiltonian-symmetry}. 

Then, we use $U_{13} (\phi, \delta)$ rotation to diagonalize $\widetilde{H}_{ \nu\text{SM}}^{(0)}$ in eq.~\eqref{tilde-H-helioP}: 
\begin{eqnarray} 
&&
U_{13} (\phi, \delta) ^{\dagger} 
\widetilde{H}_{ \nu\text{SM}}^{(0)}
U_{13} (\phi, \delta) 
= 
\frac{1}{2E}
\text{diag} (\lambda_{-}^{\nu\text{SM}}, \lambda_{0}^{\nu\text{SM}}, \lambda_{+}^{\nu\text{SM}}).
\label{eigenvalue-def}
\end{eqnarray}
where $\phi$ denotes the matter-dressed $\theta_{13}$, and the eigenvalues $\lambda_{-}$, $\lambda_{0}$, $\lambda_{+}$ are given by 
\begin{eqnarray} 
&&
\lambda_{-} ^{\nu\text{SM}}
= 
\sin^2 ( \phi - \theta_{13}) \Delta m^2_{ \text{ren} } 
+ c^2_{\phi} a  
+ \epsilon s^2_{12} \Delta m^2_{ \text{ren} }, 
\nonumber \\
&& 
\lambda_{0} ^{\nu\text{SM}}
= \epsilon c^2_{12} \Delta m^2_{ \text{ren} }, 
\nonumber \\
&& 
\lambda_{+} ^{\nu\text{SM}}
= 
\cos^2 ( \phi - \theta_{13}) \Delta m^2_{ \text{ren} } 
+ s^2_{\phi} a  
+ \epsilon s^2_{12} \Delta m^2_{ \text{ren} }.
\label{unrenom-eigenvalue}
\end{eqnarray} 
In the helio-perturbation theory~\cite{Minakata:2015gra}, $\lambda_{+}$ and $\lambda_{-}$ are always the two states which participate the atmospheric level crossing, and $\lambda_{+} = \lambda_{3}$ and $\lambda_{-} = \lambda_{2}$ ($\lambda_{-} = \lambda_{1}$) around the level crossing in the normal (inverted) mass ordering. See Fig.~3 in ref.~\cite{Minakata:2015gra}. Through the diagonalization procedure the matter mixing angle $\phi$ is determined as 
\begin{eqnarray} 
&& 
\cos 2\phi 
= 
\frac{ \cos 2\theta_{13} \Delta m^2_{ \text{ren} } - a }{ \lambda_{+}^{\nu\text{SM}} - \lambda_{-}^{\nu\text{SM}} }, 
\nonumber \\
&& 
\sin 2\phi 
= 
\frac{ \sin 2\theta_{13} \Delta m^2_{ \text{ren} } }{ \lambda_{+}^{\nu\text{SM}} - \lambda_{-}^{\nu\text{SM}} }. 
\label{phi-def}
\end{eqnarray}
We call the basis with diagonalized zeroth-order term 
as the hat basis. The first-order UV term has the form 
\begin{eqnarray} 
&& 
\hat{H}_\text{ UV }^{(1)} = 
\frac{b}{2E} 
U_{13} (\phi, \delta) ^{\dagger} 
U_{23}^{\dagger} 
A U_{23} 
U_{13} (\phi, \delta) 
\equiv \frac{b}{2E} G. 
\label{G-def}
\end{eqnarray}
For later convenience we parametrize the $G$ matrix elements by factoring out the $e^{ \pm i \delta}$ factors as 
\begin{eqnarray} 
&& 
G \equiv
\left[
\begin{array}{ccc}
H_{11} & e^{- i \delta} H_{12} & e^{- i \delta} H_{13} \\
e^{ i \delta} H_{21} & H_{22} & H_{23} \\
e^{ i \delta} H_{31} & H_{32} & H_{33} \\
\end{array}
\right]. 
\label{H-def-from-G}
\end{eqnarray}
The explicit expressions of the $H_{ij}$ are presented in Appendix~\ref{sec:H-elements}. 

\subsection{Renormalized eigenvalue basis}
\label{sec:Rhat-basis} 

As in ref.~\cite{Minakata:2022zua} we move to the ``renormalized'' hat basis in which the eigenvalues absorb the diagonal $H_{ii}$ elements 
\begin{eqnarray} 
&&
\lambda_{-} 
=
\lambda_{-}^{\nu\text{SM}} + b H_{11} - b H^{(2)}_{11}, 
\nonumber \\
&& 
\lambda_{0} 
=  
\lambda_{0} ^{\nu\text{SM}} 
+ b H_{22} - b H^{(2)}_{22}, 
\nonumber \\
&& 
\lambda_{+} 
= 
\lambda_{+} ^{\nu\text{SM}} 
+ b H_{33} - b H^{(2)}_{33}, 
\label{renom-eigenvalue}
\end{eqnarray} 
where $H_{ij}^{(2)}$ is defined by eq.~\eqref{H-def-from-G} but by replacing $A$ by 
$A^{(2)}$ in eq.~\eqref{G-def}. For an explicit form see eq.~\eqref{G2-def}. 
Restricting to the first-order $\nu$SM and UV terms the hat basis Hamiltonian takes the form, using the notation $s_{(\phi - \theta_{13})} \equiv \sin (\phi - \theta_{13})$ etc., as 
\begin{eqnarray} 
\hat{H} 
&=&
\frac{1}{2E} 
\left[
\begin{array}{ccc}
\lambda_{-} & 0 & 0 \\
0 & \lambda_{0} & 0 \\
0 & 0 & \lambda_{+} \\
\end{array}
\right] 
+
\epsilon c_{12} s_{12} 
\frac{\Delta m^2_{ \text{ren} }}{2E} 
\left[
\begin{array}{ccc}
0 & c_{(\phi - \theta_{13})} & 0 \\
c_{(\phi - \theta_{13})} & 0 & s_{(\phi - \theta_{13})} e^{ - i \delta} \\
0  & s_{(\phi - \theta_{13})} e^{ i \delta} & 0 
\end{array}
\right]
\nonumber \\
&+&
\frac{b}{2E} 
\left[
\begin{array}{ccc}
0 & e^{- i \delta} H_{12} & e^{- i \delta} H_{13} \\
e^{ i \delta} H_{21} & 0 & H_{23} \\
e^{ i \delta} H_{31} & H_{32} & 0 \\
\end{array}
\right].
\label{hatH-0th-1st-helioUV}
\end{eqnarray}

\subsection{Computation of the $V$ matrix: Zeroth order} 
\label{sec:Vmatrix-0th} 

We calculate the $V$ matrix to formulate the SF equation. At the zeroth order it is easy to obtain using the knowledge obtained above. The only point we have to pay attention is how non-unitarity affects the $V$ matrix. 
The relation between the hat (zeroth-order eigenstate in matter) and the check basis (vacuum mass eigenstate) Hamiltonian are given by
\begin{eqnarray} 
\hat{H} 
&=&
U_{13} (\phi, \delta) ^\dagger
\widetilde{H} 
U_{13} (\phi, \delta) 
= 
U_{13} (\phi, \delta) ^\dagger U_{13} U_{12} 
\check{H} 
U_{12}^{\dagger} U_{13}^{\dagger} U_{13} (\phi, \delta). 
\label{hat-H}
\end{eqnarray}
with $U_{13}$ and $U_{12}$ without arguments imply the ones in vacuum. It means, in terms of the states, 
$\hat{\nu}_{i} = [ U_{13} (\phi) ^\dagger U_{13} U_{12} ]_{ik} \check{\nu}_{k}$, or $\check{\nu}_{i} = [ U_{12}^{\dagger} U_{13}^{\dagger} U_{13}(\phi, \delta) ]_{ij} \hat{\nu}_{j} $. 
Then, the flavor state is connected to the hat-basis state as 
\begin{eqnarray} 
&& 
\nu_{\alpha} 
= \left\{ ( 1 - \alpha ) U \right\}_{\alpha i} \check{\nu}_{i} 
= 
\left[ ( 1 - \alpha ) 
U_{23} U_{13} U_{12} \right] _{\alpha i} \check{\nu}_{i} 
\nonumber \\
&=& 
\left[ ( 1 - \alpha ) 
U_{23} (\theta_{23}) 
U_{13} (\phi, \delta) 
\right] _{\alpha j} \hat{\nu}_{j} 
\equiv 
\left[ V^{(0)} + V^{(1)} _{ \text{UV} } \right] _{\alpha j} 
\hat{\nu}_{j} 
\label{flavor-hat}
\end{eqnarray}
where we have recovered the vacuum rotation angle $\theta_{23}$ for clarity.\footnote{
The formula corresponding to eq.~\eqref{flavor-hat} in DMP is: $\nu_{\alpha} = \left[ ( 1 - \widetilde{\alpha} ) U_{23} (\theta_{23}) U_{13}(\phi) U_{12} (\psi, \delta) \right] _{\alpha j} \bar{\nu}_{j}$~\cite{Minakata:2022zua}. }
Equation~\eqref{flavor-hat} reveals the $V$ matrix in the leading and first orders in the helio-UV perturbation theory. 
We shall treat the $\alpha$ term in \eqref{flavor-hat} as the first-order genuine UV term, so that the $V$ matrix is given at the zeroth and the first order UV terms as  
\begin{eqnarray} 
V^{(0)} 
&=& 
U_{23} (\theta_{23}) U_{13} (\phi, \delta), 
\nonumber \\
V^{(1)} _{ \text{UV} } 
&=& 
- \alpha U_{23} (\theta_{23}) U_{13} (\phi, \delta) 
= - \alpha V^{(0)}. 
\label{V0-V1UV-def} 
\end{eqnarray}

\subsection{First-order correction to the $V$ matrix}
\label{sec:Vmatrix-1st} 

In addition to the $\alpha$-matrix-origin first order term $V^{(1)} _{ \text{UV} }$ as given above, the other first-order correction arises from perturbative corrections due to $\hat{H}^{(1)}$. We call the former as the genuine UV part, as in the subscript, and the latter the unitary evolution part, the EV part. See ref.~\cite{Martinez-Soler:2018lcy} for these concepts. 

Since the computation for the first-order $V$ matrix is exactly parallel to the one in section~\ref{sec:V-matrix}, we just give the result, leaving the interested readers to follow the steps described there. The $V$ matrix representation of the flavor state in the form of eq.~\eqref{Vmatrix-def} to first order in the helio-UV perturbation theory is given as 
\begin{eqnarray} 
&& 
\left[
\begin{array}{c}
\nu_{e} \\
\nu_{\mu} \\
\nu_{\tau} \\
\end{array}
\right] 
= 
U_{23} (\theta_{23}) U_{13} (\phi, \delta) 
\biggl\{
1 + \mathcal{W}_{ \nu\text{SM} } ^{(1)} 
+ \mathcal{W}_{ \text{EV} } ^{(1)} 
- \mathcal{Z} _{ \text{UV} } ^{(1)} 
\biggr\} 
\left[
\begin{array}{c}
\hat{\nu}_{-} \\
\hat{\nu}_{0} \\
\hat{\nu}_{+} \\
\end{array}
\right].
\label{Vmatrix-for-SF}
\end{eqnarray}
These first-order terms in eq.~\eqref{Vmatrix-for-SF} are given by 
\begin{eqnarray} 
&&
\hspace{-6mm}
\mathcal{W}_{ \nu\text{SM} } ^{(1)} 
( \theta_{13}, \theta_{12}, \delta, \phi; \lambda_{i} ) 
= 
\epsilon c_{12} s_{12} 
\left[
\begin{array}{ccc}
0 & - c_{ ( \phi - \theta_{13} ) } \frac{ \Dmsqren }{ \lambda_{-} - \lambda_{0} } & \\
c_{ ( \phi - \theta_{13} ) } \frac{ \Dmsqren }{ \lambda_{-} - \lambda_{0} } & 0 & 
s_{ ( \phi - \theta_{13} ) } e^{ - i \delta} \frac{ \Dmsqren }{ \lambda_{+} - \lambda_{0} } \\
0 & 
- s_{ ( \phi - \theta_{13} ) } e^{ i \delta} \frac{ \Dmsqren }{ \lambda_{+} - \lambda_{0} } & 0 \\
\end{array}
\right], 
\nonumber \\
&&
\mathcal{W}_{ \text{EV} } ^{(1)} 
( \theta_{23}, \delta, \phi; \lambda_{i}, H_{ij} ) 
=
\left[
\begin{array}{ccc}
0 & - e^{- i \delta} H_{12} \frac{ b }{ \lambda_{-} - \lambda_{0} } & 
e^{- i \delta} H_{13} \frac{ b }{ \lambda_{+} - \lambda_{-} } \\
e^{ i \delta} H_{21} \frac{ b }{ \lambda_{-} - \lambda_{0} }  & 0 & 
H_{23} \frac{ b }{ \lambda_{+} - \lambda_{0} } \\
- e^{ i \delta} H_{31} \frac{ b }{ \lambda_{+} - \lambda_{-} } & 
- H_{32} \frac{ b }{ \lambda_{+} - \lambda_{0} }  & 
0 \\
\end{array}
\right], 
\nonumber \\
&&
\mathcal{Z} _{ \text{UV} } ^{(1)} 
( \theta_{23}, \delta, \phi; \alpha_{\beta \gamma} ) 
= 
\left( V^{(0)} \right)^{\dagger} \alpha V^{(0)}. 
\label{W-Z-helioP-UV}  
\end{eqnarray}
The last term, the genuine UV term, may require a comment. It was originally the $- \alpha V^{(0)}$ term in eq.~\eqref{V0-V1UV-def}, but we have defined $\mathcal{Z} _{ \text{UV} } ^{(1)}$ such that 
\begin{eqnarray} 
&&
V^{(1)} _{ \text{UV} }
= 
- \alpha V^{(0)} 
= 
- V^{(0)} \mathcal{Z} _{ \text{UV} } ^{(1)}, 
\label{Z-def}
\end{eqnarray}
to make the expression of the $V$ matrix coherent. The $\mathcal{Z} _{ \text{UV} }^{(1)}$ term in eq.~\eqref{W-Z-helioP-UV} is a simple solution of the equation~\eqref{Z-def}. 

\subsection{Computation of the probability with the $V$ matrix method}
\label{sec:probabilty-UV-EV} 

We calculate the oscillation probability with use of the $V$ matrix method, by utilizing eq.~\eqref{P-ba-general}. But since the probability in the $\nu$SM part is fully computed in ref.~\cite{Minakata:2015gra},\footnote{
The calculation in ref.~\cite{Minakata:2015gra} is done by using the ATM convention (see Appendix~\ref{sec:3-conventions}) of the $U$ matrix. In general, care is needed to translate it to the $V$ matrix under the present PDG convention. But, the change in convention does not alter the expression of the oscillation probability in the $\nu$SM part, because the rephasing cannot affect the observables. This statement is true for the UV part as well, but the $\alpha$ parameters must transform accordingly, as explained in Appendix~\ref{sec:3-conventions}. } 
here we concentrate on the UV-related parts only, the genuine non-unitary (UV) part, and unitary evolution (EV) part. See Appendix B in ref.~\cite{Minakata:2015gra} for the expressions of the probability in the $\nu$SM part.

We restrict ourselves to the $\nu_{\mu} \rightarrow \nu_{e}$ channel. That is because we give in section~\ref{sec:hamiltonian-symmetry} an all-order proof of the reparametrization symmetry we derive, which is valid in all the flavor oscillation channels. The genuine non-unitary part $P(\nu_\mu \rightarrow \nu_e)^{(1)} _{ \text{UV} }$ at first order is given by 
\begin{eqnarray} 
&&
P(\nu_\mu \rightarrow \nu_e)^{(1)} _{ \text{UV} }
\nonumber \\
&=&
2 s_{23} \sin 2\phi 
\left[ 
\cos 2\phi \mbox{Re} \left( \alpha_{\mu e} e^{ - i \delta} \right) 
- ( \alpha_{ee} + \alpha_{\mu \mu} ) s_{23} \sin 2\phi 
\right] 
\sin^2 \frac{( \lambda_{+} - \lambda_{-}) x}{4E} 
\nonumber \\
&-&
s_{23} \sin 2\phi  
\mbox{Im} \left( \alpha_{\mu e} e^{ - i \delta} \right)  
\sin \frac{( \lambda_{+} - \lambda_{-}) x}{2E}.  
\label{1st-order-Pmue-UV}
\end{eqnarray} 
For this computation, use of the original form $V^{(1)} _{ \text{UV} } = - \alpha V^{(0)}$ is more profitable. 
The unitary evolution part $P(\nu_{\mu} \rightarrow \nu_{e})^{(1)} _{ \text{EV} }$ reads 
\begin{eqnarray}
&&
P(\nu_{\mu} \rightarrow \nu_{e})^{(1)} _{ \text{EV} } 
\nonumber \\
&=& 
\sin 2\theta_{23} \sin 2\phi c_{\phi} \mbox{Re} \left( H_{21} \right) 
\frac{ b }{ \lambda_{-} - \lambda_{0} } 
\left\{
- \sin^2 \frac{( \lambda_{0} - \lambda_{-}) x}{4E} 
- \sin^2 \frac{( \lambda_{+} - \lambda_{-}) x}{4E} 
+ \sin^2 \frac{( \lambda_{+} - \lambda_{0}) x}{4E} 
\right\}
\nonumber \\
&+&
\sin 2\theta_{23} \sin 2\phi s_{\phi} \mbox{Re} \left( H_{23} \right) 
\frac{ b }{ \lambda_{+} - \lambda_{0} } 
\left\{
- \sin^2 \frac{( \lambda_{0} - \lambda_{-}) x}{4E} 
+ \sin^2 \frac{( \lambda_{+} - \lambda_{-}) x}{4E} 
+ \sin^2 \frac{( \lambda_{+} - \lambda_{0}) x}{4E} 
\right\}
\nonumber \\
&+&
4 s^2_{23} \cos 2\phi \sin 2\phi \mbox{Re} \left( H_{13} \right) 
\frac{ b }{ \lambda_{+} - \lambda_{-} } 
\sin^2 \frac{( \lambda_{+} - \lambda_{-}) x}{4E} 
\nonumber \\
&-&
2 \sin 2\theta_{23} \sin 2\phi 
\left\{ c_{\phi} \mbox{Im} \left( H_{21} \right) 
\frac{ b }{ \lambda_{-} - \lambda_{0} } 
+ s_{\phi} \mbox{Im} \left( H_{23} \right) 
\frac{ b }{ \lambda_{+} - \lambda_{0} } \right\} 
\nonumber \\
&\times& 
\sin \frac{( \lambda_{0} - \lambda_{-}) x}{4E} 
\sin \frac{( \lambda_{+} - \lambda_{-}) x}{4E} 
\sin \frac{( \lambda_{+} - \lambda_{0}) x}{4E}.
\label{1st-order-Pmue-EV}
\end{eqnarray}
These results agree with those obtained in ref.~\cite{Martinez-Soler:2018lcy}. 

\section{Symmetry Finder for the helio-UV perturbation theory} 
\label{sec:SF-helio-UV} 

We follow the SF method~\cite{Minakata:2021dqh,Minakata:2021goi,Minakata:2022zua} which is introduced in sections~\ref{sec:introducing-SF} and~\ref{sec:Symmetry-SRP}, and utilize the formalism to extract the symmetries from the helio-UV perturbation theory. For convenience of the readers who want to compare the outcome of our analysis to the one obtained for the DMP-UV-perturbation theory we have prepared Appendix~\ref{sec:DMP-UV-summary}  and Table~\ref{tab:DMP-UV-symmetry} for the DMP-UV, which is to be compared with Table~\ref{tab:helioP-UV-symmetry} for the helio-UV.

\subsection{Symmetry Finder (SF) equation}
\label{sec:SFeq}

For clarity we restrict ourselves to the reparametrization symmetry of the 1-3 (in our case $\nu_{-} \leftrightarrow \nu_{+}$) state exchange type. We start from the state which is physically equivalent with the one in eq.~\eqref{Vmatrix-for-SF}: 
\begin{eqnarray} 
&& 
\hspace{-6mm}
F \left[
\begin{array}{c}
\nu_{e} \\
\nu_{\mu} \\
\nu_{\tau} \\
\end{array}
\right] 
= 
F 
U_{23} (\theta_{23}) U_{13} (\phi, \delta) 
R^{\dagger} R 
\biggl\{
1 + \mathcal{W}_{ \nu\text{SM} } ^{(1)} 
+ \mathcal{W}_{ \text{EV} } ^{(1)} 
- \mathcal{Z} _{ \text{UV} } ^{(1)} 
\biggr\} 
R^{\dagger} R 
\left[
\begin{array}{c}
\nu_{-} \\
\nu_{0} \\
\nu_{+} \\
\end{array}
\right]. ~~
\label{SF-helioP-ansatz}
\end{eqnarray}
In eq.~\eqref{SF-helioP-ansatz} we have introduced the flavor-state rephasing matrix $F$ defined by 
\begin{eqnarray} 
&&
F \equiv 
\left[
\begin{array}{ccc}
e^{ i \tau } & 0 & 0 \\
0 & 1 & 0 \\
0 & 0 & e^{ i \sigma } \\
\end{array}
\right],
\label{F-def}
\end{eqnarray}
and a generalized $\nu_{-} \leftrightarrow \nu_{+}$ state exchange matrix $R$
\begin{eqnarray} 
&& 
R \equiv 
\left[
\begin{array}{ccc}
0 & 0 & - e^{ - i ( \delta - \alpha) } \\
0 & 1 & 0 \\
e^{ i ( \delta - \beta) } & 0 & 0 \\
\end{array}
\right], 
\hspace{8mm}
R^{\dagger} \equiv 
\left[
\begin{array}{ccc}
0 & 0 & e^{ - i ( \delta - \beta) } \\
0 & 1 & 0 \\
- e^{ i ( \delta - \alpha) } & 0 & 0 \\
\end{array}
\right],
\label{R-def}
\end{eqnarray}
where $\tau$, $\sigma$, $\alpha$, and $\beta$ denote the arbitrary phases. As we discuss the $\nu_{-} \leftrightarrow \nu_{+}$ exchange symmetry, the both $F$ and $R$ matrices in eqs.~\eqref{F-def} and \eqref{R-def} takes the nonvanishing, nontrivial (not unity) elements in the $\nu_{-} - \nu_{+}$ sub-sector.

The SF equation, the statement that the generic flavor state eq.~\eqref{SF-helioP-ansatz} should be written as a transformed state, is given by 
\begin{eqnarray} 
&&
\hspace{-10mm} 
\left[
\begin{array}{ccc}
e^{ i \tau } & 0 & 0 \\
0 & 1 & 0 \\
0 & 0 & e^{ i \sigma } \\
\end{array}
\right]
\left[
\begin{array}{c}
\nu_{e} \\
\nu_{\mu} \\
\nu_{\tau} \\
\end{array}
\right] 
=  
\left[
\begin{array}{ccc}
1 & 0 &  0  \\
0 & c_{23} & s_{23} e^{ - i \sigma } \\
0 & - s_{23} e^{ i \sigma } & c_{23} \\
\end{array}
\right] 
F U_{13} (\phi, \delta) R^{\dagger} 
\nonumber \\
&\times&
R 
\biggl\{
1 + \mathcal{W}_{ \nu\text{SM} }^{(1)} (\Phi; \lambda_{1}, \lambda_{2})
+ \mathcal{W}_{ \text{EV} }^{(1)} (\Phi, \alpha; \lambda_{1}, \lambda_{2})
- \mathcal{Z} _{ \text{UV} }^{(1)} (\Phi, \alpha)
\biggr\} 
R^{\dagger}
R 
\left[
\begin{array}{c}
\nu_{-} \\
\nu_{0} \\
\nu_{+} \\
\end{array}
\right] 
\nonumber \\
&& 
\hspace{-18mm} 
= \left[
\begin{array}{ccc}
1 & 0 &  0  \\
0 & c_{23}^{\prime} & s_{23}^{\prime} \\
0 & - s_{23}^{\prime} & c_{23}^{\prime} \\
\end{array}
\right] 
U_{13} ( \phi^{\prime}, \delta + \xi) 
%
\biggl\{
1 + \mathcal{W}_{ \nu\text{SM} }^{(1)} (\Phi^{\prime}; \lambda_{2}, \lambda_{1})
+ \mathcal{W}_{ \text{EV} }^{(1)} (\Phi^{\prime}, \alpha^{\prime}; \lambda_{2}, \lambda_{1})
- \mathcal{Z} _{ \text{UV} }^{(1)} (\Phi^{\prime}, \alpha^{\prime})
\biggr\} 
\left[
\begin{array}{c}
- e^{ - i ( \delta - \alpha) } \nu_{+} \\
\nu_{0} \\
e^{ i ( \delta - \beta) } \nu_{-} \\
\end{array}
\right]. 
\nonumber \\
\label{SF-eq-helioP}
\end{eqnarray} 

\subsection{The first and second conditions: $\nu$SM part} 
\label{sec:1st-2nd-condition-helioP} 

We solve the SF equation~\eqref{SF-eq-helioP} with the ansatz $s_{23}^{\prime} = s_{23} e^{ - i \sigma }$ which enforces $\sigma$ integral multiples of $\pi$. But, the corresponding condition for $s_{13}$ is missing. Though the ansatz for $s_{23}$ is sufficient for the decomposability of the SF equation into the first and second conditions, the restriction on $\tau$ of being integral multiples of $\pi$ is not imposed at this stage. 

The first and second conditions, the zeroth-order term and the $\nu$SM first-order term in eq.~\eqref{SF-eq-helioP} reads
\begin{eqnarray} 
F U_{13} (\phi, \delta) R^{\dagger} &=& 
U_{13} ( \phi^{\prime}, \delta + \xi), 
\nonumber \\
R \mathcal{W}_{ \nu\text{SM} }^{(1)} ( \theta_{13}, \theta_{12}, \delta, \phi; \lambda_{i} ) 
R^{\dagger} 
&=& 
\mathcal{W}_{ \nu\text{SM} }^{(1)} ( \theta_{13}^{\prime}, \theta_{12}^{\prime}, \delta + \xi, \phi^{\prime}; \lambda_{i} ). 
\label{SF-eq-1st-2nd-helioP}
\end{eqnarray}
The first condition can be boiled down to the compact form as 
\begin{eqnarray} 
&&
c_{\phi^{\prime}} 
= - s_{\phi} e^{ - i ( \alpha - \tau ) } 
= - s_{\phi} e^{ i ( \beta + \sigma ) }, 
\hspace{8mm}
s_{\phi^{\prime}} 
= c_{\phi} e^{ i ( \beta + \tau + \xi ) } 
= c_{\phi} e^{ - i ( \alpha - \sigma + \xi ) }, 
\label{1st-condition-helioP}
\end{eqnarray}
and the consistency conditions for the phases result. 
\begin{eqnarray} 
&&
\alpha + \beta - \tau + \sigma = 0 ~~~(\text{mod.} ~2\pi), 
\hspace{8mm}
\tau - \sigma + \xi = 0, ~\pm \pi. 
\label{1st-sol-consistency} 
\end{eqnarray}
The first condition \eqref{1st-condition-helioP} is identical with the one in eq.~\eqref{1st-condition-SRP} in the SRP theory. 
The explicit form of the second condition on the $\nu$SM part reads: 
\begin{eqnarray} 
&&
\epsilon c_{12} s_{12} 
\left[ 
\begin{array}{ccc}
0 & s_{ \left( \phi - \theta_{13} \right) } e^{ i \alpha } \frac{ \Delta m^2_{ \text{ren} } }{\lambda_{+} - \lambda_{0} } 
& 0 \\
- s_{ \left( \phi - \theta_{13} \right) } e^{ - i \alpha } 
\frac{ \Delta m^2_{ \text{ren} }  }{\lambda_{+} - \lambda_{0} } & 
0 & c_{ \left( \phi - \theta_{13} \right) } e^{ - i ( \delta - \beta) } \frac{ \Delta m^2_{ \text{ren} } }{\lambda_{-} - \lambda_{0} } \\
0 & 
- c_{ \left( \phi - \theta_{13} \right) } e^{ i ( \delta - \beta) } \frac{ \Delta m^2_{ \text{ren} }  }{\lambda_{-} - \lambda_{0} } & 
0
\end{array}
\right]  
\nonumber \\
&=& 
\epsilon c_{12}^{\prime} s_{12}^{\prime} 
\left[ 
\begin{array}{ccc}
0 & - c_{ ( \phi^{\prime} - \theta_{13}^{\prime} ) } 
\frac{ \Delta m^2_{ \text{ren} }  }{ \lambda_{+} - \lambda_{0} } & 0 \\
 c_{ ( \phi^{\prime} - \theta_{13}^{\prime} ) } 
 \frac{ \Delta m^2_{ \text{ren} } }{\lambda_{+} - \lambda_{0} }  & 0 & 
 s_{ ( \phi^{\prime} - \theta_{13}^{\prime} ) } e^{ - i ( \delta + \xi ) } \frac{ \Delta m^2_{ \text{ren} }  }{\lambda_{-} - \lambda_{0} } \\
0 & - s_{ ( \phi^{\prime} - \theta_{13}^{\prime} ) } e^{ i ( \delta + \xi ) } 
\frac{ \Delta m^2_{ \text{ren} } }{\lambda_{-} - \lambda_{0} } & 0
\end{array}
\right], 
\label{2nd-condition-full} 
\end{eqnarray}
where the notation is such that $c_{12}^{\prime} \equiv \cos \theta_{12}^{\prime}$, and $c_{ ( \phi^{\prime} - \theta_{13}^{\prime} ) } \equiv \cos ( \phi^{\prime} - \theta_{13}^{\prime} )$ etc. 

Here is an important note for $\tau$, $\sigma$, $\alpha$, $\beta$, and $\xi$, and their solutions. Equation~\eqref{1st-condition-helioP} tells us that $\alpha - \tau$ and $\beta + \sigma$ must be integers, where we abbreviate ``in units of $\pi$'' for the moment. Then, $\beta$ must be an integer as well. Now, the second condition~\eqref{2nd-condition-full} requires that $\alpha$ must be an integer, which implies that $\tau$ must be an integer. Look at the 1-2 or 2-1 elements. Then, by comparing the 2-3 elements at the both sides of eq.~\eqref{2nd-condition-full}  we know that $\xi$ is an integer. Thus, we have shown that $\tau$, $\sigma$, $\alpha$, $\beta$, and $\xi$ are all integers in units of $\pi$~\cite{Minakata:2021goi}. The resultant solutions of the first condition is tabulated in Table~\ref{tab:SF-solutions}, showing universal feature of the solutions as we mentioned in the SRP analysis. 

The $\nu$SM part of the first and second conditions in eqs.~\eqref{1st-condition-helioP} and~\eqref{2nd-condition-full} with $\mathcal{W}_{ \nu\text{SM} }^{(1)}$ given in eq.~\eqref{W-Z-helioP-UV} is fully analyzed in ref.~\cite{Minakata:2021goi}. It resulted in the sixteen reparametrization symmetries of the 1-3 state exchange type in the helio-perturbation theory. They are denoted as ``Symmetry X-helioP'', where X = IA, IB, IIA, IIB, IIIA, IIIB, IVA, and IVB, which are duplicated with ``non-f'' and ``f'' types, where the latter means flipping of $s_{12}$ is involved. See the first three columns of Table~\ref{tab:helioP-UV-symmetry}. 
Notice that the $\theta_{13}$ transformations, either the sign flip, or $\theta_{13} \rightarrow \theta_{13} \pm \pi$, or their combinations are involved in some of them. They arises as the solution of the second condition, as no $\theta_{13}$ is involved in the first condition. 

The decomposability of the second condition implies that the symmetries of the helio-UV theory cannot be larger than the sixteen symmetries. The question is whether all of them survive in the UV extension. 

%
\begin{table}[h!] 
\vglue -0.2cm
\begin{center}
\caption{Summary of the reparametrization symmetries in the helio-UV perturbation theory~\cite{Martinez-Soler:2018lcy}. The first column is for the symmetry type denoted as ``X'' where X = IA, IB, IIA, IIB, IIIA, IIIB, IVA, and IVB. Each X is duplicated with and without ``f'', where ``f'' implies $s_{12}$ sign flip. The first to third columns are identical to the ones in ref.~\cite{Minakata:2021goi}. The fourth column provides information of the $\alpha$ parameter transformation in the X row, and the rephasing matrix $\text{Rep(X)} _{ \text{\tiny helioP} }$ in the Xf row. The both of them are determined by the symmetry type and common to the symmetries XA, XAf, XB, and XBf (four rows). 
}
\label{tab:helioP-UV-symmetry}
\vglue 0.2cm
\begin{tabular}{c|c|c|c}
\hline 
Type & 
Vacuum parameter transf. & 
Matter parameter transf. &
$\alpha$ transf./Rep(X)
\\
\hline 
\hline 
IA & 
none & 
$\lambda_{-} \leftrightarrow \lambda_{+}$, 
$c_{\phi} \rightarrow \mp s_{\phi}$, 
$s_{\phi} \rightarrow \pm c_{\phi}$ & 
none \\
& & 
$c_{(\phi - \theta_{13})} \rightarrow \mp s_{(\phi - \theta_{13})}$, 
$s_{(\phi - \theta_{13})} \rightarrow \pm c_{(\phi - \theta_{13})}$ &
\\
\hline 
IAf & 
$\theta_{13} \rightarrow \theta_{13} \pm \pi$, $\theta_{12} \rightarrow - \theta_{12}$ & 
$\lambda_{-} \leftrightarrow \lambda_{+}$, 
$c_{\phi} \rightarrow \mp s_{\phi}$, 
$s_{\phi} \rightarrow \pm c_{\phi}$ & 
diag(1,1,1) \\
 & & 
$c_{(\phi - \theta_{13})} \rightarrow \pm s_{(\phi - \theta_{13})}$, 
$s_{(\phi - \theta_{13})} \rightarrow \mp c_{(\phi - \theta_{13})}$ & 
\\
\hline 
IB & 
$\theta_{13} \rightarrow - \theta_{13}$,  
$\delta \rightarrow \delta + \pi$. & 
$\lambda_{-} \leftrightarrow \lambda_{+}$, 
$c_{\psi} \rightarrow \pm s_{\psi}$, 
$s_{\psi} \rightarrow \pm c_{\psi}$ &
same as IA, IAf \\
 & & 
$c_{(\phi - \theta_{13})} \rightarrow \pm s_{(\phi - \theta_{13})}$, 
$s_{(\phi - \theta_{13})} \rightarrow \pm c_{(\phi - \theta_{13})}$ &
\\
\hline 
IBf & 
$\theta_{13} \rightarrow - \theta_{13} \pm \pi $, 
$\theta_{12} \rightarrow - \theta_{12}$, & 
$\lambda_{-} \leftrightarrow \lambda_{+}$, 
$c_{\psi} \rightarrow \pm s_{\psi}$, 
$s_{\psi} \rightarrow \pm c_{\psi}$ & 
same as IA, IAf \\
 & $\delta \rightarrow \delta + \pi$ & 
$c_{(\phi - \theta_{13})} \rightarrow \mp s_{(\phi - \theta_{13})}$, 
$s_{(\phi - \theta_{13})} \rightarrow \mp c_{(\phi - \theta_{13})}$ &
\\
\hline
IIA & 
$\theta_{23} \rightarrow - \theta_{23}$, 
$\theta_{13} \rightarrow - \theta_{13}$ & 
$\lambda_{-} \leftrightarrow \lambda_{+}$, 
$c_{\phi} \rightarrow \pm s_{\phi}$, 
$s_{\phi} \rightarrow \pm c_{\phi}$ &
$\alpha_{\tau e} \rightarrow - \alpha_{\tau e}$, 
\\ 
 & & 
$c_{(\phi - \theta_{13})} \rightarrow \pm s_{(\phi - \theta_{13})}$, 
$s_{(\phi - \theta_{13})} \rightarrow \pm c_{(\phi - \theta_{13})}$ &
$\alpha_{\tau \mu} \rightarrow - \alpha_{\tau \mu}$
\\
\hline 
IIAf & 
$\theta_{23} \rightarrow - \theta_{23}$, 
$\theta_{13} \rightarrow - \theta_{13} \pm \pi$ 
 & 
$\lambda_{-} \leftrightarrow \lambda_{+}$, 
$c_{\phi} \rightarrow \pm s_{\phi}$, 
$s_{\phi} \rightarrow \pm c_{\phi}$ & 
diag(1,1,-1)
\\ 
 & $\theta_{12} \rightarrow - \theta_{12}$  & 
$c_{(\phi - \theta_{13})} \rightarrow \mp s_{(\phi - \theta_{13})}$, 
$s_{(\phi - \theta_{13})} \rightarrow \mp c_{(\phi - \theta_{13})}$ \\
\hline 
IIB & 
$\theta_{23} \rightarrow - \theta_{23}$, 
$\delta \rightarrow \delta + \pi$ & 
$\lambda_{-} \leftrightarrow \lambda_{+}$, 
$c_{\phi} \rightarrow \mp s_{\phi}$, 
$s_{\phi} \rightarrow \pm c_{\phi}$ &
same as IIA, IIAf  
\\ 
 & & 
$c_{(\phi - \theta_{13})} \rightarrow \mp s_{(\phi - \theta_{13})}$, 
$s_{(\phi - \theta_{13})} \rightarrow \pm c_{(\phi - \theta_{13})}$ \\
\hline 
IIBf & 
$\theta_{23} \rightarrow - \theta_{23}$, 
$\theta_{13} \rightarrow \theta_{13} \pm \pi$, & 
$\lambda_{-} \leftrightarrow \lambda_{+}$, 
$c_{\phi} \rightarrow \mp s_{\phi}$, 
$s_{\phi} \rightarrow \pm c_{\phi}$ &
same as IIA, IIAf 
\\ 
 & $\theta_{12} \rightarrow - \theta_{12}$, $\delta \rightarrow \delta + \pi$ & 
$c_{(\phi - \theta_{13})} \rightarrow \pm s_{(\phi - \theta_{13})}$, 
$s_{(\phi - \theta_{13})} \rightarrow \mp c_{(\phi - \theta_{13})}$. \\
\hline 
IIIA & 
$\theta_{13} \rightarrow - \theta_{13} \pm \pi$, & 
$\lambda_{-} \leftrightarrow \lambda_{+}$, 
$c_{\phi} \rightarrow \pm s_{\phi}$, 
$s_{\phi} \rightarrow \pm c_{\phi}$ &
$\alpha_{\mu e} \rightarrow - \alpha_{\mu e}$ 
\\ 
 & & 
$c_{(\phi - \theta_{13})} \rightarrow \mp s_{(\phi - \theta_{13})}$, 
$s_{(\phi - \theta_{13})} \rightarrow \mp c_{(\phi - \theta_{13})}$ &
$\alpha_{\tau e} \rightarrow - \alpha_{\tau e}$ 
\\
\hline 
IIIAf & 
$\theta_{13} \rightarrow - \theta_{13}$, 
$\theta_{12} \rightarrow - \theta_{12}$. & 
$\lambda_{-} \leftrightarrow \lambda_{+}$, 
$c_{\phi} \rightarrow \pm s_{\phi}$, 
$s_{\phi} \rightarrow \pm c_{\phi}$ &
diag(-1,1,1)
\\ 
 & & 
$c_{(\phi - \theta_{13})} \rightarrow \pm s_{(\phi - \theta_{13})}$, 
$s_{(\phi - \theta_{13})} \rightarrow \pm c_{(\phi - \theta_{13})}$ \\
\hline 
IIIB & 
$\theta_{13} \rightarrow \theta_{13} \pm \pi$, 
$\delta \rightarrow \delta + \pi$. & 
$\lambda_{-} \leftrightarrow \lambda_{+}$, 
$c_{\phi} \rightarrow \mp s_{\phi}$, 
$s_{\phi} \rightarrow \pm c_{\phi}$ &
same as IIIA, IIIAf 
\\ 
 & & 
$c_{(\phi - \theta_{13})} \rightarrow \pm s_{(\phi - \theta_{13})}$, 
$s_{(\phi - \theta_{13})} \rightarrow \mp c_{(\phi - \theta_{13})}$ \\
\hline 
IIIBf & 
$\theta_{12} \rightarrow - \theta_{12}$, 
$\delta \rightarrow \delta + \pi$. & 
$\lambda_{-} \leftrightarrow \lambda_{+}$, 
$c_{\phi} \rightarrow \mp s_{\phi}$, 
$s_{\phi} \rightarrow \pm c_{\phi}$ &
same as IIIA, IIIAf 
\\ 
 & & 
$c_{(\phi - \theta_{13})} \rightarrow \mp s_{(\phi - \theta_{13})}$, 
$s_{(\phi - \theta_{13})} \rightarrow \pm c_{(\phi - \theta_{13})}$ \\
\hline 
IVA & 
$\theta_{23} \rightarrow - \theta_{23}$, 
$\theta_{13} \rightarrow \theta_{13} \pm \pi$ & 
$\lambda_{-} \leftrightarrow \lambda_{+}$, 
$c_{\phi} \rightarrow \mp s_{\phi}$, 
$s_{\phi} \rightarrow \pm c_{\phi}$ &
$\alpha_{\mu e} \rightarrow - \alpha_{\mu e}$ 
\\ 
 & & 
$c_{(\phi - \theta_{13})} \rightarrow \pm s_{(\phi - \theta_{13})}$, 
$s_{(\phi - \theta_{13})} \rightarrow \mp c_{(\phi - \theta_{13})}$ & 
$\alpha_{\tau \mu} \rightarrow - \alpha_{\tau \mu}$ 
\\
\hline 
IVAf & 
$\theta_{23} \rightarrow - \theta_{23}$, 
$\theta_{12} \rightarrow - \theta_{12}$ & 
$\lambda_{-} \leftrightarrow \lambda_{+}$, 
$c_{\phi} \rightarrow \mp s_{\phi}$, 
$s_{\phi} \rightarrow \pm c_{\phi}$ & 
diag(-1,1,-1)
\\ 
 & & 
$c_{(\phi - \theta_{13})} \rightarrow \mp s_{(\phi - \theta_{13})}$, 
$s_{(\phi - \theta_{13})} \rightarrow \pm c_{(\phi - \theta_{13})}$ \\
\hline 
IVB & 
$\theta_{23} \rightarrow - \theta_{23}$, 
$\theta_{13} \rightarrow - \theta_{13} \pm \pi$, & 
$\lambda_{-} \leftrightarrow \lambda_{+}$, 
$c_{\phi} \rightarrow \pm s_{\phi}$, 
$s_{\phi} \rightarrow \pm c_{\phi}$ &
same as IVA, IVAf 
\\ 
 &
$\delta \rightarrow \delta + \pi$. 
 &
$c_{(\phi - \theta_{13})} \rightarrow \mp s_{(\phi - \theta_{13})}$, 
$s_{(\phi - \theta_{13})} \rightarrow \mp c_{(\phi - \theta_{13})}$ \\
\hline 
IVBf & 
$\theta_{23} \rightarrow - \theta_{23}$, 
$\theta_{13} \rightarrow - \theta_{13}$, & 
$\lambda_{-} \leftrightarrow \lambda_{+}$, 
$c_{\phi} \rightarrow \pm s_{\phi}$, 
$s_{\phi} \rightarrow \pm c_{\phi}$ &
same as IVA, IVAf  
\\ 
 &
$\theta_{12} \rightarrow - \theta_{12}$, $\delta \rightarrow \delta + \pi$. 
 &
$c_{(\phi - \theta_{13})} \rightarrow \pm s_{(\phi - \theta_{13})}$, 
$s_{(\phi - \theta_{13})} \rightarrow \pm c_{(\phi - \theta_{13})}$ \\
\hline 
\end{tabular}
\end{center}
\vglue -0.4cm 
\end{table}

\subsection{The second condition: Genuine non-unitary and unitary evolution parts}
\label{sec:2nd-condition}

The first-order terms in the SF equation~\eqref{SF-eq-helioP} constitute the second condition which can be decomposed into the $\nu$SM, EV, and the UV parts. The first one is already analyzed in section~\ref{sec:1st-2nd-condition-helioP}. The latter two take the forms as 
\begin{eqnarray} 
R \mathcal{W}_{ \text{EV} } ^{(1)} ( \theta_{23}, \delta, \phi; \lambda_{i}, H_{ij} )
R^{\dagger} 
&=& 
\mathcal{W}_{ \text{EV} }^{(1)} ( \theta_{23}^{\prime}, \delta + \xi, \phi^{\prime}; \lambda_{i}^{\prime}, H_{ij}^{\prime}  ), 
\nonumber \\
R \mathcal{Z} _{ \text{UV} } ^{(1)} ( \theta_{23}, \delta, \phi; \alpha_{\beta \gamma} ) 
R^{\dagger} 
&=& 
\mathcal{Z}_{ \text{UV} }^{(1)} (\theta_{23}^{\prime}, \delta + \xi, \phi^{\prime}, \alpha_{\beta \gamma}^{\prime}).
\label{2nd-eq-helioP-UV} 
\end{eqnarray}
We analyze the genuine non-unitary and unitary evolution parts, the second and first lines in eq.~\eqref{2nd-eq-helioP-UV}, so that they are casted into the forms which are ready to solve. 

Now, the genuine non-unitary part first. It is useful to use the notation for the zeroth-order $V$ matrix as $V^{(0)} = U_{23} (\theta_{23}) U_{13} (\phi, \delta)$ as in eq.~\eqref{V0-V1UV-def} to make the equations compact. Using eq.~\eqref{W-Z-helioP-UV}, the second condition with $\mathcal{Z}_{ \text{UV} }^{(1)}$ in eq.~\eqref{2nd-eq-helioP-UV} takes the form 
\begin{eqnarray}
&&
R \left[ V^{(0)} ( \theta_{23}, \phi, \delta) \right]^{\dagger} \alpha 
V^{(0)} ( \theta_{23}, \phi, \delta) R^{\dagger} 
= 
\left[ V^{(0)} ( \theta_{23}^{\prime}, \phi^{\prime}, \delta + \xi) \right]^{\dagger} \alpha^{\prime} 
V^{(0)} ( \theta_{23}^{\prime}, \phi^{\prime}, \delta + \xi).
\nonumber \\
\label{2nd-condition-UV}
\end{eqnarray}
Then, the transformed $\alpha$ can be written in a closed form as 
\begin{eqnarray}
&&
\alpha^{\prime} 
=
V^{(0)} ( \theta_{23}^{\prime}, \phi^{\prime}, \delta + \xi) 
R 
\left[ V^{(0)} ( \theta_{23}, \phi, \delta) \right]^{\dagger} 
\alpha 
V^{(0)} ( \theta_{23}, \phi, \delta) R^{\dagger} 
\left[ V^{(0)} ( \theta_{23}^{\prime}, \phi^{\prime}, \delta + \xi) \right]^{\dagger}. 
\nonumber \\
\label{2nd-condition-UV2}
\end{eqnarray}
The right-hand side of this equation will be analyzed in the next section~\ref{sec:solution-UV}.

Next, we move to the second condition for the EV part. The second line in eq.~\eqref{2nd-eq-helioP-UV} can be written as 
\begin{eqnarray} 
&&
\left[
\begin{array}{ccc} 
0 & 
e^{ - i ( \delta - \alpha) } H_{32} \frac{ b }{ \lambda_{+} - \lambda_{0} }  & 
e^{ i ( \alpha + \beta ) } e^{ - i \delta} H_{31} \frac{ b }{ \lambda_{+} - \lambda_{-} } \\
- e^{ i ( \delta - \alpha) } H_{23} \frac{ b }{ \lambda_{+} - \lambda_{0} }  & 0 & 
e^{ i \beta } H_{21} \frac{ b }{ \lambda_{-} - \lambda_{0} } \\
- e^{ - i ( \alpha + \beta ) } e^{ i \delta} H_{13} \frac{ b }{ \lambda_{+} - \lambda_{-} } & 
- e^{ - i \beta } H_{12} \frac{ b }{ \lambda_{-} - \lambda_{0} } & 
0 \\
\end{array}
\right] 
\nonumber \\
&&
\hspace{16mm} 
= 
\left[
\begin{array}{ccc}
0 & - e^{ - i ( \delta + \xi ) } H_{12}^{\prime} \frac{ b }{ \lambda_{+} - \lambda_{0} } & 
- e^{ - i ( \delta + \xi ) } H_{13}^{\prime} \frac{ b }{ \lambda_{+} - \lambda_{-} } \\
e^{ i ( \delta + \xi ) } H_{21}^{\prime} \frac{ b }{ \lambda_{+} - \lambda_{0} }  & 0 & 
H_{23}^{\prime} \frac{ b }{ \lambda_{-} - \lambda_{0} } \\
e^{ i ( \delta + \xi ) } H_{31}^{\prime} \frac{ b }{ \lambda_{+} - \lambda_{-} } & 
- H_{32}^{\prime} \frac{ b }{ \lambda_{-} - \lambda_{0} }  & 
0 \\
\end{array}
\right].
\label{2nd-condition-EV}
\end{eqnarray}
One can show, by using the hermiticity of the $H$ matrix, that it can be written in a reduced form as 
\begin{eqnarray} 
&&
H_{21}^{\prime} 
=  - e^{ - i ( \alpha + \xi ) } H_{23} 
\nonumber \\
&& 
H_{13}^{\prime}
= 
- e^{ i ( \alpha + \beta + \xi ) } H_{31} 
\nonumber \\
&& 
H_{23}^{\prime} 
= e^{ i \beta } H_{21}.
\label{2nd-condition-EV2}
\end{eqnarray}
Notice the vastly different features of the second condition on $\mathcal{Z}_{ \text{UV} }^{(1)}$ in eq.~\eqref{2nd-condition-UV2}, and the one on $\mathcal{W}_{ \text{EV} }^{(1)}$ in eq.~\eqref{2nd-condition-EV2}. It makes consistency between them highly nontrivial. 

\section{Solution of the SF equation in the helio-UV perturbation theory}
\label{sec:SF-solution-helioP-UV} 

\subsection{Solution of the second condition: Genuine non-unitary part} 
\label{sec:solution-UV} 

We discuss first the genuine non-unitary part because we encounter with an important concept, which will be denoted as the ``key identity'' as below. If we use the simplified notation 
$[V R V^{\dagger}] \equiv V^{(0)} ( \theta_{23}^{\prime}, \phi^{\prime}, \delta + \xi) R 
\left[ V^{(0)} ( \theta_{23}, \phi, \delta) \right]^{\dagger}$, eq.~\eqref{2nd-condition-UV2} can be written as $\alpha^{\prime} = [V R V^{\dagger}] \alpha [V R V^{\dagger}]^{\dagger}$. Therefore, we first calculate the block $[V R V^{\dagger}]$ in the two steps. We define $C [13]$ as: 
\begin{eqnarray}
&&
C [13] \equiv 
U_{13} (\phi^{\prime}, \delta + \xi) R
U_{13} (\phi, \delta)^{\dagger} 
\nonumber \\
&=&
\left[
\begin{array}{ccc}
c_{\phi}^{\prime} & 0 & s_{\phi}^{\prime} e^{- i ( \delta + \xi) } \\
0 & 1 & 0 \\
- s_{\phi}^{\prime} e^{ i ( \delta + \xi) } & 0 & c_{\phi}^{\prime} \\
\end{array}
\right] 
\left[
\begin{array}{ccc}
0 & 0 & - e^{ - i ( \delta - \alpha) } \\
0 & 1 & 0 \\
e^{ i ( \delta - \beta) } & 0 & 0 \\
\end{array}
\right] 
\left[
\begin{array}{ccc}
c_{\phi} & 0 & - s_{\phi} e^{- i \delta} \\
0 & 1 & 0 \\
s_{\phi} e^{ i \delta} & 0 & c_{\phi} \\
\end{array}
\right], 
\label{C13-def}
\end{eqnarray}
so that 
\begin{eqnarray}
&&
[V R V^{\dagger}] 
\equiv 
V^{(0)} ( \theta_{23}^{\prime}, \phi^{\prime}, \delta + \xi) 
R 
\left[ V^{(0)} ( \theta_{23}, \phi, \delta) \right]^{\dagger}  
= 
\left[
\begin{array}{ccc}
1 & 0 &  0  \\
0 & c_{23}^{\prime} & s_{23}^{\prime} \\
0 & - s_{23}^{\prime} & c_{23}^{\prime} \\
\end{array}
\right] 
C [13]
\left[
\begin{array}{ccc}
1 & 0 &  0  \\
0 & c_{23} & - s_{23} \\
0 & s_{23} & c_{23} \\
\end{array}
\right]. 
\nonumber \\
\label{VRVdagger}
\end{eqnarray} 

We simply calculate $C [13]$ and $[V R V^{\dagger}]$ by inserting each solution of the SF equation in Table~\ref{tab:helioP-UV-symmetry} one by one with the values of the phase parameters $\alpha$, $\beta$, etc. corresponding to each solution as given in Table~\ref{tab:SF-solutions}. To our surprise, computation with all the solutions in Table~\ref{tab:helioP-UV-symmetry} entails an extremely simple result: 
\begin{eqnarray}
&&
C [13] 
= 
V^{(0)} ( \theta_{23}^{\prime}, \phi^{\prime}, \delta + \xi) 
R 
\left[ V^{(0)} ( \theta_{23}, \phi, \delta) \right]^{\dagger} 
= 
\text{Rep(X)} _{ \text{\tiny helioP} },
\label{identity-helioP}
\end{eqnarray}
which is a mixing-parameter independent {\em constant} despite the profound dependences on the $\nu$SM variables in $[V R V^{\dagger}]$ in the left-hand side. That is, $\text{Rep(X)} _{ \text{\tiny helioP} }$ denotes the rephasing matrix which is necessary for the Hamiltonian proof of the symmetry~\cite{Minakata:2021goi}, and is given by 
\begin{eqnarray} 
&& 
\text{Rep(II)} _{ \text{\tiny helioP} }
\equiv 
\left[
\begin{array}{ccc}
1 & 0 & 0 \\
0 & 1 & 0 \\
0 & 0 & -1
\end{array}
\right] ~~\text{for~ IIA, IIAf, IIB, IIBf}, ~~
\nonumber \\
&& 
\text{Rep(III)} _{ \text{\tiny helioP} }
\equiv 
\left[
\begin{array}{ccc}
-1 & 0 & 0 \\
0 & 1 & 0 \\
0 & 0 & 1
\end{array}
\right] ~~\text{for~ IIIA, IIIAf, IIIB, IIIBf}, 
\nonumber \\
&& 
\text{Rep(IV)} _{ \text{\tiny helioP} }
\equiv 
\left[
\begin{array}{ccc}
- 1 & 0 & 0 \\
0 & 1 & 0 \\
0 & 0 & - 1
\end{array}
\right] ~~\text{for~ IVA, IVAf, IVB, IVBf},
\label{rephasing-matrix}
\end{eqnarray}
and $\text{Rep(I)} _{ \text{\tiny helioP} }=$ diag(1,1,1) for IA, IAf, IB, IBf. 

The feature of this result is in complete parallelism with the DMP-UV theory~\cite{Minakata:2022zua}, but the DMP-UV results are not exactly the same as the helio-UV's. To distinguish our result from the DMP's we have introduced the notation $\text{Rep(X)} _{ \text{\tiny helioP} }$ with the index showing the theory dependence. See eq.~\eqref{Rep-II-III-IV-DMP} in Appendix~\ref{sec:DMP-UV-summary} for the expressions of $\text{Rep(X)} _{ \text{\tiny DMP} }$, which can be compared to $\text{Rep(X)} _{ \text{\tiny helioP} }$. Roughly speaking, the relation between the rephasing matrices of the DMP-UV and~helio-UV perturbation theories is Rep(II) $\leftrightarrow$ Rep(IV). 
Notice that our classification scheme of Symmetry X is based on the solutions of the first condition, and we do not arbitrarily alter the definitions of the symmetries  in each theory. 

It appears that the result~\eqref{identity-helioP}, in particular the second equality,  implies existence of the extremely interesting identities, which we call the ``key identity'' hereafter. 

Then, the second condition on $\mathcal{Z}^{(1)}_{ \text{UV} }$, written as the equation on the $\alpha$ matrix, eq.~\eqref{2nd-condition-UV2}, can readily be written as 
\begin{eqnarray}
&&
\alpha ^{\prime} 
=
\text{Rep(X)} _{ \text{\tiny helioP} } ~\alpha~
\text{Rep(X)} _{ \text{\tiny helioP} }^{\dagger}, 
\label{2nd-condition-UV-sol}
\end{eqnarray}
which implies that $\alpha ^{\prime} = \alpha$ for Symmetry X = I, and 
\begin{eqnarray}
&&
\alpha ^{\prime} 
= 
\left[ 
\begin{array}{ccc}
\alpha_{ee} & 0 & 0 \\
\alpha_{\mu e} & \alpha_{\mu \mu}  & 0 \\
- \alpha_{\tau e}  & - \alpha_{\tau \mu} & \alpha_{\tau \tau} \\
\end{array}
\right],
\hspace{5mm}
\left[ 
\begin{array}{ccc}
\alpha_{ee} & 0 & 0 \\
- \alpha_{\mu e} & \alpha_{\mu \mu}  & 0 \\
- \alpha_{\tau e}  & \alpha_{\tau \mu} & \alpha_{\tau \tau} \\
\end{array}
\right], 
\hspace{5mm}
\left[ 
\begin{array}{ccc}
\alpha_{ee} & 0 & 0 \\
- \alpha_{\mu e} & \alpha_{\mu \mu}  & 0 \\
\alpha_{\tau e}  & - \alpha_{\tau \mu} & \alpha_{\tau \tau} \\
\end{array}
\right], 
\label{alpha-transf-II-IV}
\end{eqnarray}
for X = II, III, and IV, in order. As in the case of DMP-UV symmetries no UV $\alpha$ parameters' transformation is present in Symmetry X=IA, IB, and their $s_{12}$ flipped counterpart. 

The resulting transformation properties of the $\alpha$ parameters and $\text{Rep(X)} _{ \text{\tiny helioP} }$ are summarized in Table~\ref{tab:helioP-UV-symmetry}. The corresponding informations in DMP, $\text{Rep(X)} _{ \text{\tiny DMP} }$ and the $\widetilde{\alpha}$ parameters' transformation are given in eq.~\eqref{Rep-II-III-IV-DMP} and Table~\ref{tab:DMP-UV-symmetry}, respectively. Notice that $\text{Rep(X)} _{ \text{\tiny helioP} }$ and hence the $\alpha$ parameters' transformation properties depend only on the symmetry type X = I, II, III, and IV, but not on the types A, Af, B, and Bf. 

\subsection{Solution of the second condition: Unitary evolution part}
\label{sec:solution-EV} 

The solutions of the first condition depend not only on the symmetry types denoted generically as XA and XB, but also on the upper and lower signs of the phase parameters $\alpha$, $\beta$, etc., as summarized in Table~\ref{tab:SF-solutions}. Using the phase parameters, one can show that the second condition~\eqref{2nd-condition-EV2} implies that $H_{ij}$ transform under Symmetry X as: 
\begin{eqnarray} 
&& 
\text{Symmetry IA, IAf, IIIB, IIIBf}: ~~~
H_{13}^{\prime} = - H_{31}, 
\hspace{10mm} 
H_{21}^{\prime} = \mp H_{23}, 
\hspace{10mm} 
H_{23}^{\prime} = \pm H_{21}, 
\nonumber \\
&&
\text{Symmetry IB, IBf, IIIA, IIIAf}: ~~~
H_{13}^{\prime} = H_{31}, 
\hspace{10mm} 
H_{21}^{\prime} = \mp H_{23}, 
\hspace{10mm} 
H_{23}^{\prime} = \mp H_{21}, 
\nonumber \\
&&
\text{Symmetry IIA, IIAf, IVB, IVBf}: ~~~
H_{13}^{\prime} = H_{31}, 
\hspace{10mm} 
H_{21}^{\prime} = \pm H_{23}, 
\hspace{10mm} 
H_{23}^{\prime} = \pm H_{21}, 
\nonumber \\
&&
\text{Symmetry IIB, IIBf, IVA, IVAf}: ~~~
H_{13}^{\prime} = - H_{31}, 
\hspace{10mm} 
H_{21}^{\prime} = \pm H_{23}, 
\hspace{10mm} 
H_{23}^{\prime} = \mp H_{21}, 
\nonumber \\
\label{Hij-transf}
\end{eqnarray}
where $\pm$ (or $\mp$) sign refers to the upper and lower signs in Table~\ref{tab:SF-solutions} and Table~\ref{tab:helioP-UV-symmetry}, 
which are synchronized between them. Notice that the transformation property of $H_{ji}$ can be obtained from the one of $H_{ij}$ by using the hermiticity $H_{ji} = (H_{ij})^*$. 

Here is a comment on exchange transformations of the eigenvalues. Since we have renormalized the eigenvalues such that the diagonal $H_{ii}$ elements are absorbed into the eigenvalues, see eq.~\eqref{renom-eigenvalue}, the second condition~\eqref{2nd-condition-EV2} does not contain the information on the $H_{ii}$ transformations. Hence it must be determined by the consistency with the eigenvalue exchange $\lambda_{-} \leftrightarrow \lambda_{+}$. That is, 
\begin{eqnarray} 
&& 
H_{11} \leftrightarrow H_{33}, 
\label{H-diag-transf}
\end{eqnarray} 
and $H_{22}$ is invariant. 

\subsection{Consistency between the UV and EV solutions and invariance of the oscillation probability}
\label{sec:consistency} 

The next crucial step is to verify the consistency between the solutions of the SF equation obtained from its genuine non-unitary part given in eq.~\eqref{2nd-condition-UV-sol} and the $H_{ij}$ transformations given in eqs.~\eqref{Hij-transf} and \eqref{H-diag-transf}. Using the explicit expressions of $H_{ij}$ in Appendix~\ref{sec:H-elements}, the consistency can be shown to hold for all the Symmetry X-helioP-UV in Table~\ref{tab:helioP-UV-symmetry}. Though this is a crucially important step, we would like to leave this exercise to the interested readers because it can be done straightforwardly. 

The remaining task is to verify the invariance of the oscillation probabilities $P(\nu_{\mu}~\rightarrow~\nu_{e})^{(1)}_{ \text{EV} }$ in eq.~\eqref{1st-order-Pmue-EV} and $P(\nu_{\mu} \rightarrow \nu_{e})^{(1)}_{ \text{UV} }$  in eq.~\eqref{1st-order-Pmue-UV}. The former is written in terms of the $\nu$SM and $H_{ij}$ parameters without any naked $\alpha$ parameters. Therefore, showing the invariance under Symmetry X can be carried out straightforwardly for all the sixteen symmetries with the transformation properties of these parameters given in Table~\ref{tab:helioP-UV-symmetry} and eq.~\eqref{Hij-transf}. 
On the other hand, $P(\nu_{\mu} \rightarrow \nu_{e})^{(1)}_{ \text{UV} }$ consists of the $\nu$SM and the naked $\alpha$ parameters. We can use the transformation properties of these variables summarized in Table~\ref{tab:helioP-UV-symmetry} to prove the invariance under the all Symmetry X. These exercises for invariance proof are again left for the interested readers. 

In this paper we do not discuss the other oscillation channels explicitly, apart from the $\nu_{\mu} \rightarrow \nu_{e}$ as in the above, because we will prove the Hamiltonian invariance in section~\ref{sec:hamiltonian-symmetry} which automatically applies to all the oscillation channels. 

\section{The helioP-UV symmetry as a Hamiltonian symmetry}
\label{sec:hamiltonian-symmetry}

In this section we show that all the helioP-UV symmetries summarized in Table~\ref{tab:helioP-UV-symmetry} leave the flavor basis Hamiltonian invariant up to the rephasing factor. It implies that all the helioP-UV symmetries hold in all orders in the helio-UV perturbation theory. Therefore, our discussion in this section will include the full Hamiltonian including the second order UV terms $A^{(2)}$ in eq.~\eqref{A-A2-def}. 

We have the following two ways to construct the flavor basis Hamiltonian, $H_{\text{\tiny VM}}$ and $H_{\text{\tiny Diag}}$.\footnote{
In refs.~\cite{Minakata:2021dqh,Minakata:2021goi,Minakata:2022zua} and arXiv v1 of this article, $H_{\text{\tiny VM}}$ and $H_{\text{\tiny Diag}}$ are denoted as $H_{\text{\tiny LHS}}$ and $H_{\text{\tiny RHS}}$, respectively. } 
For $H_{\text{\tiny VM}}$ whose subscript implies ``vacuum-matter'', which means that it is composed of the vacuum and matter terms. In unitary case in vacuum $H_{ \text{flavor} } = U \check{H} U^{\dagger}$, where $\check{H}$ is the vacuum mass eigenstate basis Hamiltonian, and $U$ the $\nu$SM flavor mixing matrix, see eq.~\eqref{MNS-PDG}. 
In non-unitary case in matter, since the flavor basis $\nu$ is related to the mass eigenstate basis $\check{\nu}$ as $\nu = N \check{\nu}$, the flavor-basis Hamiltonian which we call $H_{\text{\tiny VM}}$ can be written as 
$H_{\text{\tiny VM}} = N \check{H} N^{\dagger}$, where the check basis Hamiltonian $\check{H}$ is given in eq.~\eqref{checkH-UV-def}. 
For $H_{\text{\tiny Diag}}$ the subscript implies ``diagonalized'', which exhibits the feature that it is the Hamiltonian obtained by rotation back from the diagonalized hat-basis to the flavor basis. The way how $H_{\text{\tiny Diag}}$ is obtained will be explained in section~\ref{sec:H-RHS-transf}. Of course, they are equal to each other, $H_{\text{\tiny VM}} = H_{\text{\tiny Diag}}$. 

\subsection{Transformation property of $H_{\text{\tiny VM}}$} 
\label{sec:H-LHS-transf}

Using $N= \left( 1 - \alpha \right) U$ and $N N^{\dagger} = \left( 1 - \alpha \right) \left( 1 - \alpha \right)^{\dagger}$, $2E$ times $H_{\text{\tiny VM}} = N \check{H} N^{\dagger}$ can be written as 
\begin{eqnarray}
&&
\hspace{-4mm}
2E H_{\text{\tiny VM}} =
\left( 1 - \alpha \right) 
\left\{ 
U (\Xi)
\left[
\begin{array}{ccc}
m^2_{1} & 0 & 0 \\
0 & m^2_{2} & 0 \\
0 & 0 & m^2_{3} \\
\end{array}
\right] 
U (\Xi)^{\dagger}  
+ 
( 1 - \alpha )^{\dagger} 
\cdot 
\left[
\begin{array}{ccc}
a - b & 0 & 0 \\
0 & - b & 0 \\
0 & 0 & - b \\
\end{array}
\right] 
\cdot
\left( 1 - \alpha \right) 
\right\} 
( 1 - \alpha )^{\dagger},
\nonumber \\
\label{H-LHS}
\end{eqnarray}
where we have used a collective notation $\Xi$ for all the vacuum parameters involved. Here we have used a slightly different phase-redefined basis from the one in eq.~\eqref{checkH-UV-def} to make the vacuum Hamiltonian $\propto$ diag($m^2_{1}, m^2_{2}, m^2_{3}$) making it more symmetric, but it does not affect our symmetry discussion. 

We have shown in ref.~\cite{Minakata:2021goi} that the vacuum term transforms under Symmetry X as 
\begin{eqnarray}
&&
\left\{ 
U (\Xi)
\left[
\begin{array}{ccc}
m^2_{1} & 0 & 0 \\
0 & m^2_{2} & 0 \\
0 & 0 & m^2_{3} \\
\end{array}
\right] 
U (\Xi)^{\dagger} 
\right\} 
\rightarrow 
\text{Rep(X)} _{ \text{\tiny helioP} }
\left\{ 
U (\Xi)
\left[
\begin{array}{ccc}
m^2_{1} & 0 & 0 \\
0 & m^2_{2} & 0 \\
0 & 0 & m^2_{3} \\
\end{array}
\right] 
U (\Xi)^{\dagger} 
\right\}
\text{Rep(X)} _{ \text{\tiny helioP} }^{\dagger}.
\nonumber \\
\label{Hvac-transform}
\end{eqnarray}
where $\text{Rep(X)} _{ \text{\tiny helioP} }$ is the rephasing matrix defined in eq.~\eqref{rephasing-matrix}.
Using the transformation property $\alpha ^{\prime} = \text{Rep(X)} _{ \text{\tiny helioP} } \alpha \text{Rep(X)} _{ \text{\tiny helioP} } ^{\dagger}$ in eq.~\eqref{2nd-condition-UV-sol}, the matter term in eq.~\eqref{H-LHS} which originates from the $\nu$SM and the UV sectors of the theory obeys the same transformation property as in the vacuum term. Then, the whole $H_{\text{\tiny VM}}$ transforms under Symmetry X as 
\begin{eqnarray}
&&
H_{\text{\tiny VM}} 
\rightarrow 
\text{Rep(X)} _{ \text{\tiny helioP} }
H_{\text{\tiny VM}} 
\text{Rep(X)} _{ \text{\tiny helioP} }^{\dagger}, 
\label{H-LHS-transformed}
\end{eqnarray}
which means that $H_{\text{\tiny VM}}$ is invariant under Symmetry X up to the rephasing factor $\text{Rep(X)} _{ \text{\tiny helioP} }$. By being the real diagonal matrix with unit elements $\pm 1$, $\text{Rep(X)} _{ \text{\tiny helioP} }$ does not affect physical observables as it can be absorbed into the neutrino wave functions. 

We note that the vacuum and matter terms of $H_{\text{\tiny VM}}$ in eq.~\eqref{H-LHS} have quadratic and quartic dependences on $(1 - \alpha)$, respectively. The fact that they have the same transformation property with $\text{Rep(X)} _{ \text{\tiny helioP} }$ under Symmetry X solely relies on the $\alpha$ matrix transformation property in eq.~\eqref{2nd-condition-UV-sol}. 
On the other hand, $\text{Rep(X)} _{ \text{\tiny helioP} }$ is inherently the $\nu$SM concept, see eq.~\eqref{identity-helioP}. Therefore, there is no a priori reason why $\alpha$ must transform by it and only by it. With the ``Columbus' egg'' view one might argue that: Of course it must be the case because invariance under the symmetry requires it. Yet, it is remarkable to see that it indeed emerges from the theory via the genuine UV part of the SF equation~\eqref{2nd-condition-UV2}. It indicates an intriguing interplay between the $\nu$SM and the UV sectors in the theory. 

In passing, we note that we do not use the property that the matter density is uniform to obtain the invariance proof, the feature which prevails in the proof of invariance of $H_{\text{\tiny Diag}}$ in section~\ref{sec:H-RHS-transf}. 

\subsection{Transformation property of $H_{\text{\tiny Diag}}$} 
\label{sec:H-RHS-transf}

In this section we discuss $H_{\text{\tiny Diag}}$ to show that it is invariant under Symmetry X-helioP-UV with the same rephasing matrix as needed for $H_{\text{\tiny VM}}$. We first construct $H_{\text{\tiny Diag}}$. By using the state relation in eq.~\eqref{flavor-hat} $H_{\text{\tiny Diag}}$ is given by the hat-basis Hamiltonian $\hat{H}$ as 
\begin{eqnarray}
H_{\text{\tiny Diag}} 
&=& 
( 1 - \alpha ) U_{23} U_{13} (\phi, \delta) 
\hat{H} 
U^\dagger_{13} (\phi, \delta) U_{23}^{\dagger} 
( 1 - \alpha )^{\dagger}.
\label{H-RHS}
\end{eqnarray}
The expression of $\hat{H}$ is given in eq.~\eqref{hatH-0th-1st-helioUV} to first order in the helio-UV perturbation. In this section we proceed with this first-order Hamiltonian to prove invariance of $H_{\text{\tiny Diag}}$ under Symmetry X. In section~\ref{sec:2nd-order} we will present a simple argument to show that our proof of invariance prevails even after we include the second order effect. 

Since we innovate the way to prove the invariance $H_{\text{\tiny Diag}}$, we include the $\nu$SM part as well, though it has been fully treated in ref.~\cite{Minakata:2021goi}. From the identity~\eqref{identity-helioP} one obtains 
\begin{eqnarray}
&&
V^{(0)} ( \theta_{23}^{\prime}, \phi^{\prime}, \delta^{\prime} ) 
= 
\text{Rep(X)} _{ \text{\tiny helioP} }
V^{(0)} ( \theta_{23}, \phi, \delta) R^{\dagger}.
\label{V0-prime-V0}
\end{eqnarray}
Then, $H_{\text{\tiny Diag}}$ in eq.~\eqref{H-RHS} with use of eq.~\eqref{V0-V1UV-def} transforms under Symmetry X as 
\begin{eqnarray} 
&&
H_{\text{\tiny Diag}} 
= 
( 1 - \alpha ) 
V^{(0)} ( \theta_{23}, \phi, \delta) 
\hat{H} ( \theta_{23}, \theta_{12}, \phi, \delta; \widetilde{\alpha}_{\beta \gamma}, \lambda_{i} ) 
\left[ V^{(0)} ( \theta_{23}, \phi, \delta) \right]^{\dagger} 
( 1 - \alpha )^{\dagger} 
\nonumber \\
&\rightarrow&_{\text{\tiny Symmetry X}} ~
( 1 - \alpha^{\prime} ) 
V^{(0)} ( \theta_{23}^{\prime}, \phi^{\prime}, \delta^{\prime} ) 
\hat{H} ( \theta_{23}^{\prime}, \theta_{12}^{\prime}, \phi^{\prime}, \delta^{\prime}; \widetilde{\alpha}_{\beta \gamma}^{\prime}, \lambda_{i}^{\prime} ) 
\left[ V^{(0)} ( \theta_{23}^{\prime}, \phi^{\prime}, \delta^{\prime} ) \right]^{\dagger} 
( 1 - \alpha^{\prime} )^{\dagger} 
%
\nonumber \\
&& 
\hspace{-12mm}
= \text{Rep(X)} _{ \text{\tiny helioP} } ( 1 - \alpha ) 
V^{(0)} ( \theta_{23}, \phi, \delta) R^{\dagger} 
\hat{H} ( \theta_{23}^{\prime}, \theta_{12}^{\prime}, \phi^{\prime}, \delta^{\prime}; \widetilde{\alpha}_{\beta \gamma}^{\prime}, \lambda_{i}^{\prime} ) 
R 
\left[ V^{(0)} ( \theta_{23}, \phi, \delta) \right]^{\dagger} 
( 1 - \alpha )^{\dagger} 
\text{Rep(X)} _{ \text{\tiny helioP} } ^{\dagger}.
\nonumber \\
\label{H-RHS-UV}
\end{eqnarray}
Note that $R$ is the ``untransformed'' matrix. 

\subsection{Symmetry IIIB as an example}
\label{sec:invariance-proof}

What we should do now is to verify that the relation 
\begin{eqnarray} 
&& 
R^{\dagger} 
\hat{H} ( \theta_{23}^{\prime}, \theta_{12}^{\prime}, \phi^{\prime}, \delta^{\prime}; \alpha_{\beta \gamma}^{\prime}, \lambda_{i}^{\prime} ) 
R 
= 
\hat{H} ( \theta_{23}, \theta_{12}, \phi, \delta; \alpha_{\beta \gamma}, \lambda_{i} )
\label{RGR}
\end{eqnarray} 
holds for all the sixteen symmetries, Symmetry X-helioP-UV where X = IA, IAf, $\cdot \cdot \cdot $, IVBf. It proves invariance of $H_{\text{\tiny Diag}}$ up to the rephasing factor $\text{Rep(X)} _{ \text{\tiny helioP} }$. 

To give the readers some feeling let us examine one example, the case of Symmetry IIIB, to show how the job is done. We restrict to the first order $\nu$SM and UV parts, as the proof for the $\hat{H}^{(0)}$ part can be done trivially. Using the solutions of the first condition in Table~\ref{tab:SF-solutions} and the transformation property of the $\nu$SM variables given in Table~\ref{tab:helioP-UV-symmetry}, the left-hand side of eq.~\eqref{RGR} can be written as 
\begin{eqnarray} 
&& 
\left[
\begin{array}{ccc}
0 & 0 & \pm e^{ - i \delta } \\
0 & 1 & 0 \\
\pm e^{ i \delta } & 0 & 0 \\
\end{array}
\right] 
\biggl\{
\epsilon c_{12} s_{12} 
\frac{\Delta m^2_{ \text{ren} }}{2E} 
\left[
\begin{array}{ccc}
0 & \pm s_{(\phi - \theta_{13})} & 0   \\
\pm s_{(\phi - \theta_{13})} & 0 & \pm c_{(\phi - \theta_{13})} e^{ - i \delta} \\
0  & \pm c_{(\phi - \theta_{13})} e^{ i \delta} & 0 
\end{array}
\right] 
\nonumber \\
&+&
\frac{b}{2E} 
\left[
\begin{array}{ccc}
0 & \pm e^{- i \delta} H_{32} & e^{- i \delta} H_{31} \\
\pm e^{ i \delta} H_{23} & 0 & \pm H_{21} \\
e^{ i \delta} H_{13} & \pm H_{12} & 0 \\
\end{array}
\right] 
\biggr\}
\left[
\begin{array}{ccc}
0 & 0 & \pm e^{ - i \delta } \\
0 & 1 & 0 \\
\pm e^{ i \delta } & 0 & 0 \\
\end{array}
\right]. 
\label{LHS-RGR}
\end{eqnarray} 
The $H_{ij}$ transformation property in eq.~\eqref{Hij-transf} is used for the second (EV) term. It is easy to calculate the entity in eq.~\eqref{LHS-RGR} to show that it is identical to the first order term in $\hat{H}$ given in eq.~\eqref{hatH-0th-1st-helioUV}. Therefore, eq.~\eqref{RGR} holds for Symmetry IIIB-helioP-UV. 

What is remarkable is that the equality in eq.~\eqref{RGR} can be shown to hold for all the Symmetry X-helioP-UV, where X=IA, IAf, $\cdot \cdot \cdot $, IVBf. It means that $H_{\text{\tiny Diag}}$ transforms under Symmetry X as 
\begin{eqnarray} 
&& 
H_{\text{\tiny Diag}} 
\rightarrow 
\text{Rep(X)} _{ \text{\tiny helioP} } 
H_{\text{\tiny Diag}} 
\text{Rep(X)} _{ \text{\tiny helioP} } ^{\dagger}.
\label{H-RHS-UV-invariance}
\end{eqnarray}
That is, $H_{\text{\tiny Diag}}$ is invariant apart from the rephasing factors $\text{Rep(X)} _{ \text{\tiny helioP} }$ and $\text{Rep(X)} _{ \text{\tiny helioP} }^{\dagger}$. Notice again that $\text{Rep(X)} _{ \text{\tiny helioP} }$ is rooted in the $\nu$SM, see eq.~\eqref{identity-helioP}, but also governs the UV part of the theory. 

\subsection{Including the second-order UV effect} 
\label{sec:2nd-order} 

Now let us include the second-order UV effect into our proof of invariance. 
Let us define the second-order $G$ matrix as in eq.~\eqref{G-def}, 
\begin{eqnarray} 
&& 
\hat{H}_\text{ UV }^{(2)} = 
\frac{b}{2E} 
U_{13} (\phi, \delta) ^{\dagger} 
U_{23}^{\dagger} 
A^{(2)} U_{23} 
U_{13} (\phi, \delta) 
\equiv \frac{b}{2E} G^{(2)},
\label{G2-def}
\end{eqnarray}
and define $H^{(2)}$ matrix to parametrize $G^{(2)}$ matrix by replacing $H_{ij}$ by $H_{ij}^{(2)}$ in \eqref {H-def-from-G}. One can easily show by using the UV $\alpha$ parameter transformation property given in Table~\ref{tab:helioP-UV-symmetry} that the transformation property, i.e. the sign flipping pattern, of the $A^{(2)}$ matrix is exactly identical to that of $A$. It means that the transformation property of $H_{ij}^{(2)}$ is the same as that of $H_{ij}$ given in eq.~\eqref{Hij-transf}. 
Since inclusion of the second order UV term merely changes $H_{ij}$ to $H_{ij} - H_{ij}^{(2)}$ in eq.~\eqref{RGR}, and their transformation properties are the same, the invariance proof given in section~\ref{sec:invariance-proof} remains valid with inclusion of the second order UV effect. 

To summarize, we have shown in this section that the flavor basis Hamiltonian $H_{ \text{flavor} }$, the both $H_{\text{\tiny VM}}$ and $H_{\text{\tiny Diag}}$, transforms as $H_{ \text{flavor} } \rightarrow \text{Rep(X)} _{ \text{\tiny helioP} } H_{ \text{flavor} } \text{Rep(X)} _{ \text{\tiny helioP} } ^{\dagger}$ under Symmetry X-helioP-UV, where X=IA, IAf, IB, $\cdot \cdot \cdot $, IVBf. This establishes the property of Symmetry X as the Hamiltonian symmetry which holds in all-orders in the helioP-UV perturbation theory in all the oscillation channels. 

\section{Conclusion and discussions}
\label{sec:conclusion}

In this paper we tried to update and summarize the present status of our knowledge and understanding of the reparametrization symmetry in neutrino oscillation in matter. We have introduced and used a systematic method called Symmetry Finder (SF)~\cite{Minakata:2021dqh,Minakata:2021goi,Minakata:2022zua} to identify the symmetries and investigate their characteristic features in the several theories. A ``success and failure'' record in our symmetry search may be summarized as follows: 
\begin{itemize} 

\item 
In the $\nu$SM: 
The eight 1-2 state exchange symmetries are uncovered both in the SRP (solar-resonance perturbation) theory, see Table~\ref{tab:SRP-symmetry} in section~\ref{sec:Symmetry-SRP}, and the DMP perturbation theory, see Table~\ref{tab:DMP-UV-symmetry} and ref.~\cite{Minakata:2021dqh}. Similarly, the sixteen 1-3 state exchange symmetries are identified in the helio-perturbation theory~\cite{Minakata:2021goi}. In spite of the ``globally valid'' nature of the framework, no 1-3 exchange symmetry is identified in DMP. 

\item 
In UV(unitarity violation)-extended theories of the $\nu$SM: 
In the helio-UV and DMP-UV perturbation theories, the $\nu$SM symmetry in each theory is elevated to the UV-extended one with the additional transformations on the UV sector $\alpha$ matrix, $\alpha \rightarrow \text{Rep(X)} \alpha \text{Rep(X)} ^{\dagger}$, where Rep(X) denotes the rephasing matrix. 
The number and the state exchange type of the symmetry are kept the same as those of the corresponding $\nu$SM theory.
For the procedures and results, see sections~\ref{sec:helio-UV-P} to \ref{sec:hamiltonian-symmetry} and Table~\ref{tab:helioP-UV-symmetry} for the helio-UV symmetries, and ref.~\cite{Minakata:2022zua} and Table~\ref{tab:DMP-UV-symmetry} for the DMP-UV ones. 

\end{itemize}

We note that picture of the reparametrization symmetry is transparent in the locally valid theories. The regions of validity of SRP and the helio-perturbation theories are at around the solar and atmospheric resonances, respectively. Correspondingly they have the 1-2 and 1-3 state exchange symmetries, respectively, reflecting the main players in each region. However, it appears that this simple picture does not apply to the globally valid DMP perturbation theory. Though the framework can describe both the solar and atmospheric resonances and the 1-2 state exchange symmetry is identified~\cite{Minakata:2021dqh}, we were not able to pin down where is the 1-3 state exchange symmetry in DMP. 

As it stands, the field of reparametrization symmetry in neutrino oscillation is still in its infancy with only less than two years period of the SF search. Reflecting this status, our current understanding of the symmetry is immature in many ways. At this moment the symmetry can be discussed for a given particular neutrino oscillation framework. That is, we cannot identify ``general symmetry'' for the generic flavor-basis, or mass-basis, Hamiltonian in matter. See, however, ref.~\cite{Denton:2021vtf} for an alternative approach with possible relevance to this point. We must keep in mind that the development of the field in the future may bring us to a new unexpected regime of understanding of neutrino oscillation physics. Certainly, it is still too premature to ask what is the ultimate goal of the symmetry approach. 

What is new in this paper? In Part I, the SF analysis of the SRP theory with the self-contained $V$ matrix treatment is new. 
In Part II, the symmetry analysis in the helio-UV perturbation theory using the SF framework and the recognition of the ``key identity'' are all new. Together with the similar analysis in ref.~\cite{Minakata:2022zua} for the DMP-UV perturbation theory, each exercise offers an important consistency check to each other for everything we have learnt from the both theories, and hence it is important to carry through. 

Yet, the penetrating theme throughout this paper is to convey to the readers our state of the art understanding of the symmetry in neutrino oscillation. The summary of the obtained reparametrization symmetry so far can be found in Table~\ref{tab:DMP-UV-symmetry} for the DMP and DMP-UV perturbation theories, Table~\ref{tab:SRP-symmetry} for the SRP theory, Table~\ref{tab:helioP-UV-symmetry} for the helio- and helio-UV perturbation theories. 

\subsection{The reparametrization symmetry as a diagnostics tool} 
\label{sec:diagnostics} 

In Part II of this article and in ref.~\cite{Minakata:2022zua}, we have made an intriguing proposal: Reparametrization symmetry can be used for diagnosing neutrino theory with non-unitarity. There is a clear indication for such possibility. We have observed in ref.~\cite{Minakata:2021nii} (but not mentioned for reasons explained in arXiv v1 of ref.~\cite{Minakata:2022zua}) that the oscillation probability given in the UV extended DMP theory possesses the $\nu$SM symmetries called Symmetry IA- and IB-DMP, see Table~\ref{tab:DMP-UV-symmetry}. Importantly, these symmetries are not accompanied by the UV $\alpha$ parameters' transformation, which implies that a part of the reparametrization symmetries distinguishes between the $\nu$SM and UV variables. 

By performing the SF analyses of the DMP-UV perturbation theory in ref.~\cite{Minakata:2022zua}, and the helio-UV theory in this paper, we have confirmed that (1) the above mentioned property of Symmetry IA and IB is reproduced by the SF formalism, and (2) the remaining six symmetries IIA, IIB, $\cdot \cdot \cdot $, IVB in the DMP-UV theory, and the similar twelve symmetries in the helio-UV theory, do have the associated $\alpha$ transformations, respectively, as reported in ref.~\cite{Minakata:2022zua} and section~\ref{sec:SF-solution-helioP-UV}. Therefore, the reparametrization symmetry as a whole can recognizes and distinguishes the $\nu$SM and the UV sectors of the theory. 

In fact, the $\alpha$ parameters' transformation under Symmetry X has quite interesting features. It is governed solely by the rephasing matrix, 
$\widetilde{\alpha} ^{\prime} = \text{Rep(X)} _{ \text{\tiny DMP} } \widetilde{\alpha} \text{Rep(X)} _{ \text{\tiny DMP} }^{\dagger}$ in DMP,\footnote{
$\widetilde{\alpha}$ denotes the $\alpha$ matrix in the SOL convention, see section~\ref{sec:PDG-SOL}. } 
and $\alpha^{\prime} = \text{Rep(X)} _{ \text{\tiny helioP} } \alpha \text{Rep(X)} _{ \text{\tiny helioP} }^{\dagger}$ in the helio-UV theories. $\text{Rep(X)} _{ \text{\tiny helioP} }$ differs from $\text{Rep(X)} _{ \text{\tiny DMP} }$ only by reshuffling of Rep(X), $\text{Rep(II)} \leftrightarrow \text{Rep(IV)}$ between the helioP and DMP. See eqs.~\eqref{rephasing-matrix} and \eqref{Rep-II-III-IV-DMP}. In the both DMP and the helio-perturbation theories Rep(X) is the diagonal matrix with elements $e^{ \pm i \pi}$, the constant matrices. This is a very different transformation property from the ones of the $\nu$SM variables, which can be described as the ``discrete rotations''. 

\subsection{A conjecture for the larger symmetries} 
\label{sec:larger symmetries} 

Most probably, the most important outcome in the symmetry discussion in the $\nu$SM and its UV extension is the key identity $V^{(0)} (\Phi^{\prime}) R [V^{(0)} (\Phi) ]^{\dagger} =$ Rep(X), the helioP version in eq.~\eqref{identity-helioP} and the DMP version in eq.~\eqref{symmetry-charge} (see below), where $V^{(0)}$ denotes the zeroth-order $V$ matrix with $\Phi$ its arguments in the collective notation. Remember that the identity plays several key roles, which include determining the $\alpha$ parameters' transformation properties and offering a new path for the Hamiltonian proof of the symmetry. 

We have conjectured in ref.~\cite{Minakata:2022zua} that the whole body of the reparametrization symmetry is much larger than that we saw in the above summary. Notice that the left-hand side of the identity involves $V^{(0)} (\Phi)$ before and after the transformation, the generic quantity in a given theory. We see no obvious dependences on the types of the state exchange in it apart from the particular form of $R$ specific to our case. The right-hand sides of the identity is just a constant. It naturally leads to the conjecture that by generalizing $R$ choice the key identity accommodates a generic class of discrete rotations of the $\nu$SM variables. If true, it would solve the issue of missing 13 exchange symmetry when applied to DMP. 

\subsection{The key identity and its possible topological nature}
\label{sec:identity-DMP} 

Now, the remaining important question is the interpretation of the constant and phase-sensitive nature of Rep(X). For this purpose, let us go to the identity (its complex conjugate) in DMP~\cite{Minakata:2022zua} for definiteness 
\begin{eqnarray} 
&& 
V^{(0)} ( \theta_{23}, \psi, \phi, \delta) R^{\dagger} 
V^{(0)} ( \theta_{23}^{\prime}, \psi^{\prime}, \phi^{\prime}, \delta + \xi)^{\dagger}
= \text{Rep(X)}_{ \text{\tiny DMP} }^{\dagger}, 
\label{symmetry-charge}
\end{eqnarray}
where $R$ denotes the $R$ matrix in DMP and $V^{(0)} ( \theta_{23}, \psi, \phi, \delta)$ the zeroth order $V$ matrix. While we discuss here eq.~\eqref{symmetry-charge} in DMP, the similar identity exists in the helio-UV theory, eq.~\eqref{identity-helioP}, and our consideration below must apply it as well. 

The identity indeed reveals a quite interesting feature as noticed above. Despite that the left-hand side displays rich dependences of the untransformed and transformed $\nu$SM variables, the right-hand side consists of the constant elements $\pm 1 = e^{ \pm i \pi}$, whose character may suggest a topological origin of the identity. Since the way we understand it could affect our interpretation of the reparametrization symmetry, let us address this issue. We try to argue below that the left-hand side of eq.~\eqref{symmetry-charge} can be regarded as the symmetry charge. Nonetheless, we must say that our consideration below may still be at a speculative level. 

In $U(1)$ gauge theory with complex scalar field $\varphi$, the symmetry charge can be calculated as $Q = \int d^3x \pi \delta \varphi$, where $\delta \varphi$ denotes a variation of the field under an infinitesimal $U(1)$ transformation, $\varphi \rightarrow \varphi - i \epsilon \varphi$, and $\pi$ is the canonical conjugate of $\varphi$~\cite{Itzykson:1980rh}. $\epsilon$ displays an infinitesimal nature of the transformation, and is to be removed when we define $\delta \varphi$. While there is no reason to expect the $U(1)$ charge to be quantized, if one calculates $Q$ around the vortex solution, it indeed is quantized to an integer times the unit of charge, the Nielsen-Olesen vortex~\cite{Nielsen:1973cs}. The quantization of the scalar charge around the vortex comes from the nontrivial homotopy $\pi _{1} \left( S^1 \right) = Z$~\cite{Coleman:1985rnk}. 

Now we try to interpret the identity~\eqref{symmetry-charge} along the similar line of thought. We consider that the basic elements of the transformation is given by $[V^{(0)} ( \theta_{23}, \psi, \phi, \delta)]^{\dagger}$. Then, the quantity corresponding to $\delta \varphi$ in scalar field case is $R^{\dagger} V^{(0)} ( \theta_{23}^{\prime}, \psi^{\prime}, \phi^{\prime}, \delta + \xi)^{\dagger}$, because the subtracted untransformed part $R^{\dagger} V^{(0)} ( \theta_{23}, \psi, \phi, \delta)^{\dagger}$ gives no contribution. We must leave $R^{\dagger}$, our $\epsilon$ equivalent, because it is not small but order unity and performs state exchange. The important difference between our case and the $U(1)$ charge is that we now treat the discrete symmetry, not continuous one. The integration over space coordinate $x$ is absent because this is quantum mechanics, or zero-dimensional field theory. Lacking knowledges by the author of field theory of discrete symmetry, we cannot prove that the ``canonical conjugate'' is given by $V^{(0)} ( \theta_{23}, \psi, \phi, \delta) $, but it is at least not unnatural. 
Despite that we do not know whether the similar reasoning exists for the discrete group to guarantee integral property of $\text{Rep(X)} _{ \text{\tiny DMP} }$, such as homotopy in the vortex case, it appears to the author that it is legitimate to leave it as a conjecture given its intriguing feature and the practical utility. We believe that this point deserves further investigation. 

\appendix 

\section{Three useful conventions of the lepton flavor mixing matrix}
\label{sec:3-conventions} 

We start from the most commonly used form, the PDG convention~\cite{Zyla:2020zbs} of the $U$ matrix defined in eq.~\eqref{MNS-PDG}. Recently, we have started to use the two other conventions called the ``SOL'' and the ``ATM'' which differ only by the phase redefinitions from $U_{\text{\tiny PDG}}$. $U_{\text{\tiny SOL}}$ is defined in eq.~\eqref{SOL-def}. $U_{\text{\tiny ATM}}$ is defined by 
\begin{eqnarray} 
&&
\hspace{-2mm}
U_{\text{\tiny ATM}} 
\equiv 
\left[
\begin{array}{ccc}
1 & 0 &  0  \\
0 & 1 & 0 \\
0 & 0 & e^{ - i \delta} \\
\end{array}
\right] 
U_{\text{\tiny PDG}} 
\left[
\begin{array}{ccc}
1 & 0 &  0  \\
0 & 1 & 0 \\
0 & 0 & e^{ i \delta} \\
\end{array}
\right] 
=
\left[
\begin{array}{ccc}
1 & 0 &  0  \\
0 & c_{23} & s_{23} e^{ i \delta} \\
0 & - s_{23} e^{- i \delta} & c_{23} \\
\end{array}
\right] 
\left[
\begin{array}{ccc}
c_{13}  & 0 & s_{13}  \\
0 & 1 & 0 \\
- s_{13} & 0 & c_{13} \\
\end{array}
\right] 
\left[
\begin{array}{ccc}
c_{12} & s_{12}  &  0  \\
- s_{12} & c_{12} & 0 \\
0 & 0 & 1 \\
\end{array}
\right]. 
\nonumber \\
\label{ATM-def} 
\end{eqnarray}
The reason for our terminology of $U_{\text{\tiny ATM}}$ and $U_{\text{\tiny SOL}}$ is because the CP phase factor $e^{ \pm i \delta}$ is attached to (sine of) the ``atmospheric angle'' $\theta_{23}$ in $U_{\text{\tiny ATM}}$, and to the ``solar angle'' $\theta_{12}$ in $U_{\text{\tiny SOL}}$, respectively. In the PDG convention 
$e^{ \pm i \delta}$ is attached to $s_{13}$. $U_{\text{\tiny ATM}}$ is used to compute the probability e.g., in refs.~\cite{Minakata:2015gra,Denton:2016wmg,Martinez-Soler:2018lcy,Martinez-Soler:2019nhb}. $U_{\text{\tiny SOL}}$ is used for the same purpose in refs.~\cite{Martinez-Soler:2019noy,Minakata:2021nii}. 

It should be remembered that the oscillation probability calculated by using the PDG, ATM, and the SOL conventions is exactly identical. It is because the phase redefinition cannot alter the physical observables. Therefore, the measured values of the mixing angles and CP phase does not depend which convention is used for the $U$ matrix to compute the probability. 

On the other hand, the $\alpha$ matrix is $U$ matrix convention dependent. Once the phase convention of the $U$ matrix is changed from $U_{\text{\tiny PDG}}$ to $U_{\text{\tiny SOL}}$, a consistent definition of $N_{\text{\tiny SOL}}$ requires the $\alpha$ matrix to transform~\cite{Martinez-Soler:2018lcy}, as can be seen in 
\begin{eqnarray} 
N_{\text{\tiny SOL}} 
&\equiv& 
\left[
\begin{array}{ccc}
1 & 0 &  0  \\
0 & e^{ - i \delta} & 0 \\
0 & 0 & e^{ - i \delta} \\
\end{array}
\right] 
N_{\text{\tiny PDG}} 
\left[
\begin{array}{ccc}
1 & 0 &  0  \\
0 & e^{ i \delta} & 0 \\
0 & 0 & e^{ i \delta} \\
\end{array}
\right] 
= \left\{ 
\bf{1} - 
\left[
\begin{array}{ccc}
1 & 0 &  0  \\
0 & e^{ - i \delta} & 0 \\
0 & 0 & e^{ - i \delta} \\
\end{array}
\right] 
\alpha 
\left[
\begin{array}{ccc}
1 & 0 &  0  \\
0 & e^{ i \delta} & 0 \\
0 & 0 & e^{ i \delta} \\
\end{array}
\right] 
\right\}
U_{\text{\tiny SOL}} 
\nonumber \\
&\equiv& 
\left( \bf{1} - \alpha^{\text{\tiny SOL}}  \right) U_{\text{\tiny SOL}}.
\label{alpha-SOL}
\end{eqnarray}
Therefore, the $\alpha$ matrix is convention dependent. It takes the form in the SOL convention as 
\begin{eqnarray} 
&&
\alpha^{\text{\tiny SOL}} 
= 
\left[
\begin{array}{ccc}
1 & 0 &  0  \\
0 & e^{ - i \delta} & 0 \\
0 & 0 & e^{ - i \delta} \\
\end{array}
\right] 
\alpha 
\left[
\begin{array}{ccc}
1 & 0 &  0  \\
0 & e^{ i \delta} & 0 \\
0 & 0 & e^{ i \delta} \\
\end{array}
\right] 
=
\left[ 
\begin{array}{ccc}
\alpha_{ee} & 0 & 0 \\
e^{ - i \delta} \alpha_{\mu e} & \alpha_{\mu \mu}  & 0 \\
e^{ - i \delta} \alpha_{\tau e}  & \alpha_{\tau \mu} & \alpha_{\tau \tau} \\
\end{array}
\right] 
\equiv
\left[ 
\begin{array}{ccc}
\widetilde{\alpha}_{ee} & 0 & 0 \\
\widetilde{\alpha}_{\mu e} & \widetilde{\alpha}_{\mu \mu}  & 0 \\
\widetilde{\alpha}_{\tau e}  & \widetilde{\alpha}_{\tau \mu} & \widetilde{\alpha}_{\tau \tau} \\
\end{array}
\right],~~~
\label{alpha-SOL-def}
\end{eqnarray}
where we have introduced the simplified notation $\alpha_{\beta \gamma}^{\text{\tiny SOL}} \equiv \widetilde{\alpha}_{\beta \gamma}$ for convenience. Similarly, we have for the ATM convention 
\begin{eqnarray} 
&&
\hspace{-10mm}
\alpha^{\text{\tiny ATM}} 
= 
\left[
\begin{array}{ccc}
1 & 0 &  0  \\
0 & 1 & 0 \\
0 & 0 & e^{ - i \delta} \\
\end{array}
\right] 
\alpha 
\left[
\begin{array}{ccc}
1 & 0 &  0  \\
0 & 1 & 0 \\
0 & 0 & e^{ i \delta} \\
\end{array}
\right] 
=
\left[ 
\begin{array}{ccc}
\alpha_{ee} & 0 & 0 \\
\alpha_{\mu e} & \alpha_{\mu \mu}  & 0 \\
e^{ - i \delta} \alpha_{\tau e} & e^{ - i \delta} \alpha_{\tau \mu} & \alpha_{\tau \tau} \\
\end{array}
\right]
\equiv 
\left[ 
\begin{array}{ccc}
\alpha_{ee}^{\text{\tiny ATM}} & 0 & 0 \\
\alpha_{\mu e}^{\text{\tiny ATM}} & \alpha_{\mu \mu}^{\text{\tiny ATM}} & 0 \\
\alpha_{\tau e}^{\text{\tiny ATM}} & \alpha_{\tau \mu}^{\text{\tiny ATM}} & \alpha_{\tau \tau}^{\text{\tiny ATM}} \\
\end{array}
\right]. ~~~
\label{alpha-ATM-def}
\end{eqnarray} 

\section{DMP-UV symmetry: A brief summary}
\label{sec:DMP-UV-summary}  

To show the difference between the helio-UV and the DMP-UV symmetries we recollect just two equations from ref.~\cite{Minakata:2022zua}. In the DMP-UV theory we have $H_{ \text{VM} }$ and $H_{ \text{Diag} }$ (using our new notations) similar to the ones in eqs.~\eqref{H-LHS} and \eqref{H-RHS}. See eqs.~(9.2) and (9.5) in ref.~\cite{Minakata:2022zua}. One can show that the both $H_{ \text{VM} }$ and $H_{ \text{Diag} }$  transform under Symmetry X as
\begin{eqnarray} 
&& 
H_{\text{\tiny VM}} 
\rightarrow 
\text{Rep(X)} _{ \text{\tiny DMP} } 
H_{\text{\tiny VM}} 
\text{Rep(X)} _{ \text{\tiny DMP} } ^{\dagger}, 
\nonumber \\
&&
H_{\text{\tiny Diag}} 
\rightarrow 
\text{Rep(X)} _{ \text{\tiny DMP} } 
H_{\text{\tiny Diag}} 
\text{Rep(X)} _{ \text{\tiny DMP} } ^{\dagger}, 
\label{H-LRHS-invariance}
\end{eqnarray}
where the rephasing matrix Rep(X)$_{ \text{\tiny DMP} }$  is given by Rep(I)$_{ \text{\tiny DMP} }$ = diag (1,1,1), and 
\begin{eqnarray} 
&&
\text{Rep(II)} _{ \text{\tiny DMP} } = 
\left[
\begin{array}{ccc}
1 & 0 & 0 \\
0 & -1 & 0 \\
0 & 0 & 1 \\
\end{array}
\right],
\hspace{6mm}
\text{Rep(III)} _{ \text{\tiny DMP} } = 
\left[
\begin{array}{ccc}
- 1 & 0 & 0 \\
0 & 1 & 0 \\
0 & 0 & 1
\end{array}
\right], 
\hspace{6mm}
\text{Rep(IV)} _{ \text{\tiny DMP} } = 
\left[
\begin{array}{ccc}
- 1 & 0 & 0 \\
0 & -1 & 0 \\
0 & 0 & 1
\end{array}
\right].
\nonumber \\
\label{Rep-II-III-IV-DMP} 
\end{eqnarray}
Notice that 
$\text{Rep(II)} _{ \text{\tiny DMP} } = \text{Rep(IV)} _{ \text{\tiny helioP} }$ (up to overall sign), 
$\text{Rep(IV)} _{ \text{\tiny DMP} } = \text{Rep(II)} _{ \text{\tiny helioP} }$ (up to overall sign), and 
$\text{Rep(III)} _{ \text{\tiny DMP} } = \text{Rep(III)} _{ \text{\tiny helioP} }$. 
Since our classification scheme of Symmetry X, X=I, II, III, IV, is based on the solutions of the first condition, which are universal among DMP, SRP, and the helio-perturbation theories, we do not exchange our definitions of the Symmetry II and IV in the helio- and helio-UV perturbation theories. 
The resulting $\widetilde{\alpha}$ parameter transformation property is given in the fourth column of Table~\ref{tab:DMP-UV-symmetry}.

\section{$H$ matrix elements}
\label{sec:H-elements}

The hermitian $H$ matrix is defined in eq.~\eqref{H-def-from-G}. The explicit expressions of its elements are given by 
\begin{eqnarray} 
H_{11} 
&=&
2 c^2_{\phi} 
\alpha_{ee} \left( 1 - \frac{ \Delta_{a} }{ \Delta_{b} } \right) 
+ 2 s^2_{\phi} 
\left[ s_{23}^2 \alpha_{\mu \mu} + c_{23}^2 \alpha_{\tau \tau} + c_{23} s_{23} \mbox{Re} \left( \alpha_{\tau \mu} \right) \right] 
\nonumber \\ 
&-& 
2 c_{\phi} s_{\phi} 
\left[ 
s_{23} \mbox{Re} \left( e^{ - i \delta} \alpha_{\mu e} \right) 
+ c_{23} \mbox{Re} \left( e^{ - i \delta} \alpha_{\tau e} \right) 
\right], 
\nonumber \\
H_{22} 
&=& 
2 \left[ c^2_{23} \alpha_{\mu \mu} + s^2_{23} \alpha_{\tau \tau} 
- c_{23} s_{23} \mbox{Re} \left( \alpha_{\tau \mu} \right) \right], 
\nonumber \\
H_{33} 
&=& 
2 s^2_{\phi} 
\alpha_{ee} \left( 1 - \frac{ \Delta_{a} }{ \Delta_{b} } \right) 
+ 2 c^2_{\phi} 
\left[ s_{23}^2 \alpha_{\mu \mu} + c_{23}^2 \alpha_{\tau \tau} + c_{23} s_{23} \mbox{Re} \left( \alpha_{\tau \mu} \right) \right] 
\nonumber \\
&+&
2 c_{\phi} s_{\phi} \left[ 
s_{23} \mbox{Re} \left( \alpha_{\mu e} e^{- i \delta} \right) 
+ c_{23} \mbox{Re} \left( \alpha_{\tau e} e^{- i \delta} \right) 
\right], 
\nonumber 
\end{eqnarray}
\begin{eqnarray} 
H_{21} 
&=& 
\left[ c_{23} c_{\phi} \left( \alpha_{\mu e} e^{- i \delta} \right) 
- s_{23} c_{\phi} \left( \alpha_{\tau e} e^{- i \delta} \right) 
- s_{\phi} 
\left\{ c^2_{23} \alpha_{\tau \mu}^* - s^2_{23} \alpha_{\tau \mu}
+ 2 c_{23} s_{23} ( \alpha_{\mu \mu} - \alpha_{\tau \tau} ) \right\} \right] 
= H_{12}^*, 
\nonumber \\
H_{13} 
&=& 
\biggl[
2 c_{\phi} s_{\phi} 
\left\{ \alpha_{ee} \left( 1 - \frac{ \Delta_{a} }{ \Delta_{b} } \right) 
- \left[ s_{23}^2 \alpha_{\mu \mu} + c_{23}^2 \alpha_{\tau \tau} + c_{23} s_{23} \mbox{Re} \left( \alpha_{\tau \mu} \right) \right] 
\right\} 
\nonumber \\
&+& 
\cos 2\phi \left\{ 
s_{23} \mbox{Re} \left( \alpha_{\mu e} e^{- i \delta} \right) 
+ c_{23} \mbox{Re} \left( \alpha_{\tau e} e^{- i \delta} \right) \right\} 
- i \left\{ s_{23} \mbox{Im} \left( \alpha_{\mu e} e^{- i \delta} \right) 
+ c_{23} \mbox{Im} \left( \alpha_{\tau e} e^{- i \delta} \right) \right\} 
\biggr] 
= H_{31}^*,  
\nonumber \\
H_{23} 
&=& 
\left[ c_{23} s_{\phi} \left( \alpha_{\mu e} e^{- i \delta} \right) 
- s_{23} s_{\phi} \left( \alpha_{\tau e} e^{- i \delta} \right) 
+ c_{\phi}  
\left\{ c^2_{23} \alpha_{\tau \mu}^* - s^2_{23} \alpha_{\tau \mu} 
+ 2 c_{23} s_{23} ( \alpha_{\mu \mu} - \alpha_{\tau \tau} ) \right\} \right] 
= H_{32}^*. 
\label{Hij-explicit}
\end{eqnarray}

\end{document}